\documentclass[12pt,notitlepage,fleqn]{article}
\usepackage{graphicx}
\usepackage{amsmath}
\usepackage{indentfirst}
\usepackage{amsfonts}
\usepackage{amssymb}
\newtheorem{problem}{Hypothesis}

\begin{document}

\begin{center}
{\LARGE The Physical Foundation of the Quark Model}

The Quark Model as an approximation of the BCC Model

\bigskip

{\normalsize Jiao-Lin Xu}

\bigskip

{\small The Center for Simulational Physics, The Department of Physics and Astronomy}

{\small University of Georgia, Athens, GA 30602, USA.}

E- mail: {\small \ jxu@hal.physast.uga.edu}

\bigskip

\textbf{Abstract}
\end{center}

{\small From the Dirac sea concept,} {\small the BCC model infers that the
quarks u and d constitute a body center cubic quark lattice in the vacuum;
when a quark q}$^{\ast}$ {\small is excited from the vacuum, the nearest
primitive cell (u}$^{\prime}$ and d$^{\prime}$) {\small is accompanying
excited by the quark} {\small q}$^{\ast}$. {\small Using the energy band
theory, the model deduces the quantum numbers (I, S, C, b, and Q) and the
masses of all quarks using a united mass formula. Then, it shows that the
system of} {\small 3 excited quarks\ (q}$^{\ast}u^{\prime}d^{\prime}${\small )
is a baryon, and it deduces the baryon spectrum in terms of the sum laws. This
theoretical baryon spectrum is in accordance with the experimental results. It
also shows that there are only two elementary quarks (u and d), while the
other quarks (s, c, b, ...) are the excited states of the elementary quarks,
hence the SU(3) (u, d, and s), the SU(4) (u, d, s, and c), and the SU(5) (u,
d, s, c, and b) are the natural extensions of the SU(2) (u and d).\ The BCC
model\ \textbf{provides the physical foundation (quarks, SU(N) groups, and
that a baryon is made of 3 quarks) for the Quark Model. The Quark Model is the
SU(N) approximation of the BCC model.}\ The confinement concept is not needed
in the BCC model, because it is replaced by the accompanying excitation
concept. The SU(N)} groups{\small \ are also not} {\small necessary, as they
are replaced by the body center cubic groups. We also predict some new
baryons: }$\Lambda(2560), $ $\Sigma_{C}(2280),$ {\small \ }$\Omega^{-}%
(3720)${\small , }$\Lambda_{C}^{+}(6600)${\small , }$\Lambda_{b}^{0}%
(9960)${\small ... }\newpage

\section{Introduction}

The Quark Model \cite{QuarkModel} has already explained the baryon spectrum
and the meson spectrum in terms of the quarks. It successfully gives intrinsic
quantum numbers ($I$, $S$, $C$, $b$, and $Q$) to all baryons and mesons.
However, (1) it has not given a satisfactory mass spectrum of baryons and
mesons in a united mass formula \cite{QMFORM}. (2) The intrinsic quantum
numbers and masses of the quarks are all entered ``by hand'' \cite{Hand in}.
(3) It needs too many quarks (elementary particles: 6 flavors $\times$3
colors$\times$2 (quark and antiquark)= 36 quarks). (4) It needs too many
parameters, just as T. D. Lee has already noticed \cite{Vacuum engineering} :
``The standard model... needs $\sim$ 20 parameters: e, G, $\theta_{w},$
various masses for the three generations of leptons and quarks and the four
weak decay angles $\theta_{1},$ $\theta_{2},$ $\theta_{3,}$ and $\delta$''.
(5) Confinement is a very plausible idea but to date its rigorous proof
remains outstanding \cite{CONFINEMENT}. Meanwhile all free quark searches
since 1977 have had negative results \cite{Free QUARK}. Therefore, although
the Quark Model has a strong mathematical foundation (SU(3), SU(4), ...), its
physical foundation is not sufficient. This paper attempts
to{\small \textbf{\ provide a physical foundation for the quark model. }\ }

First, we need a mechanism.

Twenty years ago, T. D. Lee pointed out \cite{T. D. Lee}:

``We believe our vacuum, though Lorentz invariant, to be quite complicated.
Like any other physical medium, it can carry long-range-order parameters and
it may also undergo phase transitions...'' .

Recently Frank Wilczek, the J. Robert Oppenheimer Professor at the Institute
for Advanced Study in Princeton, pointed out \cite{wilczek}:

``Empty space-the vacuum-is in reality a richly structured, though highly
symmetrical, medium. Dirac's sea was an early indication of this feature,
which is deeply embedded in quantum field theory and the Standard Model.
Because the vacuum is a complicated material governed by locality and
symmetry, one can learn how to analyze it by studying other such
materials-that is, condensed matter.''

Professor Wilczek not only pointed out one of the most important and most
urgent research directions of the modern physics-studying the structure of the
vacuum, but also provided a very practical and efficient way for the
study-learning from studying condensed matter.

Applying the Lee-Wilczek idea, this paper conjectures a structure of the
vacuum (body center cubic quark lattice), which will be used as the mechanism
to generate the quark spectrum, the baryon spectrum, and the meson spectrum.

According to Dirac's sea concept \cite{diarcsea}, there is an electron Dirac
sea, a $\mu$ lepton Dirac sea, a $\tau$ lepton Dirac sea, a $u$ quark Dirac
sea, a $d$ quark Dirac sea, an $s$ quark Dirac sea, a $c$ quark Dirac sea, and
a $b$ quark Dirac sea... in the vacuum. All of these Dirac seas are in the
same space, at any location, that is, at any physical space point. These
particles will interact with one another and form the perfect vacuum material.
However, some kinds of particles do not play an important role in forming the
vacuum material. First, the main force which makes and holds the structure of
the vacuum material must be the strong interaction, not the
weak-electromagnetic interactions. Hence, in considering the structure of the
vacuum material, we leave out the Dirac seas of those particles which do not
have strong interactions ($e$, $\mu$, and $\tau$). Secondly, the vacuum
material is super stable, hence we also omit the Dirac seas which can only
make unstable baryons ($s$ quark, $c$ quark, $b$ quark). Finally, there are
only two kinds of possible particles left: the vacuum state $u$ quarks and the
vacuum state $d$ quarks. There are super strong attractive forces between the
$u$ quarks and the $d$ quarks which will make and hold the densest structure
of the vacuum material.

According to solid state physics \cite{Solidstates}, if two kinds of particles
(with radius $R_{1}<R_{2}$) satisfy the condition $1>R_{1}/R_{2}>0.73$, the
densest structure is the body center cubic crystal \cite{bodycenter}. We know
the following: first, $u$ quarks and $d$ quarks are not exactly the same, thus
$R_{u}\neq R_{d}$; second, they are very close to each other (the same isospin
with different $I_{z}$), thus $R_{u}\approx R_{d}$. Hence, if $R_{u}<R_{d}$
(or $R_{d}<R_{u}$), we have $1>R_{u}/R_{d}>0.73$ (or $1>R_{d}/R_{u}>0.73$).
Therefore, we conjecture that the vacuum state u quarks and d quarks construct
the body center cubic quark lattice in the vacuum (in this paper, it will be
regarded as \textbf{the BCC model}). \textbf{The BCC model will be the
physical foundation of the Quark Model. }

Since the system is a {\small multi}-particle system, we cannot solve the
problem exactly. We will study a series of approximations: a primitive cell
\cite{PrimitiveCell} approximation, a periodic field approximation
\cite{Periodic-F}, a combined approximation of the primitive cell and periodic
field, and an SU(N) symmetry (the Quark Model) approximation \cite{SU(N)}.

In the primitive cell approximation, we consider the excited quark and the
primitive cell \textbf{(}in which the excited quark is contained) only,
omitting the quark lattice. Although we can get some important results (such
as: a baryon is made of three quarks, the system satisfies SU(3) symmetry...),
we cannot deduce flavored quarks and flavored baryons.

In the periodic field approximation, further considering the quark lattice
periodic field (with body center cubic symmetry), we obtain the baryon
spectrum. However, we can not get the quark spectrum.

In the combined approximation of the primitive cell and periodic field, we can
\textbf{deduce the quark spectrum,} the baryon spectrum, and the meson spectrum.

In the SU(N) symmetry approximation, based on the results of the combined
approximation, we assume that the N ground quark excited states (with
different flavors) are independent quarks. The N quarks satisfy the
SU(N)\ symmetry and belong to the fundamental representation of the SU(N)
group. We can obtain similar results with the Quark Model \cite{QuarkModel}.

These approximation results show (1) quarks u and d are elementary particles,
other quarks (s, c, b...) are the energy band excited states of elementary
quarks; (2) u quarks and d quarks indeed construct the body center cubic quark
lattice in the vacuum; (3) flavored quarks originate from the symmetry of the
periodic field of the quark lattice; (4) the confinement concept is
unnecessary, since it can be replaced by the accompanying excitation concept;
(5) \textbf{the quark model is the SU(N) approximation of the BCC model; (6)
and the BCC model provides the physical foundation of the quark model.\ \ }

This paper is organized as follows: The fundamental hypotheses are presented
in \emph{Section II}. The primitive cell \cite{PrimitiveCell} approximation is
introduced in \emph{Section III}. The periodic field approximation
\cite{Periodic-F} is discussed in \emph{Section IV. }The combined
approximation of the primitive cell and the periodic field is accomplished in
\emph{Section V}. The SU(N)\ \cite{SU(N)} approximation (the Quark Model) is
introduced in \emph{Section VI}. A comparison of the results of the BCC model
with the experimental results is listed in \emph{Section VII}. The predictions
and discussions of the model are stated in \emph{Section VIII}. The
conclusions are in \emph{Section IX}.

\section{Fundamental Hypotheses}

First, the BCC model assumes that u quarks and d quarks are fundamental
particles and make a body center cubic quark lattice in the vacuum. Then, it
deduces the spectrum of excited quarks. Finally, it finds the spectrum of
baryons and mesons.

In order to explain the model accurately and concisely, we will start from the
fundamental hypotheses in an axiomatic form.

\begin{problem}
There are two kinds of elementary quarks in the vacuum state. They have the
same baryon number $B=1/3$, spin $s=1/2$, and isospin $I=1/2$. The quarks with
the third component of the isospin $I_{z}=+1/2$\ are called u quarks, and the
quarks with the third component of the isospin $I_{z}=-1/2$\ are called d
quarks. There are super strong attractive interactions among the quarks. The
super attractive forces make and hold \textbf{an infinite body center cubic
quark (u and d) lattice in the vacuum.}
\end{problem}

\begin{problem}
When a quark (q) is excited from the vacuum quark lattice (q$\rightarrow
q^{\ast})$, the primitive cell\ (in which the excited quark is contained) is
simultaneously excited by the excited quark $q^{\ast}$ also. Since there are
only two quarks, u and d, in the primitive cell, there are only 2 accompanying
excited quark, u$^{\prime}$\ and d$^{\prime},$\ in the primitive cell. We call
the excitation of the primitive cell the accompanying excitation.
\end{problem}

The accompanying excited quark u$^{\prime}$ (or d$^{\prime}$) has not left its
position in the quark lattice, so it is not free to move in the space.
However, its electric charge and baryon number are temporarily excited from
the vacuum state, under the influence of the excited quark q$^{\ast}$. Because
the excited energy of the electric charge is much smaller than the excited
energy of the quark and because it can not be separated from the excited
energy of the quark q$^{\ast}$ in the experiments, we assume that the
accompanying excitation energy is very small and can be treated as a small
perturbation energy. For the zeroth-order approximation, the accompanying
excited quark $u^{\prime}$ has
\begin{equation}
B=1/3,\text{ }S=0,\text{ }s=I=1/2,\text{ }I_{z}=1/2,\text{ }Q=2/3,\text{
}m_{u^{\prime}}=0;\label{Quantum-0f-u}%
\end{equation}
and the accompanying excited quark $d^{\prime}$ has
\begin{equation}
B=1/3,\text{ }S=0,s=I=1/2,\text{ }I_{z}=-1/2,\text{ }Q=-1/3,\text{
}m_{d^{\prime}}=0.\label{Quantum-0f-d}%
\end{equation}

The accompanying excitation is temporary for a cell. When the quark $q^{\ast}$
is excited from the vacuum, the primitive cell undergoes an accompanying
excitation which is due to the excited quark $q^{\ast}$; but when the quark
$q^{\ast}$ leaves the cell, the excitation of the cell disappears.\textbf{\ }%
Although the truly excited cells are quickly changed, one following another,
with the motion of the excited quark $q^{\ast}$, an excited primitive cell
(u$^{\prime}$ and d$^{\prime}$) always appears to accompany the excited quark
$q^{\ast},$ just as an electric field always accompanies the originated
electric charge.\textbf{\ }

\begin{problem}
Due to the effect of the vacuum quark lattice, fluctuations of energy
$\varepsilon${\LARGE \ }and intrinsic quantum numbers (such as the strange
number $S$) of an excited quark may exist. The fluctuation of the Strange
number, if it exists, is always $\Delta S=\pm1$ \cite{real value of S}. From
the fluctuation of the Strange number, we will be able to deduce new quantum
numbers, such as Charmed number $C$ and Bottom number $b$. \textbf{\ }
\end{problem}

For an excited quark q$^{\ast}$(which is accompanied by $u^{\prime}$ and
$d\prime)$ moving in the body center cubic quark lattice, the Hamiltonian can
be written as
\begin{equation}
H=H_{q^{\ast}}+H_{Cell}+H_{Latt}^{\prime}+H_{q^{\ast}-Cell}+H_{q^{\ast}%
-Latt}^{\prime}+H_{cell-Latt}^{\prime}\text{.}\label{H}%
\end{equation}
Where H$_{q^{\ast}}$ is the Hamiltonian of the excited quark q$^{\ast}$,
$H_{cell}$ is the Hamiltonian of the accompanying excited cell ($u\prime$ and
$d\prime$), H$_{Latt}^{\prime}$ is the Hamiltonian of the vacuum quark lattice
(excluding the accompanying cell), H$_{q^{\ast}-cell}$ is the interaction
Hamiltonian between the excited quark and the accompanying excited primitive
cell, $H_{q^{\ast}-Latt}$ is the interaction Hamiltonian between the excited
quark and the vacuum quark lattice (excluding the accompanying cell), and
$H_{Cell-Latt}^{\prime}$ is the interaction Hamiltonian between the
accompanying excited cell and the vacuum quark lattice (excluding the
accompanying cell).

Since we do not know the exact form of the strong interactions, we cannot
write out the expression of H. Furthermore, since the system is a
multi-particle system, even if we have the expression of H, we still cannot
solve the problem accurately. Thus, we have to use some approximations to
attack the problem. We will study a primitive cell
\ \ \ \ \ \cite{PrimitiveCell} approximation, a periodic field approximation
\cite{Periodic-F}, a combined approximation of the cell and the periodic
field, and an SU(N) symmetry approximation \cite{SU(N)}. According to the BCC
Model, the SU(N) approximation is the Quark Model. In other words, \textbf{the
Quark Model is an approximation of the BCC Model.}

First, we discuss the primitive cell approximation.

\ \ \ \ \ \ \ \ \ \ \ \ \ \ \ \ \ \ \ \ \ \ \ \ \ \ \ \ \ \ \ \ \ 

\section{The \textbf{Primitive Cell Approximation}}

In order to find a good approximation, we need to look at the whole system.
The system is made up of \textbf{the vacuum quark lattice and an excited quark
q}$^{\ast}$\textbf{\ with an accompanying excited primitive cell (}$u\prime
$\textbf{\ and }$d\prime$\textbf{).} The vacuum lattice forms the physical
vacuum background. The excited quark freely moves in the lattice, as an
electron moves in a superconductor. An accompanying excited cell is always
accompanying the excited quark, like an electric field accompanying the
originated electric charge. Since the interactions between the quarks are
short in range, the excited quark q$^{\ast}$ can excite only the nearest
quarks ($u\prime$ and $d\prime$) which are inside the primitive cell\textbf{.}
They cannot be separated, just as Coulomb's electric field cannot be separated
from the original electric charge. Therefore,\textbf{\ an observable
`particle' is not only an excited quark }$q^{\ast}$\textbf{, but a group of
three quarks (an excited quark q}$^{\ast}$\textbf{, an accompanying excited
quark u}$^{\prime}$\textbf{, and an accompanying excited quark d}$^{\prime}$\textbf{).}

The simplest approximation is \textbf{the primitive cell approximation}. In
this approximation, we consider the excited quark q$^{\ast}$ and the
accompanying excited primitive cell \textbf{(}$u\prime+d\prime$) only,
omitting the quark lattice. Thus, \textbf{there are only three quarks in the
system: the excited quark q}$^{\ast}$\textbf{\ and the two accompanying
excited quarks }$u^{\prime}$\textbf{\ and }$d^{\prime}$\textbf{. }We assume
that the quantum numbers and energy of the system are the sums of the quantum
numbers and energies of the constituent quarks \cite{Nonrelat-Model}%
\textbf{.\qquad}%

\begin{equation}
B=\sum B_{q},\ \text{\ }Q=\sum Q_{q,}\ \ I_{z}=\sum I_{zq},\text{ }M=\sum
m_{q}.\label{SUM-formula}%
\end{equation}
\ 

\subsection{The Quarks}

\ The system has three quarks: the excited quark q$^{\ast}(u^{\ast}$ or
$d^{\ast})$, and the accompanying excited quarks $u^{\prime}$ and $d^{\prime}%
$. According to \textbf{Hypothesis I}, the excited quark $q^{\ast}(u^{\ast}$
or $d^{\ast})$ has: for u$^{\ast}$%
\begin{equation}
\mathbf{\ }B=1/3\text{\textbf{, S = 0, }}s=1/2\text{\textbf{, }}I=1/2\text{,
I}_{Z}\text{= 1/2, and Q = 2/3 ;}%
\end{equation}
for d$^{\ast}$%
\begin{equation}
\mathbf{\ }B=1/3\text{\textbf{, S = 0, }}s=1/2\text{\textbf{, }}I=1/2\text{,
I}_{Z}\text{= -1/2, and Q = -1/3}.
\end{equation}
From (\ref{Quantum-0f-u}) and (\ref{Quantum-0f-d}), we can get the quantum
numbers and energies of the accompanying excited quarks $u^{\prime}$ and
$d^{\prime}$.

\subsection{Baryons}

Using the `sum formulae' (\ref{SUM-formula}), we can find the quantum numbers
of the three quark system ($q^{\ast}u^{\prime}d^{\prime}$). For q$^{\ast}=$
$u^{\ast}$, the system ($u^{\ast}u^{\prime}d^{\prime}$) has\ $B=1,$ $I=1/2,$
$I_{z}=1/2,$ $Q=1$. Comparing it with the experimental results of a $proton$
(B $=1,$ $I=1/2,$ $I_{z}=1/2,$ $Q=1$, we get
\begin{equation}
(u^{\ast}u^{\prime}d^{\prime})\rightarrow proton\text{ .}\label{Proton}%
\end{equation}
Similarly, for q$^{\ast}=$ $d^{\ast}$, we got that the system is a neutron
\begin{equation}
(d^{\ast}u^{\prime}d^{\prime})\rightarrow neutron\text{ }\label{Neutron}%
\end{equation}
with \ $B=1,$ $I=1/2,$ $I_{z}=-1/2,$ $Q=0.$

From (\ref{Quantum-0f-u}) and (\ref{Quantum-0f-d}), we obtain that
\begin{equation}
m_{u^{\prime}}=m_{d^{\prime}}=0.\label{m(u)=m(d)=0}%
\end{equation}
Using $M_{n}\thickapprox M_{p}=939$ $Mev\thickapprox940$ Mev and
M$_{p}=m_{u^{\ast}}+m_{u^{\prime}}+m_{d^{\prime}},$ we obtain the mass of the
quark u$^{\ast}$%
\begin{equation}
m_{u^{\ast}}=940Mev.\label{m(u)=939}%
\end{equation}
Similarly, we can get the mass of the quark $d^{\ast}$%
\begin{equation}
m_{d^{\ast}}=940Mev.\label{m(d)=939}%
\end{equation}

In the three quark system ($q^{\ast}u^{\prime}d^{\prime}$), from
\textbf{Hypothesis I} , quarks u$^{\prime}$ and d$^{^{\prime}}$ satisfy the
SU(2) symmetries. Moreover, quark $q^{\ast}$ is an excited state of $u$ or
$d$. Thus, the Hamiltionia H($q^{\ast}u^{\prime}d^{\prime}$) of the three
quark system ($q^{\ast}u^{\prime}d^{\prime}$) satisfies the SU(3) symmetries.

\subsection{Mesons}

Using the sum formulae (\ref{SUM-formula}), we can find the quantum numbers of
a quark and an antiquark system.

The system ($u^{\ast}\overline{d^{\ast}}$) has\ $B=0,$ $I=1,$ $I_{z}=1,$
$Q=1;$ it is a meson $\pi^{+}$.\ 

The system ($d^{\ast}\overline{u^{\ast}}$) has\ $B=0,$ $I=1,$ $I_{z}=-1,$
$Q=0;$ it is a meson $\pi^{-}$.

The system ($u^{\ast}\overline{u^{\ast}}$) has\ $B=0,$ $I=1,0;$ $I_{z}=0;$
$Q=0$.\ 

The system ($d^{\ast}\overline{d^{\ast}}$) has\ $B=0,$ $I=1,0;$ $I_{z}=0;$
$Q=0$.\ \ 

Thus, in the primitive cell approximation, we can get the mesons $\pi=(\pi
^{+},\pi^{0},\pi^{-})$%
\begin{equation}
\pi^{+}=(u^{\ast}\overline{d^{\ast}}),\text{ }\pi^{0}=\frac{1}{\sqrt{2}%
}\left(  u^{\ast}\overline{u^{\ast}}-d^{\ast}\overline{d^{\ast}}\right)
,\text{ }\pi^{-}=d^{\ast}\overline{u^{\ast}}\label{PAI}%
\end{equation}
with $B=0,$ $I=1,$ $S=C=0,$ $Q=1,0,-1;$ and the meson $\eta^{0}$%

\begin{equation}
\eta^{0}=\frac{1}{\sqrt{2}}\left(  u^{\ast}\overline{u^{\ast}}+d^{\ast
}\overline{d^{\ast}}\right) \label{Yita}%
\end{equation}
with $B=0,$ $I=0,$ $S=C=0,$ $Q=0.$

\subsection{The Resonance States}

The resonance states of the baryons (q$^{\ast}u\prime d\prime$) and mesons
(q$^{\ast}\overline{q^{\ast}})$ can be
found:$\ \ \ \ \ \ \ \ \ \ \ \ \ \ \ \ \ \ \ \ \ \ \ \ \ \ \ \ \ \ \ \ $%
\begin{equation}%
\begin{tabular}
[c]{l}%
$\left[  (u^{\ast}u\prime d\prime)(u^{\ast}\overline{d^{\ast}})\right]  $ =
$\left[  (u^{\ast}u\prime u^{\ast})(d^{\prime}\overline{d^{\ast}})\right]  $
I$_{z}=3/2,Q=+2$\\
$\left[  (u^{\ast}u\prime d\prime)(u^{\ast}\overline{u^{\ast}})\right]
=\left[  (u^{\ast}u^{\ast}d\prime)(u^{\prime}\overline{u^{\ast}})\right]  $
\ \ I$_{z}=1/2,Q=+1$\\
$\left[  (u^{\ast}u\prime d\prime)(d^{\ast}\overline{d^{\ast}})\right]
=\left[  (u^{\ast}u\prime d^{\ast})(d^{\prime}\overline{d^{\ast}})\right]  $
\ I$_{z}=1/2,Q=+1$\\
$\left[  (u^{\ast}u\prime d\prime)(d^{\ast}\overline{u^{\ast}})\right]
=\left[  (d^{\ast}u^{\ast}d\prime)(u^{\prime}\overline{u^{\ast}})\right]  $
\ \ \ I$_{z}=-1/2,Q=0$\\
$\left[  (d^{\ast}u\prime d\prime)(u^{\ast}\overline{d^{\ast}})\right]
=\left[  (u^{\ast}u^{\prime}d^{\ast})(d^{\prime}\overline{d^{\ast}})\right]  $
I$_{z}=1/2,Q=+1$\\
$\left[  (d^{\ast}u\prime d\prime)(u^{\ast}\overline{u^{\ast}})\right]
=\left[  (d^{\ast}u^{\ast}d\prime)(u^{\prime}\overline{u^{\ast}})\right]  $
\ \ I$_{z}=-1/2,Q=0$\\
$\left[  (d^{\ast}u\prime d\prime)(d^{\ast}\overline{d^{\ast}})\right]
=\left[  (d^{\ast}u\prime d^{\ast})(d^{\prime}\overline{d^{\ast}})\right]  $
I$_{z}=-1/2,Q=0$\\
$\left[  (d^{\ast}u\prime d\prime)(d^{\ast}\overline{u^{\ast}})\right]
=\left[  (d^{\ast}d^{\ast}d\prime)u\prime\overline{u^{\ast}}\right]  $
\ \ \ \ \ I$_{z}=-3/2,Q=-1$%
\end{tabular}
\label{Baryon+Meson}%
\end{equation}%

\begin{equation}%
\begin{tabular}
[c]{l}%
$\Delta^{++}=(u^{\ast}u\prime u^{\ast})(d^{\prime}\overline{d^{\ast}})$\\
$\Delta^{+}=\sqrt{\frac{1}{3}}((u^{\ast}u^{\ast}d\prime)(u^{\prime}%
\overline{u^{\ast}})+(u^{\ast}u\prime d^{\ast})(d^{\prime}\overline{d^{\ast}%
})+(u^{\ast}u^{\prime}d^{\ast})(d^{\prime}\overline{d^{\ast}})$\\
$\Delta^{0}=\sqrt{\frac{1}{3}}((d^{\ast}u^{\ast}d\prime)(u^{\prime}%
\overline{u^{\ast}})+(d^{\ast}u^{\ast}d\prime)(u^{\prime}\overline{u^{\ast}%
})+(d^{\ast}u\prime d^{\ast})(d^{\prime}\overline{d^{\ast}}))$\\
$\Delta^{-}=(d^{\ast}d^{\ast}d\prime)($u$^{\prime}\overline{u^{\ast}})$%
\end{tabular}
\end{equation}

From the above, we have a 4 fold resonace state $\Delta.$ It has B = 1, S = 0,
I = 3/2, and Q = 2, 1, 0, -1.

\subsection{The High Energy Scattering}

When the hadrons collide against each other with high energy, the accompanying
excited quarks u$^{\prime}$ and d$^{\prime}$ may be excited into excited
quarks $u^{\ast}$ and $d^{\ast}$ temporarily. Hence,\textbf{\ }there may be
two or three excited quarks in the system in a very high energy
scattering\textbf{. If one high energy particle collides against two quarks at
the same time and with the same probability, a two-jet will be born. If the
high energy particle collides against three quarks at the same time and with
the same probability, a three-jet will be born.} The high energy scattering
procedure is a very complex process. It is beyond the scope of this paper.

\subsection{Summary}

The primitive cell approximation looks like the Quark Model, however it is not
really the Quark Model. \textbf{It is only an embryo of the Quark Model.}
Although it is very simple, it can give many important results which are
useful in the BCC model.

1. A baryon is made of three quarks from (\ref{Proton}) and (\ref{Neutron}).
\textbf{This is based on the physical structure of the body center cubic quark
lattice. There are only two quarks, u and d, in each primitive cell of the
quark lattice. If an excited quark q}$^{\ast}$\textbf{\ is moving in the cell,
the two quarks (u and d) of the cell are accompanying excited by the excited
quark q}$^{\ast}$\textbf{. Thus, the system (}$q^{\ast}u\prime d\prime
$\textbf{) is made of three quarks now. We have already shown that the system
is a baryon (\ref{Proton}) and (\ref{Neutron}).}

2. A meson is made of a quark and an antiquark from (\ref{PAI}) and
(\ref{Yita}).

3. Any excited quark q$^{\ast}$ is always accompanied by two accompanying
excited quarks, u$^{\prime}$ and d$^{\prime}$. Since they cannot be separated,
a free quark can never be seen.

4. The three quark system satisfies the SU(3) symmetry.

5. The quantum numbers of the baryons and the mesons can be\ determined by the
sum formulae (\ref{SUM-formula}).

6. The three quarks are in different states. One of them (q$^{\ast}$) is in a
completely excited state from the vacuum lattice. The other two (u$^{\prime}$
and d$^{\prime}$) are in the accompanying excited states, and they are in
difirent isospin states (I$_{z}(u^{\prime})=+1/2,$ I$_{z}(d^{\prime})=-1/2)$.
Thus, the three quarks are in different states, the system (q$^{\ast}u\prime
d\prime)$ obeys the \textbf{Pauli exclusion principle} \cite{Pauli and Nanbu}.

In the primitive cell approximation, \textbf{the most important result is that
a baryon is made of three quarks. It is based on the physical structure of the
body center cubic quark lattice. }

However, since the cell approximation omits the periodic field of the quark
lattice, it can not deduce the strange number, the charmed number, the bottom
number, or the masses of the quarks.\ As we shall see in the next section,
these quantum numbers are products of the periodic field of the quark lattice.\ \ \ \ \ 

\section{The Periodic Field Approximation}

In the periodic field approximation, we assume:

(1) The ideal quark lattice (see (\ref{H})) will be regarded as the physical
vacuum background,
\begin{equation}
H_{Latt}\approx0.
\end{equation}

(2) The interaction Hamiltonian between the excited quark $q^{\ast}$ and the
ideal quark lattice will be replaced by a periodic field with the body center
cubic symmetries,
\begin{equation}
H_{q^{\ast}-Latt}\rightarrow V(\vec{r}).\label{V(r)}%
\end{equation}

(3) The quantum numbers and energy of the three quark system (q$^{\ast
}u^{\prime}d^{\prime})$ are represented by a point particle which will be
called the Lee Particle \ \cite{Periodic Field}\textbf{\ }in the periodic
field approximation. In other words, \textbf{\ the quantum numbers and energy
of the three quark system are concentrated in the excited particle (Lee
Particle).} We have
\begin{equation}%
\begin{tabular}
[c]{l}%
u$_{Lee}^{\ast}:$ B = 1, Q = 1,\\
d$_{Lee}^{\ast}:$ B = 1, Q = 0.
\end{tabular}
\label{Baryon u and d}%
\end{equation}
for free Lee Particles. Therefore, in the periodic approximation, \textbf{an
excited quark q}$_{Lee\mathbf{\ }}^{\ast}$\textbf{(Lee Particle)
\cite{Periodic Field}\ represents a baryon.}

(4) \textbf{In the periodic approximation, an excited quark q}$_{Lee}^{\ast} $
\textbf{is regarded as a baryon. The excited quarks which are in different
excited states will be different baryons.}

Finally, the Hamiltonian (\ref{H})\ is simplified into the periodic
approximation H$_{per.}$\
\begin{equation}
H_{per.}=H_{q_{Lee}^{\ast}}+\ V(\vec{r}).\label{Periodic-Field-H}%
\end{equation}
Thus, in the periodic field approximation, the problem is simplified into an
excited Lee Particle (the group of an excited quark q$^{\ast}$ and two
accompanying excited quarks u$^{\prime}$ and d$^{\prime}$) moving in the
periodic field. The problem has already been dicussed in \cite{Periodic
Field}. In that paper, we used the point particle (the Lee Particle)
approximation to represent the primitive cell. We have deduced a baryon
spectrum which is in agreement with the experimental results \cite{particle}%
\textbf{.} Readers can find the results in Table 1-Table 6 of that paper.

In the periodic field approximation, \textbf{the most important result is that
the strange numbers (S = -1, S = -2, and S = -3), the charmed number, and the
bottom number all come from the body center cubic symmetries of the vacuum
quark lattice. In fact, the strange particles (}$\Lambda$\textbf{, }$\Sigma
$\textbf{, }$\Xi$, \textbf{and }$\Omega$\textbf{), the charmed particles
(}$\Lambda_{c}$)\textbf{, and the bottom particles (}$\Lambda_{b}
$)\textbf{\ are not new particles which are completely different from the
nucleons. They are only higher energy band excited states of the Lee Particle.}

However, the periodic field approximation can not deduce the quark spectrum.
Thus, it is not easy to deduce the meson spectrum.\ In the next section, we
will see that in the combined approximation\textbf{\ }of the cell and the
periodic field, we can deduce the quark spectrum, the baryon spectrum, and the
meson spectrum.\ \ \ \ \ \ \ \ \ \ \ \ \ \ \ \ \ \ \ \ \ \ \ \ \ \ \ \ \ \ \ 

\section{The combined approximation\textbf{\ of the Cell and the Periodic
Field}}

In the primitive cell approximation, we omitted the periodic field. In the
periodic approximation, we united the excited quark $q^{\ast}$ and the
primitive cell ($u^{\prime}d^{\prime}$) to get the Lee Particle (the point
approximation)\cite{Periodic Field}. In the combined approximation, we will
not only consider the periodic field, but also consider the accompanying
excited cell.

1. The excited quark q$^{\ast}$ moves in the perfect periodic field with body
center cubic symmetries, the accompanying excited cell ($u^{\prime}d^{\prime}%
$) is always accompanying the quark q$^{\ast}$.

2. According to the sum-formulas\textbf{\ }(\ref{SUM-formula}), the quantum
numbers and the energy of the three quark system ($q^{\ast}u^{\prime}%
d^{\prime}$) are the sums of the quantum numbers and the energies of the
constituent quarks inside the system.

In this approximation, we will deduce the spectrum of the quarks first. Then,
using the sum-laws, we deduce the spectrum of the baryons.\textbf{\qquad\qquad}

\subsection{The Spectrum of the Quarks}

First of all, we will find the energy bands of the excited quark $q^{\ast}$.
Then, we will find the quantum numbers and energies of the energy bands.
Finally, we will find the spectrum of the excited quarks.

\subsubsection{The Motion Equation of the Quark}

When a quark is excited from the vacuum, it is moving in the vacuum. Since the
quark is a Fermion, its motion equation should be the Dirac equation. Taking
into account that (according to the renormalization theory \cite{renormal} )
the bare mass of the quark is much larger than the empirical values of the
excited energies of the quark $q^{\ast}$, we use the Schr\"{o}dinger equation
instead of the Dirac equation (our results will show that this is a very good
approximation). From (\ref{Periodic-Field-H}), using 
$H_{q}=- {\frac {{\hbar
}^{2}}{2m_{q}}}{\nabla}^{2}$, we have:
\begin{equation}
{ \frac {{\hbar
}^{2}}{2m_{q}}}\nabla^{2}\Psi+(\varepsilon-V(\vec{r}))\Psi=0,
\end{equation}
where $V(\vec{r})$ denotes the strong interaction periodic field\ of the quark
lattice with body center cubic symmetries, and $m_{q}$ is the bare mass of the
quark $q^{\ast}$.

\subsubsection{Finding the Energy Bands of the quark}

Using the energy band theory \cite{eband} and the free particle approximation
\cite{freeparticle} (taking $V(\vec{r})=V_{0}$ constant and making the wave
functions satisfy the body center cubic periodic symmetries), we have
\begin{equation}
{\frac {{\hbar
}^{2}}{2m_{q}}}\nabla^{2}\Psi+(\varepsilon-V_{0})\Psi=0,\label{motion}%
\end{equation}
where $V_{0}$ is a constant potential. The solution of Eq. (\ref{motion}) is a
plane wave
\begin{equation}
\Psi_{\vec{k}}(\vec{r})=\exp\{-i(2\pi/a_{x})[(n_{1}-\xi)x+(n_{2}-\eta
)y+(n_{3}-\zeta)z]\},\label{wave}%
\end{equation}
where the wave vector $\vec{k}=(2\pi/a_{x})(\xi,\eta,\zeta)$, $a_{x}$ is the
periodic constant of the quark lattice, and $n_{1}$, $n_{2}$, $n_{3}$ are
integers satisfying the condition
\begin{equation}
n_{1}+n_{2}+n_{3}=\pm\text{ even number or }0.\label{condition}%
\end{equation}
Condition (\ref{condition}) implies that the vector $\vec{n}=(n_{1}%
,n_{2},n_{3})$ can only take certain values. For example, $\vec{n}$ can not
take $(0,0,1)$ or $\left(  1,1,-1\right)  $, but can take $(0,0,2)$ and
$(1,-1,2)$.

The zeroth-order approximation of the energy \cite{freeparticle} is
\begin{equation}
\varepsilon^{(0)}(\vec{k},\vec{n})=V_{0}+\alpha E(\vec{k},\vec{n}%
),\label{mass}%
\end{equation}%
\begin{equation}
\alpha=h^{2}/2m_{q}a_{x}^{2},\label{C-ALPHAR}%
\end{equation}%
\begin{equation}
E(\vec{k},\vec{n})=(n_{1}-\xi)^{2}+(n_{2}-\eta)^{2}+(n_{3}-\zeta
)^{2}.\label{energy}%
\end{equation}

Now we will demonstrate how to find the energy bands.

The first Brillouin zone \cite{Brillouin} of the body center cubic lattice is
shown in Fig. 1. In Fig. 1 (depicted from \cite{eband} (Fig. 1) and
\cite{Brillouin} (Fig. 8.10)), the $(\xi,\eta,\zeta)$ coordinates of the
symmetry points are:
\begin{gather}
\Gamma=(\text{0, 0, 0}),\text{ }H=(\text{0, 0, 1}),\text{ }P=(\text{1/2, 1/2,
1/2}),\nonumber\\
N=(\text{1/2, 1/2, 0}),\text{ }M=(\text{1, 0, 0}),
\end{gather}
and the $(\xi,\eta,\zeta)$ coordinates of the symmetry axes are:
\begin{align}
\Delta &  =(\text{0, 0, }\zeta),\text{\ }0<\zeta<1;\text{ \ \ \ \ \ \ }%
\Lambda=(\xi\text{, }\xi\text{, }\xi),\text{ }0<\xi<1/2;\nonumber\\
\Sigma &  =(\xi\text{, }\xi\text{, 0}),\text{ }0<\xi<1/2;\text{ \ \ \ }%
D=(\text{1/2, 1/2, }\xi),\text{ }0<\xi<1/2;\nonumber\\
G  &  =(\xi\text{, 1-}\xi\text{, 0}),\text{ }1/2<\xi<1;\text{ \ }F=(\xi\text{,
}\xi\text{, 1-}\xi),\text{ }0<\xi<1/2.
\end{align}

For any valid value of the vector $\vec{n}$, substituting the $(\xi,\eta
,\zeta)$ coordinates of the symmetry points or the symmetry axes into
Eq.(\ref{energy}) and Eq.(\ref{wave}), we can get the $E(\vec{k},\vec{n})$
values and the wave functions at the symmetry points and on the symmetry axes.
In order to show how to calculate the energy bands, we give the calculation of
some low energy bands in the symmetry axis $\Delta$ as an example (the results
are illustrated in Fig. 2(a)).

First, from (\ref{energy}) and (\ref{wave}) we find the formulae for the
$E(\vec{k},\vec{n})$ values and the wave functions at the end points $\Gamma$
and $H$ of the symmetry axis $\Delta$, as well as on the symmetry axis
$\Delta$ itself:
\begin{equation}
E_{\Gamma}=n_{1}^{2}+n_{2}^{2}+n_{3}^{2},\label{gammae}%
\end{equation}%
\begin{equation}
\Psi_{\Gamma}=\exp\{-i(2\pi/a_{x})[n_{1}x+n_{2}y+n_{3}z]\}.\label{gammaksi}%
\end{equation}%
\begin{equation}
E_{H}=n_{1}^{2}+n_{2}^{2}+(n_{3}-1)^{2},\label{he}%
\end{equation}%
\begin{equation}
\Psi_{H}=\exp\{-i(2\pi/a_{x})[n_{1}x+n_{2}y+(n_{3}-1)z]\}.\label{hksi}%
\end{equation}%
\begin{equation}
E_{\Delta}=n_{1}^{2}+n_{2}^{2}+(n_{3}-\zeta)^{2},\label{deltae}%
\end{equation}%
\begin{equation}
\Psi_{\Delta}=\exp\{-i(2\pi/a_{x})[n_{1}x+n_{2}y+(n_{3}-\zeta
)z]\}.\label{deltaksi}%
\end{equation}

Then, using (\ref{gammae})--(\ref{deltaksi}), beginning from the lowest
possible energy, we can obtain the corresponding integer vectors $\vec
{n}=(n_{1},n_{2},n_{3})$ (satisfying (\ref{condition})) and the wave functions:

\begin{enumerate}
\item  The lowest $E(\vec{k},\vec{n})$ is at $(\xi,\eta,\zeta)=0$ (the point
$\Gamma$) and with only one value of $\vec{n}=(0,0,0)$ (see (\ref{gammae}) and
(\ref{gammaksi})):
\begin{equation}
\vec{n}=(\text{0, 0, 0})\text{, \ \ \ \ \ }E_{\Gamma}=0\text{, \ \ \ \ \ }%
\Psi_{\Gamma}=1.\label{GROUND E-W}%
\end{equation}

\item  Starting from $E_{\Gamma}=0$, along the axis $\Delta$, there is one
energy band (the lowest energy band $E_{\Delta}=\zeta^{2}$, with $n_{1}%
=n_{2}=n_{3}=0$ (see (\ref{deltae}) and (\ref{deltaksi})) ended at the point
$E_{H}=1$:
\begin{gather}
\vec{n}=(\text{0, 0, 0})\text{, \ \ \ \ \ }E_{\Gamma}=0\rightarrow E_{\Delta
}=\zeta^{2}\rightarrow E_{H}=1\text{, }\nonumber\\
\Psi_{\Delta}=\exp[i(2\pi/a_{x})(\zeta z)]\text{. \ \ \ \ \ \ \ \ \ }%
\end{gather}

\item  At the end point $H$ of the energy band $E_{\Gamma}=0\rightarrow
E_{\Delta}=\zeta^{2}\rightarrow E_{H}=1$, the energy $E_{H}=1$. Also at point
$H$, $E_{H}=1$ when $n=(\pm1,0,1)$, $(0,\pm1,1)$, and $(0,0,2)$ (see
(\ref{he}) and (\ref{hksi})):
\begin{equation}
E_{H}=1\text{, \ \ }\Psi_{H}=[e^{[i(2\pi/a_{x})(\pm x)]},e^{[i(2\pi/a_{x})(\pm
y)]},e^{[i(2\pi/a_{x})(\pm z)]}]\text{.}%
\end{equation}

\item  Starting from $E_{H}=1$, along the axis $\Delta$, there are three
energy bands ended at the points $E_{\Gamma}=0$, $E_{\Gamma}=2$, and
$E_{\Gamma}=4$, respectively:
\begin{gather}
\vec{n}=(\text{0,0,0})\text{, \ \ \ }E_{H}=1\rightarrow E_{\Delta}=\zeta
^{2}\rightarrow E_{\Gamma}=0,\text{ }\nonumber\\
\Psi_{\Delta}=\exp[i(2\pi/a_{x})(\zeta z)]\text{. \ \ \ \ \ \ \ \ \ }%
\end{gather}%
\begin{gather}
\vec{n}=(\text{0,0,2})\text{, \ \ }E_{H}=1\rightarrow E_{\Delta}%
=\text{(2-}\zeta\text{)}^{2}\rightarrow E_{\Gamma}=4\text{,}\nonumber\\
\text{ }\Psi_{\Delta}=\exp{[i(2\pi/a_{x})(2-\zeta)z)]}\text{.
\ \ \ \ \ \ \ \ }%
\end{gather}%
\begin{gather}
\vec{n}=(\pm\text{1,0,1})(\text{0,}\pm\text{1,1})\text{, \ \ }E_{H}%
=1\rightarrow E_{\Delta}=\text{1+(1-}\zeta\text{)}^{2}\rightarrow E_{\Gamma
}=2\text{,}\nonumber\\
\text{ }\Psi_{\Delta}=e^{{\{-i(2\pi/a_{x})[\pm x+(1-\zeta)z]\}}}%
,e^{{\{-i(2\pi/a_{x})[\pm y+(1-\zeta)z]\}}}\text{. \ \ \ \ \ }%
\end{gather}

\item  The energy bands with $4$ sets of values $\vec{n}$ $\ (\vec{n}=(\pm
$1,0,1$),$ $($0,$\pm$1,1$))$ ended at $E_{\Gamma}=2$. From (\ref{gammae}),
$E_{\Gamma}=2$ also when $\vec{n}$ takes other $8$ sets of values: $\vec
{n}=(1,\pm1,0)$, $(-1,\pm1,0)$, and $(\pm1,0,-1)$, $(0,\pm1,-1)$. Putting the
$12$ sets of $\vec{n}$ values into Eq. (\ref{gammaksi}), we can obtain $12$
plane wave functions:
\begin{equation}
E_{\Gamma}=2\text{, }\Psi_{\Gamma}=[e^{i(2\pi/a_{x})(\pm x\pm y)}%
,e^{i(2\pi/a_{x})(\pm y\pm z)},e^{i(2\pi/a_{x})(\pm z\pm x)}]\text{.}%
\end{equation}

\item  Starting from $E_{\Gamma}=2$, along the axis $\Delta$, there are three
energy bands ended at the points $E_{H}=1$, $E_{H}=3$, and $E_{H}=5$,
respectively:
\begin{equation}
\vec{n}=(\pm\text{1,0,1})(\text{0,}\pm\text{1,1})\text{,\ }E_{\Gamma
}=2\rightarrow E_{\Delta}=\text{1+(1-}\zeta\text{)}^{2}\rightarrow
E_{H}=1\text{,}%
\end{equation}%
\begin{equation}
\vec{n}=(\text{1,}\pm\text{1,0})(\text{-1,}\pm\text{1,0})\text{,\ }E_{\Gamma
}=2\rightarrow E_{\Delta}=\text{2+}\zeta^{2}\rightarrow E_{H}=3\text{,}%
\end{equation}%
\begin{equation}
\vec{n}=(\pm\text{1,0,-1})(\text{0,}\pm\text{1,-1})\text{,\ }E_{\Gamma
}=2\rightarrow E_{\Delta}=\text{1+(}\zeta\text{+1)}^{2}\rightarrow
E_{H}=5\text{.}%
\end{equation}
\end{enumerate}

Continuing the process, we can find all the energy bands and the corresponding
wave functions. The wave functions are not needed for the zeroth order
approximation, so we only show the energy bands in Fig. 2-4. There are six
small figures in Fig. 2-4. Each of them shows the energy bands in one of the
six axes in Fig.1. Each small figure is a schematic one where the straight
lines (show the energy bands) should be parabolic curves. The numbers above
the lines are the values of $\vec{n}$ = ($n_{1}$, $n_{2}$, $n_{3}$). The
numbers under the lines are the fold numbers of the energy bands with the same
energy (the zeroth-order approximation). The numbers beside both ends of an
energy band (the intersection of the energy band line and the vertical lines)
represent the highest and lowest E($\vec{k}$,$\vec{n} $) values (see Eq.
(\ref{energy})) of the band. Putting the values of the E($\vec{k}$,$\vec{n}$)
into Eq. (\ref{mass}), we get the zeroth-order energy approximation values (in Mev).

\subsubsection{The Quantum Numbers and Energies of the Quarks}

If there were no strong interaction periodic field of the quark lattice, we
would only see quarks u$^{\ast}$ and d$^{\ast}$(inside N, $\Delta,\pi,$ and
$\eta$). We could not see any flavered quarks (inside flavored baryons
$\Lambda,$ $\Sigma,$ $\Xi,$ $\Omega,$ $\Lambda_{c},$ $\Lambda_{b}$...) because
they would not exist. Due to the periodic field, although the free quarks
u$^{\ast}$ and d$^{\ast}$ are still the essential states, there is a slight
chance that the quarks are excited to the symmetry points (see Fig.1) of the
periodic field. Once at the symmetry points, the excited quarks will show
special symmetric properties. Due to the periodic field, the parabolic energy
curve of the free excited quark will be changed into energy bands (see Eq.
(\ref{mass}) and Fig. 2-5). Also, there will be energy gaps on the surfaces of
the Brillouin zones which originate from the periodic field \cite{GAPS}. The
gaps will give the excited quarks longer lifetimes and special properties,
which are different from those of the free quarks $u^{\ast}$ and $d^{\ast}$.
Because of these properties, physicists naturally regard them as new excited
quarks which are different from the excited quarks $u^{\ast}$and $d^{\ast}$.

There are two necessary conditions for the $q^{\ast}$ to be regarded as a new
excited quark state: first, the lowest energy point of the energy band must be
at a high symmetric point (the points $\Gamma$, $P$, $N$, $H$, or $M $) (see
Fig. 1);\ second, there is an energy gap between the lowest energy point of
the energy band and the ground state ($E=0$ , see Fig. 2-5).

From Fig. 2-5, we can see that all energy bands satisfy the first condition.
However, the six energy bands with $\vec{n}$ = ($0$,$0$,$0$) of the first
Brillouin zone\ do not satisfy the second condition. In other words, the six
energy bands are all in the first Brillouin zone---a part of the parabolic
energy curved surface of the free excited quark $q^{\ast}(u^{\ast}$and
$d^{\ast}$). Therefore, they represent the unflavored (S = C = b = 0) excited
quark $q_{N}^{\ast}(940)(u^{\ast}$or $d^{\ast}$):%

\begin{equation}%
\begin{array}
[c]{clcccc}%
\text{Axis} & \text{Energy Band} & \vec{n} & I & q^{\ast}(m) & \text{Q}%
_{u}\text{, \ \ Q}_{d}\\
\Delta & \text{E}_{\Gamma}\text{=0}\rightarrow\text{E}_{H}\text{=1} &
(0,0,0) & 1/2 & q_{N}^{\ast}(940) & \text{2/3, -1/3}\\
\Lambda & \text{E}_{\Gamma}\text{=0}\rightarrow\text{E}_{P}\text{=3/4} &
(0,0,0) & 1/2 & q_{N}^{\ast}(940) & \text{2/3, -1/3}\\
\Sigma & \text{E}_{\Gamma}\text{=0}\rightarrow\text{E}_{N}\text{=1/2} &
(0,0,0) & 1/2 & q_{N}^{\ast}(940) & \text{2/3, -1/3}\\
D & \text{E}_{N}\text{=1/2}\rightarrow\text{E}_{P}\text{=3/4} & (0,0,0) &
1/2 & q_{N}^{\ast}(940) & \text{2/3, -1/3}\\
F & \text{E}_{P}\text{=3/4}\rightarrow\text{E}_{H}\text{=1} & (0,0,0) & 1/2 &
q_{N}^{\ast}(940) & \text{2/3, -1/3}\\
G & \text{E}_{N}\text{=1/2}\rightarrow\text{E}_{M}\text{=1} & (0,0,0) & 1/2 &
q_{N}^{\ast}(940) & \text{2/3, -1/3}%
\end{array}
\label{Bands of First}%
\end{equation}

In order to find the quantum numbers of the\textbf{\ }energy bands
(\textbf{except for the 6 energy bands of the first Brillouin zone}) of the
excited quark, we will make a new hypothesis in addition to the ones in
Section II\textbf{.}

\begin{problem}
\textbf{\ The quantum numbers and masses of the excited quarks are determined
as follows (except for the 6 energy bands of the first Brillouin zone):}
\end{problem}

\begin{enumerate}
\item \textbf{Baryon number }$B$\textbf{. When a quark is in vacuum state, B =
0. But if it is excited from the vacuum state, it has }
\begin{equation}
B=1/3.\label{baryon}%
\end{equation}

\item \textbf{Isospin number }$I$\textbf{: the isospin }$I$\textbf{\ is
determined by the energy band degeneracy }$d$\textbf{\ \cite{eband}, where }
\begin{equation}
d=2I+1.\label{isomax}%
\end{equation}
The concept of isospin was introduced in the early 30's by Heisenberg
\cite{SU(2)} to describe the approximate charge-independent nature of the
strong interaction between protons and neutrons. For a given I, I$_{z}$ can
vary from -I to I, making a total of 2I + 1 states. Under isospin rotations,
the quantum number I is preserved; however, these \ 2I +1 \ states of
different I$_{z}$ transform among one another, and therefore they are
degenerate with respect to the strong interaction. Thus we can get the isospin
I from the degeneracy d in terms of (\ref{isomax}).

\qquad In some cases, the degeneracy $d$ should be divided into
sub-degeneracies before using the formula. First, if the `degeneracy' energy
bands are in the first and second Brillouin Zones, the `degeneracy' will be
divided into two sub degeneracies. Secondly, if $d$ is larger than the rotary
fold $R$ of the symmetry axis:
\begin{equation}
d>R\text{,}\label{degeneracy}%
\end{equation}
then we assume that the degeneracy will be divided into $\gamma$
sub-degeneracies, where
\begin{equation}
\gamma=d/R\text{.}\label{subdegen}%
\end{equation}
For the three axes which pass through the center point $\Gamma$ (the axis
$\Delta(\Gamma-H)$, the axis $\Lambda(\Gamma-P)$, the axis $\Sigma(\Gamma
-N)$), the energy bands in the same degeneracy group have symmetric $\vec
{n}=(n_{1},n_{2},n_{3})$ values (see Fig. 2(a), 2(b) and 3(a)). Hence, if the
sub-degeneracy $d_{sub}\leq R$, it will not be divided further. About
symmetric $\vec{n}$, we give a definition:\textbf{\ a group of }$\vec{n}%
$\textbf{\ = (}$n_{1},n_{2},n_{3}$\textbf{) values is said to be symmetric if
any two }$\vec{n}$\textbf{\ values in the group can transform into each other
by various permutation (change component order) and by changing the sign
``}$\pm"$\textbf{\ (multiplied by ``-1'' ) of the components (one, two, or
three)}. For example, $(-2,-1,3)$ and $(-3,2,1)$ are symmetric, $(-3,0,2)$ and
$(-3,0,1)$ are asymmetric. For the other three symmetric axes, see Appendix A.

\qquad\ After finding the sub-degeneracy, $d_{sub}$, we can use ($d_{sub}$ =
2I+1) to find the isospin $I$.

\item \textbf{Strange number }$S$\textbf{: the Strange number }$S$\textbf{\ is
determined by the rotary fold }$R$\textbf{\ of the symmetry axis \cite{eband}
with }
\begin{equation}
S=R-4,\label{strange}%
\end{equation}

\textbf{where the number }$4$\textbf{\ is the highest possible rotary fold
number.} To be specific, from Eq. (\ref{strange}) and Fig. 1, we get
\begin{equation}
\Delta(\Gamma-H)\text{ is a }4\text{-fold rotation axis, }R=4\rightarrow
S=0;\label{DELTA-S}%
\end{equation}%
\begin{equation}
\Lambda(\Gamma-P)\text{ is a }3\text{-fold rotation axis, }R=3\rightarrow
S=-1;\label{LAMTA-S}%
\end{equation}%
\begin{equation}
\Sigma(\Gamma-N)\text{ is a }2\text{-fold rotation axis, }R=2\rightarrow
S=-2.\label{SIGMA-S}%
\end{equation}
For the other three symmetry axes $D(P-N)$, $F(P-H)$, and $G(M-N),$ which are
on the surface of the first Brillouin zone (see Fig. 1), we determine the
strange numbers as follows:
\begin{equation}
D(P-N)\text{ is parallel to axis }\Delta\text{, }S_{D}=S_{\Delta
}=0;\label{D-S}%
\end{equation}%
\begin{equation}
F\text{ is parallel to an axis equivalent to }\Lambda\text{, }S_{F}%
=S_{\Lambda}=-1;\label{F-S}%
\end{equation}%
\begin{equation}
G\text{ is parallel to an axis equivalent to }\Sigma\text{, }S_{G}=S_{\Sigma
}=-2\text{.}\label{G-S}%
\end{equation}

\item \textbf{Electric charge }$Q$\textbf{: after obtaining }$B,S$%
\textbf{\ and }$I$\textbf{, we can find the charge }$Q$\textbf{\ from the
Gell-Mann-Nishijiman relationship \cite{GellMann}: }
\begin{equation}
Q=I_{z}+\frac{\text{1}}{2}(S+B).\label{charge}%
\end{equation}

\item \textbf{Charmed number }$C$\textbf{\ \cite{charmed} and Bottom number
}$b$ \textbf{\cite{bottom}}\ \textbf{: if a degeneracy d of an energy band is
smaller than the rotary fold R}
\begin{equation}
\ d\ <\ R\ and\ R-d\ \neq2,\label{Condi-Sbar}%
\end{equation}
\textbf{\ then formula (\ref{strange}) will be changed as }
\begin{equation}
\bar{S}=R-4.\label{Strangebar}%
\end{equation}
\textbf{\ From Hypothesis III (}$\Delta S=\pm1$\textbf{), the real value of
}$S$\textbf{\ is }
\begin{equation}
S=\bar{S}+\Delta S=S_{Axis}\pm1,\text{ \ \ \ \ \ if }d\ <\ R\text{ and
}R-d\neq2.\label{strangeflu}%
\end{equation}
The ``Strange number'' ,$S,$ \ in (\ref{strangeflu}) is not completely the
same as the strange number in (\ref{strange}). In order to compare it with the
experimental results, we would like to give it a new name under certain
circumstances. Based on \textbf{Hypothesis III}, the new names will be the
\textbf{Charmed} number and the \textbf{Bottom} number:
\begin{gather}
\text{if }S=+1\text{ which originates from the fluctuation }\Delta S=+1\text{,
}\nonumber\\
\text{then we call it the \textbf{Charmed} number }C\text{ }(C=+1)\text{;}%
\label{charmed}%
\end{gather}%
\begin{gather}
\text{if }S=-1\text{ which originates from the fluctuation }\Delta S=+1\text{,
}\nonumber\\
\text{and if there is an energy fluctuation,}\nonumber\\
\text{then we call it the \textbf{Bottom} number }b\text{ }(b=-1)\text{.}%
\label{bottom}%
\end{gather}
Thus, (\ref{charge}) needs to be generalized to
\begin{equation}
\text{Q=I}_{z}\text{+}\frac{\text{1}}{2}\text{(B+S}_{G}\text{)=I}_{z}%
\text{+}\frac{\text{1}}{2}\text{(B+S+C+b),}\label{chargeflu}%
\end{equation}
where we define the generalized strange number as
\begin{equation}
S_{G}=S+C+b.\label{strangegen}%
\end{equation}

\item \textbf{Charmed strange baryon }$\Xi_{C}$\textbf{\ \cite{Kesi-C} and
}$\Omega_{C}$\textbf{\ \cite{OMIGA-C}: if the energy band degeneracy }%
$d$\textbf{\ is larger than the rotary fold }$R$\textbf{, the degeneracy will
be divided. Sometimes degeneracies should be divided more than once. After the
first division, the sub-degeneracy energy bands have }$S_{Sub}=\bar{S}+\Delta
S$\textbf{. For the second division of a degeneracy bands, we have: }
\begin{gather}
\text{if the second division has fluctuation }\Delta S=+1\text{,
\ }\nonumber\\
\text{ then }S_{Sub}\text{ may be unchanged and we may have }\nonumber\\
\text{ a Charmed number }C\text{ from }C=\Delta S=+1.\label{C&S}%
\end{gather}
\textbf{Therefore, we can obtain charmed strange baryons }$\Xi_{C}%
$\textbf{\ and }$\Omega_{C}$\textbf{.}

\item \textbf{We assume that the excited quark's mass is the minimum of the
energy band which represents the excited quark.}
\begin{equation}
m_{q^{\ast}}=\text{Minimum (}\varepsilon^{(0)}(\vec{k},\vec{n})\text{)}%
=\text{Minimum (}V_{0}+\alpha E(\vec{k},\vec{n})\text{\textbf{)}%
}\label{QuarkMass}%
\end{equation}

\textbf{\ Using the static masses (static energy) }$M_{nucleon}$\textbf{\ of
the nucleons we can determine }$V_{0}$\textbf{\ in formula (\ref{QuarkMass}).
The three quark systems (q}$^{\ast}u^{\prime}d^{\prime}$\textbf{) are baryons.
The mass of the baryon is determined by }
\begin{equation}
M_{B}=m_{q^{\ast}}+m_{u^{\prime}}+m_{d^{\prime}}.\label{Sum-Mass}%
\end{equation}
\textbf{Since }$m_{u^{\prime}}=m_{d^{\prime}}=0$%
\textbf{\ from\ (\ref{Quantum-0f-u}) and (\ref{Quantum-0f-d}), we have }
\begin{equation}
M_{B}=m_{q^{\ast}}=Minimum(\text{V}_{0}\text{ + }\alpha\text{E(}\vec
{k}\text{,}\vec{n}\text{)) .}\label{M-baryon}%
\end{equation}
The lowest mass of the baryon is the static mass of the nucleons
$M_{nucleon}=$ V$_{0}$.\ From the experiments, the static mass of the nucleons
$M_{nucleon}=939$ Mev \cite{particle}. Thus, we have
\begin{equation}
V_{0}=M_{nucleon}=939\text{ Mev}\thickapprox940\text{ Mev.}\label{V0}%
\end{equation}
For the sake of convenience, we take $V_{0}=940$ Mev in (\ref{V0}). Fitting
the theoretical mass spectrum into the empirical mass spectrum of the baryons,
we can also determine the $\alpha$ value in (\ref{QuarkMass}):
\begin{equation}
\alpha=h^{2}/2m_{q}a_{x}^{2}=360\text{ Mev.}\label{Alpha}%
\end{equation}

\item \textbf{The fluctuation of the strange number will be accompanied by an
energy change (Hypothesis III). We assume that the change of the energy
(perturbation energy) is proportional to }$(-\Delta S)$\textbf{\ and a number,
}$J,$\textbf{\ representing the energy level with a phenomenological formula:
}
\begin{equation}
\Delta\varepsilon\text{ =(-1)}^{R-4}\text{100(1-}\delta\text{(J)+(}%
\delta\text{(R-4)-1)}\times\text{J)(-}\Delta\text{S),}\label{Pertubation-E}%
\end{equation}
\textbf{where }$R$\textbf{\ is the rotary number of the axis, }$\delta
(J)$\textbf{\ is a Dirac function (when }$J=0,$ $\delta(J)=1$%
\textbf{\ \ and\ when }$J\neq0,$ $\delta(J)=0$\textbf{. Hence, }$\delta
(J)$\textbf{\ makes that }$J=0\rightarrow\Delta\varepsilon=0$\textbf{.), and
}$J$\textbf{\ is an order number of the energy band with }$\Delta S\neq
0$\textbf{. }Applying (\ref{Pertubation-E}) to the symmetry axes, we have:
\end{enumerate}

for the\ axis $\Delta$, $R-4=0$,
\begin{equation}
\Delta\varepsilon=-100\times\Delta S\text{ \ J = 1, 2, ...}\label{flua}%
\end{equation}
{}

for the\ axes $\Lambda$ and $F$, $4-R=1$,%

\begin{equation}
\Delta\varepsilon=-100\times(J-1)\Delta S\text{ \ J = 1, 2, ...}\label{flub}%
\end{equation}
\ \qquad\qquad\qquad\qquad\ \ \ \qquad\qquad\qquad

for the\ axes $\Sigma$, $G$, and $D$, $R-4=2$%

\begin{equation}
\Delta\varepsilon=100\times(J-1)\Delta S\text{ \ J = 1, 2, ... }\label{fluc}%
\end{equation}
\qquad\qquad

Thus, the zeroth-order mass formula of the quarks (\ref{QuarkMass}) shall be
changed to
\begin{equation}
m_{q}=\text{Minimum(}V_{0}+\alpha E(\overrightarrow{k},\overrightarrow
{n})+\Delta\varepsilon\text{)}\label{UNITED-M}%
\end{equation}
\textbf{This formula (\ref{UNITED-M}) is the united mass formula} which can
give all masses of all quarks. \textbf{It is the united mass formula} which
can give all masses of all baryons.\ \ 

Using the above formulae for quantum numbers and energy of the quarks, we can
find the quark spectrum. We will start from the axis $\Delta.$

\subsubsection{The axis $\Delta(\Gamma-H)$\textbf{\qquad}}

From (\ref{DELTA-S}), we have S = 0. For low energy levels, there are $8$ and
$4$ fold degenerate energy bands and single bands on the axis. Since the axis
has $R=4$, from (\ref{degeneracy}) and (\ref{subdegen}), the energy bands of
degeneracy $8$ will be divided into two 4 fold degenerate bands.

1. The four fold degenerate bands on the axis $\Delta(\Gamma-H)$

For 4 fold degenerate bands, using (\ref{isomax}), we get $I=3/2$, and using
(\ref{charge}), we have $Q=5/3$, $2/3$, $-1/3$, $-4/3$. Thus, each 4 fold
degenerate band represents a 4 fold quark family q$_{\Delta}^{\ast}$ with
\begin{equation}
\text{B = 1/3, S = 0, I = 3/2, Q = 5/3, 2/3, -1/3, -4/3.}\label{Data-Quark}%
\end{equation}
Using Fig. 2(a) and Fig. 5(a), we can get $E_{\Gamma},$ $E_{H},$ and $\vec{n}
$ values. Then, putting the values of $E_{\Gamma}$ and $E_{H}$ into the energy
formula (\ref{QuarkMass}), we can find $m_{q^{\ast}}=\varepsilon^{(0)} $.
Thus, we have
\begin{equation}%
\begin{array}
[c]{llll}%
E_{H}=1 & \vec{n}=(\text{101,-101,011,0-11}) & \varepsilon^{(0)}=1300 &
q_{\Delta}^{\ast}(1300)\\
E_{\Gamma}=2\text{ } & \vec{n}=(\text{110,1-10,-110,-1-10}) & \varepsilon
^{(0)}=1660 & q_{\Delta}^{\ast}(1660)\\
E_{\Gamma}=2 & \vec{n}=(\text{10-1,-10-1,01-1,0-1-1}) & \varepsilon
^{(0)}=1660 & q_{\Delta}^{\ast}(1660)\\
E_{H}=3 & \vec{n}=(\text{112,1-12,-112,-1-12}) & \varepsilon^{(0)}=2020 &
q_{\Delta}^{\ast}(2020)\\
E_{\Gamma}=4 & \vec{n}=(\text{200,-200,020,0-20}) & \varepsilon^{(0)}=2380 &
q_{\Delta}^{\ast}(2380)\\
E_{H}=5 & \vec{n}=(\text{121,1-21,-121,--1-21}, & \varepsilon^{(0)}=2740 &
q_{\Delta}^{\ast}(2740)\\
& \text{ \ \ \ \ \ \ \ 211,2-11,-211,-2-11}) & \varepsilon^{(0)}=2740 &
q_{\Delta}^{\ast}(2740)\\
E_{H}=5 & \vec{n}=(\text{202,-202,022,0-22}) & \varepsilon^{(0)}=2740 &
q_{\Delta}^{\ast}(2740)\\
E_{H}=5 & \vec{n}=(\text{013,0-13,103,-103}) & \varepsilon^{(0)}=2740 &
q_{\Delta}^{\ast}(2740)\\
\ldots &  &  &
\end{array}
\label{Dalta-Quark}%
\end{equation}

\begin{enumerate}
\item 2. The single bands on the axis $\Delta(\Gamma-H)$
\end{enumerate}

For the single bands$,$ d = 1
$<$%
R = 4 and R-d = 3 $\neq2$. According to (\ref{Condi-Sbar}), we should use
(\ref{strangeflu}) instead of (\ref{strange}). Therefore, we have
\begin{equation}
S_{\text{Single}}=\bar{S}_{\Delta}\pm\Delta S=0\pm1\text{,}\label{single}%
\end{equation}
where $\Delta S=\pm1$ from \textbf{Hypothesis III}. The best way to guarantee
the validity of Eq. (\ref{Strangebar}) in any small region is to assume that
$\Delta S$ takes $+1$ and $-1$ alternately from the lowest energy band to
higher ones. In fact, the $\vec{n}$ values are really alternately taking
positive and negative values: $E_{H}=1,$ $\vec{n}=($0, 0, 2$);$ $E_{\Gamma
}=4,$ $\vec{n}=($0, 0, -2$);$ $E_{H}=9,$ $\vec{n}=($0, 0, 4$);$ $E_{\Gamma
}=16,$ $\vec{n}=($0, 0, -4$);$ $E_{H}=25,$ $\vec{n}=($0, 0, 6$);$ $E_{\Gamma
}=36,$ $\vec{n}=($0, 0, -6$)$ ... Using the fact, we can find a
\textbf{phenomenological formula:}%

\begin{equation}
\Delta S=-(1+S_{axis})Sign(\vec{n}),\text{ }Sign(\vec{n})=\frac{n_{1}%
+n_{2}+n_{3}}{\left|  n_{1}\right|  +\left|  n_{2}\right|  +\left|
n_{3}\right|  }.\label{Sign+-0}%
\end{equation}
For the single states on the axis $\Delta,$ we have $S_{axis}=0.$ Thus, from
(\ref{Sign+-0}), we get
\begin{equation}
\Delta S=-(1+S_{axis})Sign(\vec{n})=-Sign(\vec{n}).\label{Sign-Deta}%
\end{equation}
For the single states on the axis $\Sigma,$ since $S_{axis}=-2,$ we have
\begin{equation}
\Delta S=-(1+S_{axis})Sign(\vec{n})=Sign(\vec{n}).\label{Sign-Segma}%
\end{equation}

At E$_{\Gamma}=0,$ from (\ref{Bands of First}), we know\textbf{\ the lowest
energy band with n = (0, 0, 0)} represents the free excited quark family
$q_{N}^{\ast}(940)$.

At $E_{H}=1,$ the second lowest single energy band, with $\vec{n}=(0,0,2)$ and
$J=1$, has the strange number $\Delta S=$ $-1$ from (\ref{Sign-Deta}). Thus,
$\Delta\varepsilon=100$ Mev from (\ref{flua})$\rightarrow$ the energy
$\varepsilon=$ $940$ $+360\times1$ $+\Delta\varepsilon$ $=1400$ Mev from
(\ref{UNITED-M}), and $I=0,$ $Q=-1/3$ from (\ref{charge}). Thus, it
reprensents a strange quark q$_{S}^{\ast}(1400).$

At $E_{\Gamma}=4,$ for the third lowest band with $\vec{n}=(0,0,-2)$, we get
$\Delta S=+1$ from (\ref{Sign-Deta}). Thus, $\mathbf{S=\bar{S}}_{\Delta
}\mathbf{+1=1}$\textbf{. }The lowest $E$ of the band is at $E_{\Gamma}=4,$ the
energy $\varepsilon=940+360\times4+\Delta\varepsilon=2380-100=2280$ from
(\ref{UNITED-M}) and (\ref{flua}). Here $S=+1$ originates from the fluctuation
$\Delta S=+1$ and there is an energy fluctuation of $\Delta\varepsilon=-100$.
From\textbf{\ }(\ref{charmed}), we know that the energy band has a charmed
number $C=+1$. \textbf{It represents a new excited state of the quark, q}%
$_{C}^{\ast}(2280),$ \textbf{with}
\begin{equation}
B=1/3\text{, I =0, S = 0, C = 1, Q = 2/3, m}_{C}=2280\text{ }%
Mev.\label{Q-Numbers-of-C}%
\end{equation}
In order to make it compatible with the results of the experiments, \textbf{we
will call it the Charmed quark} \cite{charmed}. It is very important to pay
attention to \textbf{the Charmed quark\ born here, on the single energy band,
and from the fluctuation } $\Delta S=+1$ and $\Delta\varepsilon=-100$ Mev.

Continuing the above procedure, from Fig. 5 (b), (\ref{UNITED-M}),
(\ref{Sign-Deta}) and (\ref{flua}), we have (notice that point H and point
$\Gamma,$ the two end points of the axis $\Delta,$ have the same symmetries):
\begin{equation}%
\begin{array}
[c]{llllllll}%
& \text{n}_{1}\text{,n}_{2}\text{,n}_{3} & \Delta S & J & \text{ \ \ }%
\Delta\varepsilon & \text{ \ }S & C & q^{\ast}(m)\\
E_{H}=1 & 0,0,2 & -1 & 1 & +100 & -1 & 0 & q_{S}^{\ast}(1400)\\
E_{\Gamma}=4 & 0,0,-2 & +1 & 2 & -100 & \text{ \ }0 & 1 & q_{C}^{\ast}(2280)\\
E_{H}=9 & 0,0,4 & -1 & 3 & +100 & -1 & 0 & q_{S}^{\ast}(4280)\\
E_{\Gamma}=16 & 0,0,-4 & +1 & 4 & -100 & \text{ \ }0 & 1 & q_{C}^{\ast
}(6600)\\
E_{H}=25 & 0,0,6 & -1 & 5 & +100 & -1 & 0 & q_{S}^{\ast}(10040)\\
E_{\Gamma}=36 & 0,0,-6 & +1 & 6 & -100 & \text{ \ }0 & 1 & q_{C}^{\ast
}(13800)\\
\ldots &  &  &  &  &  &  &
\end{array}
\label{DELTA-ONE}%
\end{equation}

\subsubsection{The axis $\Lambda(\Gamma-P)$}

From Fig. 2(b), we see that there is a single energy band with $\vec
{n}=(0,0,0)$, and all other bands are $3$ fold degenerate energy bands ($d=3$)
and $6$ fold degenerate bands ($d=6$).

1. At E$_{\Gamma}=0,$ from (\ref{Bands of First}), we know\textbf{\ the lowest
energy band with n = (0, 0, 0)} represents the free excited quark family
$q_{N}^{\ast}(940)$.

2. From (\ref{degeneracy}) and (\ref{subdegen}), the $6$ fold degenerate
energy bands will be divided into two energy bands with $3$ fold degeneracy.
For the 3 fold degenerate energy band, $R=3$ and $S=-1$ from (\ref{LAMTA-S}%
).\ Using (\ref{isomax}) and (\ref{charge}), we have $I=1$, and q = 2/3, -1/3,
-4/3. Thus, we get a 3 fold quark family q$_{\Sigma}^{\ast}$ with B = 1/3, S =
-1,\ I = 1, and Q = 2/3, -1/3, -4/3. Similar to (\ref{Dalta-Quark}), using
Fig. 2(b), we get
\begin{equation}%
\begin{array}
[c]{llcc}%
& \text{ \ \ \ \ \ \ \ \ }\vec{n} & \varepsilon^{(0)} & q(m)\\
E_{P}=3/4 & (\text{101,011,110}) & 1210 & q_{\Sigma}^{\ast}(1210)\\
E_{\Gamma}=2 & (\text{1-10,-110,01-1,} & 1660 & q_{\Sigma}^{\ast}(1660)\\
& \text{ \ \ \ \ \ \ \ 0-11,10-1,-101}) & 1660 & q_{\Sigma}^{\ast}(1660)\\
E_{\Gamma}=2 & (\text{-10-1,0-1-1,-1-10}) & 1660 & q_{\Sigma}^{\ast}(1660)\\
E_{P}=11/4 & (\text{020,002,200}) & 1930 & q_{\Sigma}^{\ast}(1930)\\
E_{P}=11/4 & (\text{121,211,112}) & 1930 & q_{\Sigma}^{\ast}(1930)\\
E_{\Gamma}=4 & (\text{0-20,-200,00-2}) & 2380 & q_{\Sigma}^{\ast}(2380)\\
E_{P}=19/4 & (\text{1-12,-112,21-1,} & 2650 & q_{\Sigma}^{\ast}(2650)\\
& \text{ \ \ \ \ \ \ \ 2-11,12-1,-121}) & 2650 & q_{\Sigma}^{\ast}(2650)\\
E_{P}=19/4 & (\text{202,022,220}) & 2650 & q_{\Sigma}^{\ast}(2650)\\
E_{\Gamma}=6 & (\text{-211,2-1-1,-1-12}, & 3100 & q_{\Sigma}^{\ast}(3100)\\
& \text{\ \ \ \ \ \ \ 11-2,-12-1,1-21}) & 3100 & q_{\Sigma}^{\ast}(3100)\\
& \text{(-1-21,1-2-1,-11-2,} & 3100 & q_{\Sigma}^{\ast}(3100)\\
& \text{\ \ \ \ \ \ 1-1-2,-21-1,-2-11}) & 3100 & q_{\Sigma}^{\ast}(3100)\\
& \text{(-1-2-1,-1-1-2,-2-1-1)} & 3100 & q_{\Sigma}^{\ast}(3100)\\
\ldots &  &  &
\end{array}
\label{Sigma-Quark}%
\end{equation}

\subsubsection{The axis $\Sigma(\Gamma-N)$}

The axis $\Sigma$ is a 2 fold rotation axis, from (\ref{SIGMA-S}) S = -2. For
low energy levels, there are $2$ fold degenerate energy bands, $4$ fold
degenerate energy bands, and single energy bands on the axis (see Fig. 3(a)).

1. The two fold degenerate energy bands on the axis $\Sigma(\Gamma-N)$

For the two fold degenerate energy bands, each of them represents a quark
family q$_{\Xi}^{\ast}$ with $B=1/3,I=1/2$ from (\ref{isomax})$,S=-2,Q=$ -1/3,
-4/3 from (\ref{charge}). \ Similar to (\ref{Dalta-Quark}), we have
\begin{equation}%
\begin{array}
[c]{llccc}%
\text{ \ \ \ \ \ }E & \text{\ }\vec{n}\text{ =(n}_{1}\text{,n}_{2}%
\text{,n}_{3}\text{)} & S & \varepsilon^{(0)} & q(m)\\
E_{\Gamma}=2 & (\text{1-10,-110}) & -2 & 1660 & q_{\Xi}^{\ast}(1660)\\
E_{N}=5/2 & (\text{200,020}) & -2 & 1840 & q_{\Xi}^{\ast}(1840)\\
E_{\Gamma}=4 & (\text{002,00-2}) & -2 & 2380 & q_{\Xi}^{\ast}(2380)\\
& (\text{-200,0-20}) & -2 & 2380 & q_{\Xi}^{\ast}(2380)\\
E_{N}=9/2 & (\text{112,11-2}) & -2 & 2560 & q_{\Xi}^{\ast}(2560)\\
E_{\Gamma}=6 & (\text{-1-12,-1-1-2)} & -2 & 3100 & q_{\Xi}^{\ast}(3100)\\
\ldots &  &  &  &
\end{array}
\label{2 Fold on SIGMA}%
\end{equation}

2. According to (\ref{subdegen}), each $4$ degenerate energy band on the
symmetry axis $\Sigma$ will be divided into two $2$ fold degenerate bands.
From (\ref{2 Fold on SIGMA}), each of them represents a quark family $q_{\Xi
}^{\ast}$ with $B=1/3,I=1/2,S=-2,Q=$ -1/3, -4/3.\ Thus, we have
\begin{equation}%
\begin{array}
[c]{llc}%
\text{ \ \ \ \ \ }E & \text{\ }\vec{n}\text{ =(n}_{1}\text{,n}_{2}%
\text{,n}_{3}\text{)} & q(m=\varepsilon^{(0)})\\
E_{N}=3/2 & \vec{n}=(\text{101,10-1,011,01-1}) & 2\times q_{\Xi}^{\ast
}(1480)\\
E_{\Gamma}=2 & \vec{n}=(\text{-101,-10-1,0-11,0-1-1}) & 2\times q_{\Xi}^{\ast
}(1660)\\
E_{N}=7/2 & \vec{n}=(\text{121,12-1,211,21-1}) & 2\times q_{\Xi}^{\ast
}(2200)\\
E_{N}=11/2 & \vec{n}=(\text{-121,-12-1,2-11,2-1-1}) & 2\times q_{\Xi}^{\ast
}(2920)\\
E_{\Gamma}=6 & \vec{n}=(\text{1-12,1-1-2,-112,-11-2}) & 2\times q_{\Xi}^{\ast
}(3100)\\
\ldots &  &
\end{array}
\label{4 fold on Segma}%
\end{equation}
3. The single energy bands on the axis $\Sigma(\Gamma-N)$

For the single energy bands, d = 1
$<$%
R = 2 and R-d =1 $\neq2$.\ According to \textbf{Hypothesis IV}. 5,
(\ref{strangeflu}), we have to use (\ref{strangeflu}) instead of
(\ref{strange}):
\begin{equation}
S_{\text{Single}}=\bar{S}_{\Sigma}\pm\Delta S=-2\pm1\text{.}%
\label{sigmasingle}%
\end{equation}
The strange number will take $-1$ and $-3$ alternately from lower to higher
energy bands.

For the energy fluctuation on the axis $\Sigma$, from (\ref{fluc}), we have
\begin{equation}
\Delta\varepsilon=100(J-1)\Delta S\text{ Mev, J = 1, 2, 3...}%
\label{Data-E-Sigma}%
\end{equation}
Since the end points $\Gamma$ and N of the axis $\Sigma$ have different
symmetries, J will take 1, 2, ... from the lowest energy band to higher ones
for each of the two end points respectively.

At E$_{\Gamma}=0,$ $J_{\Gamma}=0,$ from (\ref{Bands of First})\textbf{, the
lowest energy band with n = (0, 0, 0) } represents the free quark family
$q_{N}^{\ast}(940)$.

At $E_{N}=1/2$,$\ J_{N}=0,$ the second lowest energy band with $\vec
{n}=(1,1,0)$ has $S=-1(\Delta S=+1)$ from (\ref{Sign-Segma}). Using
(\ref{UNITED-M}), the energy of the excited state of the quark is
$\varepsilon=1120$ Mev. It is very important to pay attention to \textbf{the
strange quark,\ q}$_{S}(1120),$ \textbf{born on the single energy band of the
axis }$\Sigma$\textbf{\ from the fluctuation } $\Delta S=+1$. It has
\begin{equation}
B=1/3,S=-1,I=0,Q=-1/3,\text{ and m}_{S}=1120.\label{Strange-Quark}%
\end{equation}

At $E_{\Gamma}=2,$ $J_{\Gamma}=1$, the third lowest band with $\vec{n}=($-1,
-1, 0$)$ should have $\Delta S=-1$ from (\ref{Sign-Segma}). Thus, it is the
quark q$_{\Omega}(1660)$ with $S=-3$, $I=0$, $Q=-4/3,$ and $\varepsilon=1660$
Mev from (\ref{UNITED-M}) .

At $E_{N}=9/2$, the fourth one ($\vec{n}=($2, 2, 0$)$) has $\Delta S=+1$ from
(\ref{Sign-Segma}), $S=-2+1=-1,$ and $J_{N}=1,$ $\Delta\varepsilon=0$ from
(\ref{Data-E-Sigma}). The total energy $\varepsilon=2560$ Mev from
(\ref{UNITED-M}). Since $\Delta\varepsilon=0,$ from (\ref{bottom}), we get
that this energy band represents an excited quark q$_{S}(2560)$ with
\begin{equation}
B=1/3,\text{ }S=-1,\text{ }I=0,\text{ }Q=-1/3,\text{ and }M=2560\text{
}Mev.\label{B(2560)}%
\end{equation}

At $E_{\Gamma}=8,$ the fifth one ($\vec{n}=($-2, -2, 0$)$) has B = 1/3, $S=-3
$, $I=0$ and $Q=-4/3$, so it represents an excited state quark q$_{\Omega
}(3720)$ with B = 1/3, $S=-3$, $I=0$, $Q=-4/3.$

At $E_{N}=25/2,$ the sixth one ($\vec{n}=($3, 3, 0$))$ has $J_{N}=2$ and
$S=-2+1=-1$ from (\ref{Sign-Segma}), as well as $\varepsilon=5440+100=5540$
from (\ref{UNITED-M}). According to \textbf{Hypothesis IV}. 5 (\ref{bottom}),
we know that the energy band has a bottom number $b=-1$. It represents an
excited quark state q$_{b}(5540)$ with $I=0$, $b=-1$, $Q=-1/3$, and
$\varepsilon=5540$. In order to make it compatible with the quark model, we
call it the bottom quark. It has
\begin{equation}
B=1/3,S=C=0,b=-1,Q=-1/3,m=5540.\label{Q-Numbers-of -Quark b}%
\end{equation}
It is very important to pay attention to the\textbf{\ bottom quark}
\textbf{born on the single energy band from the fluctuation }$\Delta S=+1$ and
$\Delta\varepsilon=100$~Mev. Using Fig. 5(c), we find the baryons:
\begin{equation}%
\begin{array}
[c]{llllll}%
\text{ \ \ \ \ \ }E & \text{\ (n}_{1}\text{,n}_{2}\text{,n}_{3}\text{)} &
\text{ }S_{G} &  & \Delta\varepsilon & \text{ \ \ }q(m)\\
E_{N}=1/2 & (\text{1,1,0}) & -1 & J_{N}=0 & \text{ \ \ \ \ }0 & q_{S}^{\ast
}(1120)\\
E_{\Gamma}=2 & (\text{-1,-1,0}) & -3 & J_{\Gamma}=1 & \text{ \ \ \ \ }0 &
q_{\Omega}^{\ast}(1660)\\
E_{N}=9/2 & (\text{2,2,0}) & -1 & J_{N}=1 & \text{ \ \ \ \ }0 & q_{S}^{\ast
}(2560)\\
E_{\Gamma}=8 & (\text{-2,-2,0}) & -3 & J_{\Gamma}=2 & -100 & q_{\Omega}^{\ast
}(3720)\\
E_{N}=25/2 & (\text{3,3,0}) & -1 & J_{N}=2 & +100 & q_{b}^{\ast}(5540)\\
E_{\Gamma}=18 & (\text{-3,-3,0}) & -3 & J_{\Gamma}=3 & -200 & q_{\Omega}%
^{\ast}(7220)\\
E_{N}=49/2 & (\text{4,4,0}) & -1 & J_{N}=3 & +200 & q_{b}^{\ast}(9960)\\
\ldots &  &  &  &  &
\end{array}
\label{Q-SIGMA1}%
\end{equation}
\ \ \ \ \ \ \ \ \ 

Continuing the above procedure (see Appendix B), we can use Fig. 2-5 to find
the whole excited states of the quark (the quark spectrum).%

\begin{equation}%
\begin{tabular}
[c]{llllllllllll}%
q$^{Q}$ & q$_{N}^{\frac{2}{3}}$ & q$_{N}^{\frac{-1}{3}}$ & q$_{\Delta}%
^{\frac{5}{3}}$ & q$_{\Delta}^{\frac{2}{3}}$ & q$_{\Delta}^{\frac{-1}{3}}$ &
q$_{\Delta}^{\frac{-4}{3}}$ & q$_{S}^{\frac{-1}{3}}$ & q$_{\Sigma}^{\frac
{2}{3}}$ & q$_{\Sigma}^{\frac{-1}{3}}$ & q$_{\Sigma}^{\frac{-4}{3}}$ &
q$_{\Omega}^{\frac{-4}{3}}$\\
S & \ 0 & \ 0 & \ 0 & \ 0 & \ \ 0 & \ \ 0 & \ -1 & \ -1 & \ \ -1 & \ -1 & -3\\
C & \ 0 & \ 0 & \ 0 & \ 0 & \ \ 0 & \ \ 0 & \ \ 0 & \ \ 0 & \ \ 0 & \ 0 &
\ 0\\
b & \ 0 & \ 0 & \ 0 & \ 0 & \ \ 0 & \ \ 0 & \ \ 0 & \ 0 & \ \ 0 & \ 0 & \ 0\\
I & 1/2 & 1/2 & 3/2 & 3/2 & 3/2 & 3/2 & \ \ 0 & \ \ 1 & \ \ 1 & \ 1 & \ 0\\
I$_{Z}$ & 1/2 & -1/2 & 3/2 & 1/2 & -1/2 & -3/2 & \ \ 0 & \ \ 1 & \ \ 0 & -1 &
\ 0\\
Q & 2/3 & -1/3 & 5/3 & 2/3 & -1/3 & -4/3 & \ \ -1/3 & 2/3 & -1/3 & -4/3 & -4/3\\
&  &  &  &  &  &  &  &  &  &  & \\
q$^{Q}$ & q$_{\Xi}^{\frac{-1}{3}}$ & q$_{\Xi}^{\frac{-4}{3}}$ & \ q$_{C}%
^{\frac{2}{3}}$ & \ q$_{b}^{\frac{-1}{3}}$ & q$_{\Omega_{C}}^{\frac{-1}{3}}$ &
\ q$_{\Xi_{c}}^{\frac{2}{3}}$ & q$_{\Xi_{C}}^{\frac{-1}{3}}$ & q$_{\Sigma_{C}%
}^{\frac{5}{3}}$ & q$_{\Sigma_{C}}^{\frac{2}{3}}$ & q$_{\Sigma_{C}}^{\frac
{-1}{3}}$ & \\
S & -2 & -2 & \ 0 & \ 0 & \ -2 & \ -1 & \ -1 & \ 0 & \ 0 & \ 0 & \\
C & \ 0 & \ 0 & \ 1 & \ 0 & \ 1 & \ 1 & \ 1 & \ 1 & \ 1 & \ 1 & \\
b & \ 0 & \ 0 & \ \ 0 & -1 & \ 0 & \ 0 & \ 0 & \ 0 & \ 0 & \ 0 & \\
I & 1/2 & 1/2 & \ 0 & \ 0 & \ 0 & 1/2 & 1/2 & \ 1 & \ 0 & \ -1 & \\
I$_{Z}$ & 1/2 & -1/2 & \ 0 & \ 0 & \ 0 & 1/2 & -1/2 & \ 1 & \ 0 & \ -1 & \\
Q & -1/3 & -4/3 & 2/3 & -1/3 & -1/3 & 2/3 & -1/3 & 5/3 & 2/3 & -1/3 &
\end{tabular}
\label{Quark-Number}%
\end{equation}%

\begin{equation}%
\begin{tabular}
[c]{lllll}%
q$_{N}^{\ast}($m) & q$_{\Delta}^{\ast}($m) & q$_{S}^{\ast}($m) & q$_{\Sigma
}^{\ast}($m) & q$_{\Xi}^{\ast}($m)\\
q$_{N}^{\ast}($940) & q$_{\Delta}^{\ast}($1300) & q$_{S}^{\ast}($1120) &
q$_{\Sigma}^{\ast}($1210) & q$_{\Xi}^{\ast}($1300)\\
q$_{N}^{\ast}($1210) & q$_{\Delta}^{\ast}($1660) & q$_{S}^{\ast}($1400) &
q$_{\Sigma}^{\ast}($1300) & q$_{\Xi}^{\ast}($1300)\\
q$_{N}^{\ast}($1480) & q$_{\Delta}^{\ast}($1660) & q$_{S}^{\ast}($2020) &
q$_{\Sigma}^{\ast}($1660) & q$_{\Xi}^{\ast}($1480)\\
q$_{N}^{\ast}($1840) & q$_{\Delta}^{\ast}($2020) & q$_{S}^{\ast}($2460) &
q$_{\Sigma}^{\ast}($1660) & q$_{\Xi}^{\ast}($1480)\\
q$_{N}^{\ast}($1840) & q$_{\Delta}^{\ast}($2380) & q$_{S}^{\ast}($2560) &
q$_{\Sigma}^{\ast}($1660) & q$_{\Xi}^{\ast}($1480)\\
q$_{N}^{\ast}($1930) & q$_{\Delta}^{\ast}($2740) & q$_{S}^{\ast}($2650) &
q$_{\Sigma}^{\ast}($1930) & q$_{\Xi}^{\ast}($1660)\\
q$_{N}^{\ast}($1930) & q$_{\Delta}^{\ast}($2740) & q$_{S}^{\ast}($2650) &
q$_{\Sigma}^{\ast}($1930) & q$_{\Xi}^{\ast}($1660)\\
q$_{N}^{\ast}($1930) & q$_{\Delta}^{\ast}($2740) & q$_{S}^{\ast}($2650) &
q$_{\Sigma}^{\ast}($1930) & q$_{\Xi}^{\ast}($1840)\\
q$_{N}^{\ast}($1930) & q$_{\Delta}^{\ast}($2740) & q$_{S}^{\ast}($2740) &
q$_{\Sigma}^{\ast}($1930) & q$_{\Xi}^{\ast}($1840)\\
q$_{N}^{\ast}($2200) &  &  & q$_{\Sigma}^{\ast}($1930) & q$_{\Xi}^{\ast}%
($1930)\\
q$_{N}^{\ast}($2200) &  &  & q$_{\Sigma}^{\ast}($2020) & q$_{\Xi}^{\ast}%
($2020)\\
q$_{N}^{\ast}($2550) &  &  & q$_{\Sigma}^{\ast}($2020) & q$_{\Xi}^{\ast}%
($2020)\\
q$_{N}^{\ast}($2560) &  &  & q$_{\Sigma}^{\ast}($2380) & q$_{\Xi}^{\ast}%
($2200)\\
q$_{N}^{\ast}(2650)$ &  &  & q$_{\Sigma}^{\ast}($2560) & q$_{\Xi}^{\ast}%
($2200)\\
q$_{N}^{\ast}($2650) &  &  & q$_{\Sigma}^{\ast}($2560) & 2$\times$q$_{\Xi
}^{\ast}($2380)\\
q$_{N}^{\ast}($2650) &  &  & q$_{\Sigma}^{\ast}($2560) & 3$\times$q$_{\Xi
}^{\ast}($2560)\\
2q$_{N}^{\ast}($2740) &  &  & q$_{\Sigma}^{\ast}($2650) & 8$\times$q$_{\Xi
}^{\ast}($2740)\\
&  &  & q$_{\Sigma}^{\ast}($2650) & \\
&  &  & 3$\times$q$_{\Sigma}^{\ast}($2740) & \\
&  &  &  &
\end{tabular}
\label{Quark-MassA}%
\end{equation}%

\begin{equation}%
\begin{tabular}
[c]{lllllll}%
q$_{\Omega}^{\ast}(m)$ & q$_{\Omega}^{\ast}(1660)$ & q$_{\Omega}^{\ast}(2460)$%
& q$_{\Omega}^{\ast}(3080)$ & q$_{\Omega}^{\ast}(3720)$ & q$_{\Omega_{c}%
}^{\ast}(7220)$ & \\
q$_{C}^{\ast}(m)$ & q$_{C}^{\ast}(2280)$ & q$_{C}^{\ast}(2450)$ & q$_{C}%
^{\ast}(2540)$ & q$_{C}^{\ast}(2970)$ & q$_{C}^{\ast}(6600)$ & \\
q$_{\Sigma_{C}}^{\ast}(m)$ & q$_{\Sigma_{C}}^{\ast}(2450)$ & q$_{\Sigma_{C}%
}^{\ast}(2540)$ & q$_{\Sigma_{C}}^{\ast}(2970)$ &  &  & \\
q$_{\Xi_{c}}^{\ast}(m)$ & q$_{\Xi_{c}}^{\ast}(2550)$ & q$_{\Xi_{c}}^{\ast
}(3170)$ &  &  &  & \\
q$_{\Omega_{c}}^{\ast}(m)$ & q$_{\Omega_{c}}^{\ast}(2660)$ & q$_{\Omega_{c}%
}^{\ast}(3480)$ &  &  &  & \\
q$_{b}^{\ast}(m)$ & q$_{b}^{\ast}(5540)$ & q$_{b}^{\ast}(9960)$ &  &  &  & \\
&  &  &  &  &  &
\end{tabular}
\label{Quark-MassB}%
\end{equation}
In (\ref{Quark-Number}), we list the quantum numbers (S, C, b, I, I$_{Z,}$ and
Q) of the quarks. In (\ref{Quark-MassA}) and (\ref{Quark-MassB}), we give the
masses of the quarks.

\subsection{The Spectrum of the Baryons}

We have already found the quantum numbers and energies of the quarks
((\ref{Quark-Number}), (\ref{Quark-MassA}), (\ref{Quark-MassB})). Now we will
recognize the baryons ($q^{\ast}u^{\prime}d)$. Using the sum laws
(\ref{SUM-formula}), we can find the quantum numbers and energies of the three
quark systems (baryons). Since the quantum numbers and energies of the
accompanying excited quarks, $u\prime$ and $d^{\prime},$ are already given in
(\ref{Quantum-0f-u}) and (\ref{Quantum-0f-d}), we will focus our attention on
the quantum numbers and energies of the excited quark $q^{\ast}$.

\subsubsection{The energy and the quantum numbers}

\begin{enumerate}
\item  The energy of the system ($q^{\ast}u^{\prime}d^{\prime}$) equals the
energy of the excited quark q$^{\ast}$%
\begin{equation}
M_{_{(}q^{\ast}u^{\prime}d^{\prime})}=m_{q^{\ast}}.\label{baryon-Mass}%
\end{equation}

\item  The quantum numbers of the system are the sum of the constituent
quarks.
\begin{equation}
\text{B \ = 1, S = S}_{q^{\ast}}\text{, C = C}_{q^{\ast}},\text{ b =
b}_{q^{\ast}}\text{ Q =}\sum\text{Q}_{q}\text{,}\label{Baryon-Q-Number}%
\end{equation}
from (\ref{Quantum-0f-u}) and (\ref{Quantum-0f-d}).

\item  The isospin of the system ($q^{\ast}u^{\prime}d^{\prime}$) is found by
\begin{equation}
\overrightarrow{I_{B}}=\overrightarrow{I_{q^{\ast}}}+\overrightarrow
{I_{u^{\prime}}}+\overrightarrow{I_{d^{\prime}}}\label{Baryon - Isospin}%
\end{equation}

\item  The top limit I$_{\max}$ of the isospin of the baryons on the symmetry
axis is determined by
\begin{equation}
2I_{\max}(axis)+1=4+S.\label{I-Top on Axis}%
\end{equation}
From(\ref{DELTA-S})- (\ref{G-S}), we have
\begin{equation}
I_{\max}=3/2\text{ for the axis }\Delta\text{ and the axis }%
D,\label{I-Top of S=0}%
\end{equation}%
\begin{equation}
I_{\max}=1\text{ for the axis }\Lambda\text{ and the axis }%
F,\label{I-Top of S=-1}%
\end{equation}%
\begin{equation}
I_{\max}=1/2\text{ for the axis }\Sigma\text{ and the axis }G\text{.}%
\label{I-Top of S=-2}%
\end{equation}

\item  The top limit (I$_{\max}$) of the isospin of the baryons at the
symmetry point is determined by the highest dimension ($D_{high}$) of the
irreducible representations of the double point group at the point:
\begin{equation}
2I_{\max}+1=D_{high}.\label{I-Top at Point}%
\end{equation}
Since the highest dimension $D_{hig}=4$ for the double point group $\Gamma,$
the group H, the group M, and the group P, we get
\begin{equation}
I_{\max}=3/2,\text{ }at\text{ }\Gamma,H,M,P\text{;}\label{I-Top at H,M,P}%
\end{equation}
and from that the highest dimension $D_{high}=2$ for the double point group
$N,$ we know
\begin{equation}
I_{\max}=1/2,\text{ at the point }N.\label{I-Top at N}%
\end{equation}
\end{enumerate}

\subsubsection{The Baryons on the axis $\Delta(\Gamma-H)$}

1. The 4 fold degenerate bands on the axis $\Delta(\Gamma-$H)

From (\ref{Dalta-Quark}), each 4 fold band reprensets a 4 fold quark family
$q_{\Delta}^{\ast}$ with S = 0, I = 3/2, and Q = 5/3, 2/3, -1/3, -4/3. Adding
the accompanying excited quarks $u^{\prime}$ and $d^{\prime}$, from
(\ref{Baryon - Isospin}), the three quark system ($q_{\Delta}^{\ast}u^{\prime
}d^{\prime}$) has $I=$ $5/2,$ 3/2, 1/2 .\ \ Since the top limit I = 3/2 from
(\ref{I-Top of S=0}), we can have $I=$ 3/2, 1/2 only, omitting I = 5/2. From
(\ref{charge}), we have: for I = 1/2, Q = 1, 0; for I = 3/2, Q = 2, 1, 0, -1.
Thus from (\ref{Baryon-Q-Number}), the system has S = 0, I = 3/2, Q = 2, 1, 0
-1; or S =0, I = 1/2, Q = +1, 0. Comparing the results with the experimental
results \cite{particle} that the baryon family $\Delta(\Delta^{++},\Delta
^{+},\Delta^{0},\Delta^{-})$ has $S=0$, $I=3/2$, $Q=2$, $1$, $0$, $-1$; and
the baryon family $N(N^{+},N^{0})$ has $S=0$, $I=1/2$, and $Q=1$, $0$, we
discover that the system represents the baryon families $\Delta$ and $N$ $($i.
e. for each 4 fold band, we get a $\Delta$ family and a N family). Using
(\ref{baryon-Mass}) and (\ref{Dalta-Quark}), we have
\begin{equation}%
\begin{array}
[c]{lllll}%
E_{H}=1 & \vec{n}=(\text{101,-101,011,0-11}) & q_{\Delta}^{\ast}(1300) &
\Delta(1300); & N(1300)\\
E_{\Gamma}=2\text{ } & \vec{n}=(\text{110,1-10,-110,-1-10}) & q_{\Delta}%
^{\ast}(1660) & \Delta(1660); & N(1660)\\
E_{\Gamma}=2 & \vec{n}=(\text{10-1,-10-1,01-1,0-1-1}) & q_{\Delta}^{\ast
}(1660) & \Delta(1660); & N(1660)\\
E_{H}=3 & \vec{n}=(\text{112,1-12,-112,-1-12}) & q_{\Delta}^{\ast}(2020) &
\Delta(2020); & N(2020)\\
E_{\Gamma}=4 & \vec{n}=(\text{200,-200,020,0-20}) & q_{\Delta}^{\ast}(2380) &
\Delta(2380); & N(2380)\\
E_{H}=5 & \vec{n}=(\text{121,1-21,-121,--1-21}, & q_{\Delta}^{\ast}(2740) &
\Delta(2740); & N(2740)\\
& \text{ \ \ \ \ \ \ \ 211,2-11,-211,-2-11}) & q_{\Delta}^{\ast}(2740) &
\Delta(2740); & N(2740)\\
E_{H}=5 & \vec{n}=(\text{202,-202,022,0-22}) & q_{\Delta}^{\ast}(2740) &
\Delta(2740); & N(2740)\\
E_{H}=5 & \vec{n}=(\text{013,0-13,103,-103}) & q_{\Delta}^{\ast}(2740) &
\Delta(2740); & N(2740)\\
\ldots &  &  &  &
\end{array}
\label{DELTA_4}%
\end{equation}

2. The single bands on the axis $\Delta(\Gamma-H)$

From (\ref{DELTA-ONE}), we have the excited quark q$_{S}^{\ast}$ (with S = -1,
I = 0, Q = -1/3), and the excited quark q$_{C}^{\ast}$ (with $B=1/3$, I =0, S
= 0, C = 1, Q = 2/3). Adding quarks $u^{\prime}$ and $d^{\prime},$ from
(\ref{Baryon-Q-Number}), the three quark system ($u^{\prime}d^{\prime}%
q_{S}^{\ast}$) has S = -1, I = 0, Q = 0 and S = -1, I = 1, Q = +1, 0, -1. They
are baryons $\Lambda$ and $\Sigma.$ The other three quark system ($u^{\prime
}d^{\prime}q_{C}^{\ast}$) has B =1, C = +1, I = 0, Q = +1, and B = 1, C = +1,
I = 1, Q = +2, 1, 0. They are baryons $\Lambda_{C}$ and $\Sigma_{C}$. For
example, at $E_{\Gamma}=4,$the three quark system ($u^{\prime}d^{\prime}%
q_{C}^{\ast}$) is $\Lambda_{C}^{+}(2280)$ and $\Sigma_{C}(2280)$. In order to
make it compatible with the results of the experiments, \textbf{we will call
them the Charmed baryon} $\Lambda_{C}^{+}(2280)$ \cite{charmed}, and the
charmed\ baryon $\Sigma_{C}(2280)$. Using (\ref{DELTA-ONE}) and
(\ref{baryon-Mass}) we have
\begin{equation}%
\begin{array}
[c]{llllll}%
E_{H}=1 & \Delta S=-1 & J=1 & q_{S}^{\ast}(1400) & \Lambda(1400) &
\Sigma(1400)\\
E_{\Gamma}=4 & \Delta S=+1 & J=2 & q_{C}^{\ast}(2280) & \Lambda_{C}%
^{+}(2280) & \Sigma_{C}(2280)\\
E_{H}=9 & \Delta S=-1 & J=3 & q_{S}^{\ast}(4280) & \Lambda(4280) &
\Sigma(4280)\\
E_{\Gamma}=16 & \Delta S=+1 & J=4 & q_{C}^{\ast}(6600) & \Lambda_{C}%
^{+}(6600) & \Sigma_{C}(6600)\\
E_{H}=25 & \Delta S=-1 & J=5 & q_{S}^{\ast}(10040) & \Lambda(10040) &
\Sigma(10040)\\
E_{\Gamma}=36 & \Delta S=+1 & J=6 & q_{C}^{\ast}(13800) & \Lambda_{C}%
^{+}(13800) & \Sigma_{C}(13800)\\
\ldots &  &  &  &  &
\end{array}
\label{Baryon-Data-One}%
\end{equation}
\ 

\subsubsection{ Baryons on the Axis $\Lambda$\ ($\Gamma-P)$\ \ \ \ \ \ \ \ \ \ \ \ \ \ \ \ \ \ \ \ \ \ \ }

From (\ref{Sigma-Quark}), we get quarks $q_{\Sigma}^{\ast}...$ Adding quarks
$u^{\prime}$ and $d^{\prime},$ the three quark system ($q_{\Sigma}^{\ast
}u^{\prime}d^{\prime}$) has the possible isospin values are I = 2, 1, 0, from
(\ref{Baryon - Isospin}). From (\ref{I-Top of S=-1}), we get I = 1, 0 only.
Using (\ref{Baryon-Q-Number}), for I = 1, we have B = 1, S = -1, I = 1, Q =
+1, 0, -1, the system represents a baryon family $\Sigma$; for I = 0, the
system has B =1, S = -1, I = Q = 0, it means a baryon $\Lambda$. Uisng
(\ref{baryon-Mass}), we get
\begin{equation}%
\begin{array}
[c]{llccc}%
E_{P}=3/4 & \vec{n}=(\text{101,011,110}) & q_{\Sigma}^{\ast}(1210) &
\Sigma(1210)\text{;} & \Lambda(1210)\\
E_{\Gamma}=2 & \vec{n}=(\text{1-10,-110,01-1,} & q_{\Sigma}^{\ast}(1660) &
\Sigma(1660)\text{;} & \Lambda(1660)\\
& \text{ \ \ \ \ \ \ \ 0-11,10-1,-101}) & q_{\Sigma}^{\ast}(1660) &
\Sigma(1660)\text{;} & \Lambda(1660)\\
E_{\Gamma}=2 & \vec{n}=(\text{-10-1,0-1-1,-1-10}) & q_{\Sigma}^{\ast}(1660) &
\Sigma(1660)\text{;} & \Lambda(1660)\\
E_{P}=11/4 & \vec{n}=(\text{020,002,200}) & q_{\Sigma}^{\ast}(1930) &
\Sigma(1930)\text{;} & \Lambda(1930)\\
E_{P}=11/4 & \vec{n}=(\text{121,211,112}) & q_{\Sigma}^{\ast}(1930) &
\Sigma(1930)\text{;} & \Lambda(1930)\\
E_{\Gamma}=4 & \vec{n}=(\text{0-20,-200,00-2}) & q_{\Sigma}^{\ast}(2380) &
\Sigma(2380)\text{;} & \Lambda(2380)\\
E_{P}=19/4 & \vec{n}=(\text{1-12,-112,21-1,} & q_{\Sigma}^{\ast}(2650) &
\Sigma(2650)\text{;} & \Lambda(2650)\\
& \text{ \ \ \ \ \ \ \ 2-11,12-1,-121}) & q_{\Sigma}^{\ast}(2650) &
\Sigma(2650)\text{;} & \Lambda(2650)\\
E_{P}=19/4 & \vec{n}=(\text{202,022,220}) & q_{\Sigma}^{\ast}(2650) &
\Sigma(2650)\text{;} & \Lambda(2650)\\
E_{\Gamma}=6 & \vec{n}=(\text{-211,2-1-1,2-1-1}, & q_{\Sigma}^{\ast}(3100) &
\Sigma(3100)\text{;} & \Lambda(3100)\\
& \text{\ \ \ \ \ \ \ 11-2,-12-11-21}) & q_{\Sigma}^{\ast}(3100) &
\Sigma(3100)\text{;} & \Lambda(3100)\\
& \vec{n}\text{=(-1-21,1-2-1,-11-2,} & q_{\Sigma}^{\ast}(3100) &
\Sigma(3100)\text{;} & \Lambda(3100)\\
& \text{\ \ \ \ \ \ 1-1-2,-21-1,-2-11}) & q_{\Sigma}^{\ast}(3100) &
\Sigma(3100)\text{;} & \Lambda(3100)\\
& \vec{n}\text{=(-1-2-1,-1-1-2,-2-1-1)} & q_{\Sigma}^{\ast}(3100) &
\Sigma(3100)\text{;} & \Lambda(3100)\\
\ldots &  &  &  &
\end{array}
\label{Baryon-SEGMA}%
\end{equation}

\subsubsection{The baryons on the axis $\Sigma(\Gamma-N)$}

1. The two fold energy bands on the axis $\Sigma(\Gamma-N)$

For the 2 fold energy bands, from (\ref{2 Fold on SIGMA}), we get the quark
families q$_{\Xi}^{\ast}$ with S = -2, I = 1/2, Q = -1/3, -4/3. \ Adding
quarks $u^{\prime}$ and $d^{\prime},$ from (\ref{Baryon-Q-Number}), the three
quark system ($u^{\prime}d^{\prime}q_{\Xi}^{\ast}$) is the baryon $\Xi$ with S
= -2, I = 1/2, Q = 0, -1. Using (\ref{baryon-Mass}), we have
\begin{equation}%
\begin{array}
[c]{llll}%
E_{\Gamma}=2 & \vec{n}=(\text{1-10,-110}) & q_{\Xi}^{\ast}(1660) & \Xi(1660)\\
E_{N}=5/2 & \vec{n}=(\text{200,020}) & q_{\Xi}^{\ast}(1840) & \Xi(1840)\\
E_{\Gamma}=4 & \vec{n}=(\text{002,00-2}) & q_{\Xi}^{\ast}(2380) & \Xi(2380)\\
& \vec{n}=(\text{-200,0-20}) & q_{\Xi}^{\ast}(2380) & \Xi(2380)\\
E_{N}=9/2 & \vec{n}=(\text{112,11-2}) & q_{\Xi}^{\ast}(2560) & \Xi(2560)\\
E_{\Gamma}=6 & \vec{n}=(-\text{1-12,-1-1-2}) & q_{\Xi}^{\ast}(3100) &
\Xi(3100)
\end{array}
\label{Baryon-Sigema-2}%
\end{equation}

2. The four fold degenerate energy bands on the axis $\Sigma(\Gamma-N)$

From (\ref{4 fold on Segma}), each $4$ fold energy band represent 2 quark
families $2\times q_{\Xi}^{\ast}$. Adding quarks $u^{\prime}$ and $d^{\prime
},$ from (\ref{Baryon-Q-Number}), we get 2 baryon families 2$\times\Xi.$ Uisng
(\ref{baryon-Mass}), we have
\begin{equation}%
\begin{array}
[c]{llcc}%
E_{N}=3/2 & \vec{n}=(\text{101,10-1,011,01-1}) & 2\times q_{\Xi}^{\ast
}(1480) & 2\text{ }\Xi(1480)\\
E_{\Gamma}=2 & \vec{n}=(\text{-101,-10-1,0-11,0-1-1}) & 2\times q_{\Xi}^{\ast
}(1660) & 2\text{ }\Xi(1660)\\
E_{N}=7/2 & \vec{n}=(\text{121,12-1,211,21-1}) & 2\times q_{\Xi}^{\ast
}(2200) & 2\text{ }\Xi(2200)\\
E_{N}=11/2 & \vec{n}=(\text{-121,-12-1,2-11,2-1-1}) & 2\times q_{\Xi}^{\ast
}(2920) & 2\text{ }\Xi(2920)\\
E_{\Gamma}=6 & \vec{n}=(\text{1-12,1-1-2,-112,-11-2}) & 2\times q_{\Xi}^{\ast
}(3100) & 2\text{ }\Xi(3100)\\
\ldots &  &  &
\end{array}
\label{Baryon-Sigema-4}%
\end{equation}

3. The single energy bands on the axis $\Sigma(\Gamma-N)$

From (\ref{Q-SIGMA1}), we get quarks $q_{S}^{\ast}$ and $q_{\Omega}^{\ast} $.
Adding quarks $u^{\prime}$and $d^{\prime},$ the three quark system
($u^{\prime}d^{\prime}q^{\ast}$) has $I=$ 1, 0. Since the top limit $I=1/2$ of
the axis $\Sigma$ from (\ref{I-Top of S=-2}), we get I = 0 only, omitting $I=$
$1$. From (\ref{Baryon-Q-Number}), for $q_{S}^{\ast}$, the system ($u^{\prime
}d^{\prime}q_{S}^{\ast}$) has B = 1, I = 0, S = -1 (Q = 0), it is a strange
baryon $\Lambda;$ for $q_{\Omega}^{\ast}$, the three quark system ($u^{\prime
}d^{\prime}q_{\Omega}^{\ast}$) has B = 1, S = -3, I = 0, Q = -1, it is a
baryon $\Omega.$

At $E_{N}=25/2,$ the sixth band is q$_{b}^{\ast}(5540)$ with $B=1/3,$ $S=C=0,$
$b=-1,$ $Q=-1/3,$ $m=5540.$\ Adding quarks $u^{\prime}$ and $d^{\prime},$ the
three quark system ($u^{\prime}d^{\prime}q_{b}^{\ast}$) has B = 1, S = C = 0,
b = - I, I = 0, and Q = 0 from (\ref{Baryon-Q-Number}). In order to make it
compatible with the results of the experiments, we call this baryon the
\textbf{Bottom baryon }$\mathbf{\Lambda}_{b}\mathbf{(5540)}$ \cite{bottom}.
Uisng (\ref{baryon-Mass}), we have:
\begin{equation}%
\begin{array}
[c]{llll}%
E_{N}=1/2 & \vec{n}=(\text{110}) & q_{S}^{\ast}(1120) & \Lambda(1120)\\
E_{\Gamma}=2 & \vec{n}=(\text{-1-10}) & q_{\Omega}^{\ast}(1660) & \Omega
^{-}(1660)\\
E_{N}=9/2 & \vec{n}=(\text{220}) & q_{S}^{\ast}(2560) & \Lambda(2560)\\
E_{\Gamma}=8 & \vec{n}=(\text{-2-20}) & q_{\Omega}^{\ast}(3720) & \Omega
^{-}(3720)\\
E_{N}=25/2 & \vec{n}=(\text{330}) & q_{b}^{\ast}(5540) & \Lambda_{b}%
^{0}(5540)\\
E_{\Gamma}=18 & \vec{n}=(\text{-3-30}) & q_{\Omega}^{\ast}(7220) & \Omega
^{-}(7220)\\
E_{N}=49/2 & \vec{n}=(\text{440}) & q_{b}^{\ast}(9960) & \Lambda_{b}%
^{0}(9960)\\
\ldots &  &  &
\end{array}
\label{SIGMA_1}%
\end{equation}

\subsubsection{The baryons on the axis D(P-N)}

1. The 4 fold degeneracy energy bands on the axis D(P-N)

From (\ref{Quark-D-4}), \ for each $4$ fold degenerate energy band, we get 2
quark families 2$\times$ q$_{N}^{\ast}$ with B = 1/3, S = C = b = 0, I = 1/2,
Q = 2/3, -1/3. Adding quarks $u^{\prime}$ and $d^{\prime},$ the three quark
system ($q_{N}^{\ast}u^{\prime}d^{\prime}$) has the possible isospin values I
= 1/2 and 3/2 at the point P from (\ref{Baryon - Isospin}). But the possible
isospin I = 1/2 only from (\ref{I-Top at N}) at the point N. From
\ (\ref{Quark-D-4}) and \ (\ref{Baryon-Q-Number}) we can get
\begin{equation}%
\begin{array}
[c]{llll}%
E_{N}=5/2 & \vec{n}=(\text{1-10,-110,020,200}) & 2\text{ }q_{N}^{\ast}(1840) &
2\text{ }N(1840)\\
E_{P}=11/4 & \vec{n}=(\text{-101,0-11,211,121}) & 2\text{ }q_{N}^{\ast
}(1930) & 2\text{ }N(1930)\\
&  &  & 2\text{ }\Delta(1930)\\
E_{N}=7/2 & \vec{n}=(\text{12-1,21-1,-10-1,0-1-1}) & 2\text{\ }q_{N}^{\ast
}(2200) & 2\text{ }N(2200)\\
E_{P}=19/4 & \vec{n}=(\text{-112,1-12,202,022}) & 2\text{ }q_{N}^{\ast
}(2650) & 2\text{ }N(2650)\\
&  &  & 2\text{ }\Delta(2650)\\
\ldots &  &  &
\end{array}
\end{equation}

2. The two fold energy bands on the axis $D(P-N)$

From (\ref{Quark-D-2}), we get $q_{N}^{\ast},$ $q_{S}^{\ast}$, and
$q_{C}^{\ast}$. Adding quarks u$^{\prime}$ and d$^{\prime}$, we get the
baryons N (and $\Delta$ at the point P), $\Lambda$, and $\Lambda_{C}$ from
(\ref{baryon-Mass}) and (\ref{Baryon-Q-Number}).

At $E_{P}=3/4,$ the two energy bands represent a quark family $q_{N}^{\ast
}(1210)\Rightarrow$the baryon families $N(1210)$ and $\Delta(1210)$.
\textbf{The baryon }$\Delta(1210)$\textbf{\ is the ground state of the
}$\Delta$ \textbf{baryons.}

At $E_{N}=9/2$, $\vec{n}=(220,-1-10)$ and $\vec{n}=(11-2,00-2)$. For the first
2 energy bands they represent the quark family\ $q_{N}^{\ast}(2560)\Rightarrow
$the baryon family $N(2560)$. For the second 2 energy bands, one represents a
charmed quark $q_{C}^{\ast}(2760)$ (with $I=0$ and $Q=+1$)$\Rightarrow$ the
Charmed baryon $\Lambda_{C}^{+}(2760)$, the other represents $q_{S}^{\ast
}(2360)\Rightarrow$ a baryon $\Lambda(2360)$.

Therefore, using (\ref{baryon-Mass}) and (\ref{Baryon-Q-Number}), we have%

\begin{equation}%
\begin{array}
[c]{lllll}%
E_{N}=1/2 & \vec{n}=(\text{000,110}) &  &  & \\
\text{\ \ }J_{N}=0 & \vec{n}=(\text{000}) & q_{N}^{\ast}(940) & N(940) & \\
\text{ \ }\Delta\varepsilon=0 & \vec{n}=(\text{110}) & q_{S}^{\ast}(1120) &
\Lambda(1120) & \\
E_{P}=3/4 & \vec{n}=(\text{101,011}) & q_{N}^{\ast}(1210) & N(1210) &
\Delta(1210)\\
E_{N}=3/2 & \vec{n}=(\text{10-1,01-1}) & q_{N}^{\ast}(1480) & N(1480) & \\
E_{P}=11/4 & \vec{n}=(\text{002,112}) & q_{N}^{\ast}(1930) & N(1930) &
\Delta(1930)\\
E_{N}=9/2 &  & \varepsilon^{(0)}=2560 &  & \\
\text{ \ }J_{N}=1 & \vec{n}=(\text{220,-1-10}) & q_{N}^{\ast}(2560) &
N(2560) & \\
\text{ \ }J_{N}=2 & \vec{n}=(\text{11-2,00-2}) &  &  & \\
& \text{ \ }\Delta S=+1 & q_{C}^{\ast}(2660) & \Lambda_{C}^{+}(2660) & \\
& \text{ \ }\Delta S=-1 & q_{S}^{\ast}(2460) & \Lambda(2460) & \\
E_{P}=19/4 & \vec{n}=(\text{-121,2-11}) & q_{N}^{\ast}(2650)) & N(2650) &
\Delta(2650)\\
E_{N}=11/2 & \vec{n}=(\text{2-1-1,-12-1}) & q_{N}^{\ast}(2920) & N(2920) & \\
\ldots &  &  &  &
\end{array}
\end{equation}

\subsubsection{The Axis $G(M-N)$}

There are $2$, $4$, and $6$ fold energy bands on the axis (see Fig. 4(b)).

1. The two fold energy bands on the axis $G(M-N)$

From (\ref{Quark-G-2}), for the $2$ fold energy band (see Fig. 4(b)), we have
$q_{N}^{\ast},$ $q_{S}^{\ast}$, $q_{\Xi}^{\ast}$. Adding quarks u$^{\prime}$
and d$^{\prime}$, we get the baryons N, $\Lambda$, and $\Xi$ from
(\ref{baryon-Mass}), (\ref{Baryon-Q-Number}), as follows$\ $
\begin{equation}%
\begin{array}
[c]{llll}%
E_{N}=1/2 & \vec{n}=(\text{000,110}) &  & \\
\text{\ \ \ }J_{N}=0 & \vec{n}=(\text{000}) & q_{N}^{\ast}(940) & N(940)\\
& \vec{n}=(\text{110}) & q_{S}^{\ast}(1120) & \Lambda(1120)\\
E_{M}=1 & \vec{n}=(\text{101,10-1}) & q_{\Xi}^{\ast}(1300) & \Xi(1300)\\
& \vec{n}=(\text{200,1-10}) & q_{\Xi}^{\ast}(1300) & \Xi(1300)\\
E_{N}=3/2 & \vec{n}=(\text{011,01-1}) & q_{\Xi}^{\ast}(1480) & \Xi(1480)\\
E_{N}=5/2 & \vec{n}=(\text{020,-110}) & q_{\Xi}^{\ast}(1840) & \Xi(1840)\\
E_{M}=3 & \vec{n}=(\text{2-11,2-1-1}) & q_{\Xi}^{\ast}(2020) & \Xi(2020)\\
E_{M}=5 & \vec{n}=(\text{3-10,2-20}) & q_{\Xi}^{\ast}(2740) & \Xi(2740)\\
E_{N}=11/2 & \vec{n}=(\text{-121,-12-1}) & q_{\Xi}^{\ast}(2920) & \Xi(2920)\\
\ldots &  &  &
\end{array}
\end{equation}
\ 

2. The four fold energy bands on the axis $G(M-N)$

From (\ref{Quark-G-4}), for each $4$ fold energy band, we have $2\times
q_{\Xi}^{\ast}$. Adding quarks u$^{\prime}$ and d$^{\prime}$, we get the two
baryon families ( 2$\times\Xi)$ from (\ref{baryon-Mass}) and
(\ref{Baryon-Q-Number}):
\begin{equation}%
\begin{array}
[c]{lllll}%
E_{M}=3 & \vec{n}=(\text{0-11,0-1-1,} & \text{211,21-1}) & 2\times q_{\Xi
}^{\ast}(2020) & 2\text{ }\Xi(2020)\\
E_{N}=7/2 & \vec{n}=(\text{-101,-10-1,} & \text{121,12-1}) & 2\times q_{\Xi
}^{\ast}(2200) & 2\text{ }\Xi(2200)\\
E_{M}=5 & \vec{n}=(\text{301,30-1,} & \text{1-21,1-2-1}) & 2\times q_{\Xi
}^{\ast}(2740) & 2\text{ }\Xi(2740)\\
... &  &  &  &
\end{array}
\end{equation}

Continuing the above procedure, using Fig. 2-5, from (\ref{baryon-Mass}) and
(\ref{Baryon-Q-Number}), we can find the whole baryon spectrum. Our results
are shown in Tables $1$ though $6$.\ \ \ \ \ \ \ \ \ \ \ \ \ \ \ 

\subsection{The Spectrum of Mesons}

In Section V.A. we have found the quark spectrum (\ref{Quark-Number}),
(\ref{Quark-MassA}), (\ref{Quark-MassB}). Using the definition that a meson =
$(q\overline{q)}$ and a phenomenological mass formula for the mesons, we can
find the meson spectrum (see nest paper: Jiao-Lin Xu and Xin Yu, The Meson Spectrum)

\newpage

\ \ \ \ \ \ \ \ \ \ \ \ \ \ \ \ \ \ \ \ \ \ \ \ \ \ \ \ 

\section{The \textbf{SU(N) Approximation (The Quark Model)}}

In order to find the relationship between the Quark Model and the BCC model,
we study the SU(N) symmetry approximation. In the approximation, we will see
that the BCC model provides the physical foundation for the Quark Model. We
have found \textbf{the quark spectrum }(\ref{Quark-Number}),
(\ref{Quark-MassA}), (\ref{Quark-MassB}) in Section V. In the SU(N) (N =3, 4,
5) approximation, based on the quark spectrum, we assume the following:

1\textbf{. The lowest energy quark excited state (ground state) of the each
flavored quark excited states are regarded as the elemental particle (quark),
omitting all other higher energy excited quarks of the quark spectrum
(\ref{Quark-MassA}) and (\ref{Quark-MassB}).}

There are 4 kinds of the flavored quarks in the BCC model. They are unflavored
quarks, q$_{N}$ and q$_{\Delta}$; the strange quarks, q$_{S}$, q$_{\Sigma},$
q$_{\Xi}$, and q$_{\Omega}$; the charmed quark q$_{C}$; and the bottom quark
q$_{b}$.

2. \ \textbf{The N quarks satisfy the SU(N)\ symmetry and belong to the
fundamental representations of the SU(N) group in the strong interactions.
Since the N quarks are the excited states of the fnndamental quarks and the
fundamental quarks satisfy the SU(2) symmetry, so SU(3), SU(4), and SU(5) are
the natural extensions of SU(2).}\ 

3. \textbf{A baryon is made of three quarks (qqq state). }

4. \textbf{A meson is made of a quark and an antiquark (}$q\overline{q}%
$\textbf{\ state). }

\textbf{5. The quantum numbers of the baryons and the mesons can
be\ determined by the sum formulae (\ref{SUM-formula}).\ }\ \ \ \ \ \ \ \ \ \ \ \ \ \ \ \ \ \ \ \ \ \ \ \ \ \ \ \ \ 

\subsection{\textbf{The quarks }}

1. There are 2 kinds of unflavored quarks, q$_{N}$ and q$_{\Delta},$ with S =
C = b =0 from (\ref{Quark-Number}). According to the\textbf{\ quark spectrum
}(\ref{Quark-MassA}), the quark family q$_{N}(940)$ (with $B=1/3,$ S = 0,
$s=I=1/2,$ and $Q=2/3,-1/3)$ is the ground state of the unflavored quarks.

2. For S $\neq0$, C = b = 0, from (\ref{Quark-Number}), there are 4 kinds of
strange quark excited states: the strange quark q$_{S},$ the strange quark
family q$_{\Sigma}$, the strange quark family q$_{\Xi}$, and the strange quark
q$_{\Omega}$. From (\ref{Quark-MassA}), the strange quark q$_{S}(1120)$ (with
B = 1/3, S = -1, s = 1/2, I = 0, and Q = -1/3) is their ground state.

3. For C $\neq0$ and S = b = 0, from (\ref{Quark-Number}), there is only one
kind of charmed quark excited states. The Charmed Quark q$_{C}(2280)$ (with B
= 1/3, C = +1, s = 1/2, I = 0, Q = 2/3) is the ground state.

4. For b $\neq0$ and S = C = 0, from (\ref{Quark-Number}), there is only one
kind of Bottom Quark q$_{b}$. From (\ref{Quark-MassB}), the Bottom Quark,
q$_{b}(5540),$ is the ground state.\ 

In the SU(N) approximation, since the three quarks are in the symmetry
positions inside the baryon (p and n), so\ they have the same masses in a
proton (uud) and a neutron (udd). Thus, $m_{u}=m_{d}=$ $M_{N}(940)/3=313$%
(Mev)$.$ Using the baryons $\Lambda=$ (uds) and $\Sigma=($uus, uds, dds), and
the average mass of $\Lambda$ and $\Sigma,$ $\overline{M_{\Lambda,\Sigma}}=$
$1/2(1116+1193)=$ $1155$(Mev)$,$ we get m$_{S}=529.$ Similarly, using
$\Lambda_{C}=(udc)$ and $\Lambda_{b}=(udb)$, we can get m$_{C}=1654$(Mev) and
m$_{b}=5014$(Mev)$.$ To sum up, the five quarks have the following quantum
numbers and masses:%

\begin{equation}%
\begin{tabular}
[c]{lllllllll}%
q$\setminus\#$ & B & I & I$_{z}$ & S & C & b & \ Q & M$_{Mev}$\\
\ q$_{u}$ & $\frac{1}{3}$ & 1/2 & $1/2$ & 0 & 0 & 0 & 2/3 & {\small 313}\\
\ q$_{d}$ & $\frac{1}{3}$ & 1/2 & -1/2 & 0 & 0 & 0 & -1/3 & {\small 313}\\
$\ $q$_{s}$ & $\frac{1}{3}$ & 0 & 0 & -1 & 0 & 0 & -1/3 & {\small 529}\\
\ q$_{c}$ & $\frac{1}{3}$ & 0 & 0 & 0 & 1 & 0 & 2/3 & {\small 1654}\\
\ q$_{b}$ & $\frac{1}{3}$ & 0 & 0 & 0 & 0 & -1 & \ -1/3 & 5014
\end{tabular}
\label{5-Quarks}%
\end{equation}
\ Comparing (\ref{5-Quarks}) with Table 12.1 of \cite{Table 12. 1}, we find
that the 5 quarks in (\ref{5-Quarks}) are the same quarks in Table 12.1 of the
Quark\ Model (in the quantum numbers B, I, I$_{z},$ S, C, b).
\textbf{Therefore, the 5 quarks (q}$_{u}$, \textbf{q}$_{d}$, \textbf{q}$_{s}$,
\textbf{q}, and \textbf{q}$_{b}$)\textbf{\ of the SU(N) are the same quarks as
the Quark Model. In other words, }
\begin{equation}
\text{\textbf{q}}_{u}\text{ = \textbf{u}, \textbf{q}}_{d}\text{ = \textbf{d},
\textbf{q}}_{s}\text{ = \textbf{s}, \textbf{q}}_{c}\text{ = \textbf{c}, and
\textbf{q}}_{b}\text{ = \textbf{b}.}\label{5-SAME-QUARKS}%
\end{equation}

\subsection{\textbf{The fundamental representations of the SU(N)\ groups.}}

According to \textbf{HYPOTHESIS I, }the fundamental quarks, u and d, satisfy
the SU(2) symmetries. From the combined approximation, the quarks q$_{S}$,
q$_{C}$, and q$_{b}$ are the excited states of the fundamental quark q (u and
d). Therefore, the SU(3) (for the quarks u, d, and s), the SU(4) (for the
quarks u, d, s, and c), and the SU(5) (for the quarks u, d, s, c, and b)
symmetries are the natural extensions of the SU(2).\ \ \ \ \ \ \ \ \ \ \ \ \ \ \ \ \ \ \ \ \ \ \ \ \ \ 

For the 5 quarks, we assume that their wave functions are
\begin{equation}
\Psi_{u}\text{=}\Psi\text{(B=1/3, I=1/2, I}_{z}\text{=1/2, S=C=b=0, Q=2/3,
m=313),}\label{Quark-u}%
\end{equation}%
\begin{equation}
\Psi_{d}\text{=}\Psi\text{(B=1/3, I=1/2, I}_{z}\text{=-1/2, S=C=b=0, Q=-1/3,
m=313),}\label{Quark-d}%
\end{equation}%
\begin{equation}
\Psi\text{s=}\Psi\text{(B=1/3, I=0, I}_{z}\text{=0, S=-1, C=b=0, Q=-1/3,
m=529),}\label{Quark-s}%
\end{equation}%
\begin{equation}
\Psi_{c}\text{=}\Psi\text{(B=1/3, I=0, I}_{z}\text{=0, C=1, S=b=0, Q=2/3,
m=1654),}\label{Quark-c}%
\end{equation}%
\begin{equation}
\Psi_{b}\text{=}\Psi\text{(B=1/3, I=0, I}_{z}\text{=0, b=-1, S=C=0, Q=-1/3,
m=5014).}\label{Quark-b}%
\end{equation}

Fundamental representation \textbf{3} of the SU(3) can be written as a column matrix%

\begin{equation}
\mathbf{3}=\left(
\begin{tabular}
[c]{l}%
$\Psi_{u}$\\
$\Psi_{d}$\\
$\Psi_{s}$%
\end{tabular}
\right)  {\LARGE .}\label{SU(3)-G}%
\end{equation}

Fundamental representation \textbf{4} of the SU(4) can be written as a column matrix%

\begin{equation}
\mathbf{4}=\left(
\begin{tabular}
[c]{l}%
$\Psi_{u}$\\
$\Psi_{d}$\\
$\Psi_{s}$\\
$\Psi_{c}$%
\end{tabular}
\right)  {\LARGE .}\label{SU(4)-G}%
\end{equation}

Fundamental representation \textbf{5} of the SU(5) can be written as a column matrix%

\begin{equation}
\mathbf{5}=\left(
\begin{tabular}
[c]{l}%
$\Psi_{u}$\\
$\Psi_{d}$\\
$\Psi_{s}$\\
$\Psi_{c}$\\
$\Psi_{b}$%
\end{tabular}
\right)  {\LARGE .}\label{SU(5)-G}%
\end{equation}

\subsection{\textbf{The baryons (qqq states). }}

From the primitive cell approximation, we can draw a reasonable assumption
that \textbf{a baryon is made of three quarks (qqq).}

The ``ordinary'' baryons are made up of u, d, and s quarks. The three quarks
satisfy SU(3) symmetry which requires that the baryons belong to the
multiplets on the right side of%

\begin{equation}
\mathbf{3}\otimes\mathbf{3}\otimes\mathbf{3}=\mathbf{10}_{s}\oplus
\mathbf{8}_{M}\oplus\mathbf{8}_{M}\oplus\mathbf{1}\label{3x3x3}%
\end{equation}%
\begin{equation}%
\begin{tabular}
[c]{l}%
\begin{tabular}
[c]{lllll}%
$\mathbf{10:}$ & $\Delta^{++},\Delta^{+},\Delta^{0},\Delta^{-};$ & $\Sigma
^{+},\Sigma^{0},\Sigma^{-};$ & $\Xi^{-},\Xi^{0};$ & $\Omega^{-}.$%
\end{tabular}
\\%
\begin{tabular}
[c]{llllll}%
$\mathbf{8:}$ \  & $p$, $n;$ & $\Lambda,$ & $\Sigma^{+},\Sigma^{0},\Sigma
^{-};$ & $\Xi^{-},\Xi^{0}.$ &
\end{tabular}
\\%
\begin{tabular}
[c]{ll}%
$\mathbf{1:}$ \  & $\Lambda.$%
\end{tabular}
\end{tabular}
\label{8 +10+1}%
\end{equation}
From (\ref{8 +10+1}), we can see that the intrinsic quantum numbers of the
SU(3) approximation are in accordance with the experimental results. However,
the SU(3) approximation can not give a mass spectrum of the baryons. It can
only give some relationships inside a multiplet. Using (\ref{SUM-formula}), we
can deduce the formula%

\begin{equation}
M_{j}-M_{i}=\sum m_{q_{j}}-\sum m_{q_{i}}.
\end{equation}
\qquad\qquad\qquad\ \qquad\ \ 

For an $8$-$multiplet,$we have%

\begin{equation}%
\begin{tabular}
[c]{l}%
$M_{p}-M_{\Lambda}=[(m_{u}+m_{u}+m_{d})-(m_{u}+m_{d}+m_{s})],$\\
$M_{n}-M_{\Lambda}=[(m_{u}+m_{d}+m_{d})-(m_{u}+m_{d}+m_{s})],$\\
$M_{\Xi^{0}}-M_{\Lambda}=[(m_{u}+m_{s}+m_{s})-(m_{u}+m_{d}+m_{s})],$\\
$M_{\Xi^{-}}-M_{\Sigma^{+}}=[(m_{d}+m_{s}+m_{s})-(m_{u}+m_{u}+m_{s})].$\\
$M_{\Xi^{-}}-M_{\Sigma^{0}}=[(m_{d}+m_{s}+m_{s})-(m_{u}+m_{d}+m_{s})].$\\
$M_{\Xi^{-}}-M_{\Sigma^{-}}=[(m_{d}+m_{s}+m_{s})-(m_{d}+m_{d}+m_{s})].$%
\end{tabular}
\end{equation}
Using 1/2 ($M_{p}+M_{n}$) $=M_{N},$ 1/2($M_{\Xi^{0}}+M_{\Xi^{-}}$) $=M_{\Xi},$
and 1/3($M_{\Sigma^{+}}+M_{\Sigma^{0}}+M_{\Sigma^{-}}$) $=M_{\Sigma},$ and
$m_{u}=m_{d}$, we get the GMO mass relation \cite{QMFORM} for the baryon octet%

\begin{equation}
1/2(M_{N}+M_{\Xi})=1/4(3M_{\Lambda}+M_{\Sigma})\label{GMO-8}%
\end{equation}

For a $\mathbf{10}$-$multiplet$, using 1/2($M_{\Xi^{0}}+M_{\Xi^{-}}$)
$=M_{\Xi},$ 1/3($M_{\Sigma^{+}}+M_{\Sigma^{0}}+M_{\Sigma^{-}}$) $=M_{\Sigma},$
1/4($M_{\Delta^{++}}+M_{\Delta^{+}}+M_{\Delta^{0}}+M_{\Delta^{-}})=M_{\Delta
},$ and $m_{u}=m_{d}$, we have $\ \ \ \ $%
\begin{equation}%
\begin{tabular}
[c]{l}%
$M_{\Omega}-M_{\Xi}=(m_{s}$+$m_{s}$+$m_{s})$-$(m_{u}$+$m_{s}$+$m_{s}%
)=q_{s}-q_{u}$,\\
$M_{\Xi}-M_{\Sigma}=(m_{u}$+$m_{s}$+$m_{s})$-$(m_{u}$+$m_{d}$+$m_{s}%
)=q_{s}-q_{u},$\\
$M_{\Sigma}-M_{\Delta}=(m_{u}$+$m_{d}$+$m_{s})$-$(m_{u}$+$m_{u}$+$m_{d}%
)=q_{s}-q_{u}$%
\end{tabular}
\label{qw}%
\end{equation}
Thus, we get the GMO mass relation \cite{QMFORM} for Decuplet
\begin{equation}
M_{\Omega}-M_{\Xi}=M_{\Xi}-M_{\Sigma}=M_{\Sigma}-M_{\Delta}\label{GMO-10}%
\end{equation}

The addition of the c quark to the light quarks extends the flavor symmetry to
SU(4). Similar to $3\otimes3\otimes3,$ we have%

\begin{equation}
\mathbf{4}\otimes\mathbf{4}\otimes\mathbf{4}=\mathbf{4}\oplus\mathbf{20}%
\oplus\mathbf{20}\oplus\mathbf{20}\label{4x4x4}%
\end{equation}
Fig. 12.2 of \cite{Figure 12. 2}\ (a) and (b) show the SU(4) baryon
multiplets. The SU(3) octet is on the ground floor of the SU(4) $multiplet$
$\mathbf{20}$ in Fig. 12.2 (a). The SU(3) decuplet is on the ground floor of
the SU(4) $multiplet$ $\mathbf{20}$ in Fig. 12.2 (b). All discovered charmed
baryons, $\Lambda_{C},$ $\Omega_{C},$ $\Xi_{C},$ and $\Sigma_{c},$ have only
one charmed quark.

The addition of a bottom quark extends the flavor symmetry to SU(5). It will
produce many high energy baryons which have not been discovered by the experiments.

For the ``ordinary'' baryons, flavor and spin may be combined in an
approximate flavor-spin SU(6) in which the six basic states are $d\uparrow
,d\downarrow,u\uparrow,u\downarrow,s\uparrow,$ and $s\downarrow$
($\uparrow,\downarrow=$ spin up, down). Then the baryons belong to the
multiplets on the right side of%

\begin{equation}
\mathbf{6}\otimes\mathbf{6}\otimes\mathbf{6}=\mathbf{56}_{s}\oplus
\mathbf{70}_{M}\oplus\mathbf{70}_{M}\oplus\mathbf{20}_{A.}%
\label{SU(6)-Multiplets}%
\end{equation}
These SU(6) multiplets decompose into flavor SU(3) multiplets as follows:

\qquad%
\begin{equation}
\mathbf{56}=\text{ }^{4}\mathbf{10}\text{ }\oplus\text{ }^{2}\mathbf{8}%
\end{equation}%
\begin{equation}
\mathbf{70}=\text{ }^{2}\mathbf{10}\oplus\text{ }^{4}\mathbf{8}\text{ }%
\oplus\text{ }^{2}\mathbf{8}\text{ }\oplus\text{ }^{2}\mathbf{1}%
\end{equation}%
\begin{equation}
\mathbf{20}=\text{ }^{2}\mathbf{8}\text{ }\oplus\text{ }^{4}\mathbf{1}%
\end{equation}
where the superscript $(2s+1)$ gives the net spins of the quarks for each
baryon in the SU(3) multiplets. The J$^{P}=(1/2)^{+}$ octet containing the
nucleon N(939) and $J^{P}=(3/2)^{+}$ decuplet containing $\Delta(1232)$
together make up the ``ground-state'' $\mathbf{56}$-plet in which the orbital
angular momenta are zero. $\mathbf{70}$ and $\mathbf{20}$ require some
excitation of the spacial part of the state function.

Since we have omitted the higher energy excited states of the quarks, we
cannot get the higher energy baryons. Similar to the Quark Model, the states
with nonzero orbital angular momenta are classified in SU(6)$\otimes SO(3)$
\cite{SO(3)} supermultiplets. The SU(6)$\otimes$SO(3) provides a suitable
framework for describing baryon state functions.

It is useful to classify the baryons into bands that have the same number N of
quanta of excitation. Each band consists of a number of supermultiplets,
specified by (D, L$_{N}^{P}$), where D is the dimensionality of the SU(6)
representation, L is the total orbital angular momentum, J is the total
angular momentum, and P is the total parity.

Table 12.4 of \cite{Table 12. 4.}\ shows the Quark Model (the SU(N=3)
approximation) assignments for many of the known baryons in terms of a
flavor-spin SU(6) basis. Only the dominant representation is listed.
Assignments for some states, especially for $\Lambda(1810),$ $\Lambda(2350),$
$\Xi(1820),$ and $\Xi(2030)$ are merely educated guesses.
\begin{equation}%
\begin{tabular}
[c]{lllll}%
J$^{P}$ & (D, L$_{N}^{P})$ & S & \ \ \ \ \ \ \ \ \ \ \ \ \ Octet members &
Singlets\\
1/2$^{+}$ & (56,0$_{0}^{+}$) & 1/2 & N(939), \ $\ \Lambda$(1116), $\Sigma
$(1193), $\Xi$(1318) & \\
1/2$^{+}$ & (56,0$_{2}^{+}$) & 1/2 & N(1440), $\Lambda$(1600), $\Sigma$(1660),
$\Xi(?)\ \ $\  & \\
1/2$^{-}$ & (70,1$_{1}^{-}$) & 1/2 & N(1535), $\Lambda$(1670), $\Sigma$(1620),
$\Xi$(?) & $\Lambda$(1405)\\
3/2$^{-}$ & (70,1$_{1}^{-}$) & 1/2 & N(1520), $\Lambda$(1690), $\Sigma$(1670),
$\Xi$(1820) & $\Lambda$(1520)\\
1/2$^{-}$ & (70,1$_{1}^{-}$) & 3/2 & N(1650), $\Lambda$(1800), $\Sigma$(1750),
$\Xi$(?) & \\
3/2$^{-}$ & (70,1$_{1}^{-}$) & 3/2 & N(1700), $\Lambda$(?), \ \ \ \ \ $\Sigma
$(?), $\ \ \ \ \Xi$(?) & \\
5/2$^{-}$ & (70,1$_{1}^{-}$) & 3/2 & N(1675), $\Lambda$(1830), $\Sigma$(1775),
$\Xi$(?) & \\
1/2$^{+}$ & (70,0$_{2}^{+}$) & 1/2 & N(1710), $\Lambda$(1810), $\Sigma$(1880),
$\Xi$(?) & $\Lambda$(?)\\
3/2$^{+}$ & (56,2$_{2}^{+}$) & 1/2 & N(1720), $\Lambda$(1890), $\Sigma$(?),
\ \ \ \ \ $\Xi$(?) & \\
5/2$^{+}$ & (56,2$_{2}^{+}$) & 1/2 & N(1680), $\Lambda$(1820), $\Sigma$(1915),
$\Xi$(2030) & \\
7/2$^{-}$ & (70,3$_{3}^{-}$) & 1/2 & N(2190), $\Lambda$(?), $\ \ \ \ \ \Sigma
$(?), \ \ \ \ $\Xi$(?) & $\Lambda$(2100)\\
9/2$^{-}$ & (70,3$_{3}^{-}$) & 3/2 & N(2250), $\Lambda$(?), $\ \ \ \ \ \Sigma
(?)$, \ \ \ \ $\Xi$(?) & \\
9/2$^{+}$ & (56,4$_{4}^{+}$) & 1/2 & N(2220), $\Lambda$(2350), $\Sigma(?)$,
\ \ \ \ \ $\Xi$(?) &
\end{tabular}
\label{Octet Members}%
\end{equation}%

\begin{equation}%
\begin{tabular}
[c]{llll}%
J$^{P}$ & (D, L$_{N}^{P})$ & S & Decuplet members\\
3/2$^{+}$ & (56,0$_{0}^{+}$) & 3/2 & $\Delta$(1232), $\Sigma$(1385), $\Xi
$(1530), $\Omega$(1672)\\
1/2$^{-}$ & (70,1$_{1}^{-}$) & 1/2 & $\Delta$(1620), $\Sigma$(?),
\ \ \ \ \ $\ \Xi$(?), $\ \ \ \ \ \ \ \Omega$(?)\\
3/2$^{-}$ & (70,1$_{1}^{-}$) & 1/2 & $\Delta$(1700), $\Sigma$(?),
\ \ \ \ \ $\ \Xi$(?), $\ \ \ \ \ \ \ \Omega$(?)\\
5/2$^{+}$ & (56,2$_{2}^{+}$) & 3/2 & $\Delta$(1905), $\Sigma$(?),
\ \ \ \ \ $\ \Xi$(?), $\ \ \ \ \ \ \ \Omega$(?)\\
7/2$^{+}$ & (56,2$_{2}^{+}$) & 3/2 & $\Delta$(1950), $\Sigma$(2030), $\ \Xi
$(?), $\ \ \ \ \ \ \ \Omega$(?)\\
11/2$^{+}$ & (56,4$_{4}^{+}$) & 3/2 & $\Delta$(2420), $\Sigma$(?),
\ \ \ \ \ $\ \Xi$(?), $\ \ \ \ \ \ \ \Omega$(?)
\end{tabular}
\label{Decuplet members}%
\end{equation}

If SU(3) is the main symmetry group (u, d, and s), then the baryons which are
composed of the three quarks (u, d, and s) will be classed into Octets and
Decuplets. However, in fact

1. There are six possible Decuplets, but only one is complete from
(\ref{Decuplet members}).

2. Four out of the five other possible Decuplets that each needs four members
(SU(3)), there is only one member in the 4 Decuplets. One member means nothing.

3. There are twelve possible Octets, but only three Octets are complete from
(\ref{Octet Members}).

Therefore, the above experimetal results show that the SU(3) is not the most
important symmetry group for the baryons which are made up of the ``ordinary''
quarks u(313), d(313), and s(529). Moreover, SU(4) symmetries (based on
u(313), d(313), s(529), c(1654)) and SU(5) symmetries (based on u(313),
d(313), s(529), c(1654), b(5014)) are very badly broken by the mass of charmed
quark c and the mass of bottom quark b. Considering the\ fact that the Quark
Model cannot deduce the mass spectrum of the baryons, \textbf{we shall replace
the SU(N) groups by the body center cubic groups (the space group and the
point groups).}

\subsection{\textbf{The mesons (the }$q\overline{q}$\textbf{\ states)}}

According to the primitive cell approximation, a meson is made up of a quark
and an antiquark (the $q\overline{q}$ states). The nine possible
$q\overline{q}$ combinations containing q$_{u}$, q$_{d}$, and q$_{s}$ quarks
group themselves into an octet and a singlet:
\begin{equation}
\mathbf{3}\text{ }\mathbf{\otimes}\text{ }\overline{3}\mathbf{=8\oplus
1}\label{8+1}%
\end{equation}%
\begin{equation}%
\begin{tabular}
[c]{lllll}%
\textbf{8:} & $K^{+}$, $K^{0};$ & $\pi^{+}$,$\pi^{0}$,$\pi^{-};$ &
$\overline{K^{0}}$,$K^{-};$ & $\eta_{8}.$\\
\textbf{1:} & $\eta_{1}.$ &  &  &
\end{tabular}
\label{Meson 8 fold state}%
\end{equation}
A fourth quark such as Charmed quark q$_{c}$ can be included by extending the
symmetry to SU(4), as shown in Fig. 12.1 of \textbf{\cite{Table 12. 1}}. Fig.
12.1 (a) shows the SU(4) 16-plets for the pseudoscalar mesons made of u, d, s,
and c quarks. Fig. 12.1 (b) shows the SU(4) 16-plets for the vector mesons
made of u, d, s, and c quarks. The octets and singlets of light mesons
(\ref{Meson 8 fold state}) occupy the central planes, to which the
c$\overline{c}$ states have been added. The neutral mesons at the center of
the planes are mixtures of u$\overline{u}$, d$\overline{d}$, s$\overline{s}$,
and c$\overline{c}$. The Bottom quark extends the symmetry to SU(5). Thus, we
can get the same result as the Quark Model \cite{QuarkModel}. For example, we
give out the mesons with the angular momentum L = 0 and the spin s = 0 (spin
single S wave mesons-q$\overline{q},$ $\uparrow\downarrow=$ spin up, down) as follows%

\begin{equation}%
\begin{tabular}
[c]{llllll}%
1$^{1}S_{0}$ & q$_{u}\uparrow$ & q$_{d}\uparrow$ & q$_{s}\uparrow$ &
q$_{c}\uparrow$ & q$_{b}\uparrow$\\
$\overline{q_{u}}\downarrow$ & (q$_{u}\overline{q_{u}})_{0}$ & $\pi^{-}$ &
K$^{-}$ & D$^{0}$ & B$^{-}$\\
$\overline{q_{d}}\downarrow$ & $\pi^{+}$ & (q$_{d}\overline{q_{d}})_{0}$ &
$\overline{K^{0}}$ & D$^{+}$ & $B^{0}$\\
$\overline{q_{s}}\downarrow$ & K$^{+}$ & K$^{0}$ & (q$_{s}\overline{q_{s}%
})_{0}$ & D$_{s}^{+}$ & B$_{s}^{0}$\\
$\overline{q_{c}}\downarrow$ & $\overline{D^{0}}$ & D$^{-}$ & D$_{s}^{-}$ &
$\eta_{c}$ & (q$_{b}\overline{q_{c}})_{0}$\\
$\overline{q_{b}}\downarrow$ & B$^{+}$ & $\overline{B^{0}}$ & $\overline
{B_{s}^{0}}$ & (q$_{c}\overline{q_{b}})_{0}$ & (q$_{b}\overline{q_{b}})_{0}$%
\end{tabular}
{\LARGE .}\label{S(J=0) Wave-MESON}%
\end{equation}
(q$_{u}\overline{q_{u}})_{0},$ (q$_{d}\overline{q_{d}})_{0}$, and
(q$_{s}\overline{q_{s}})_{0}$ can be combined into $\pi^{0},$ $\eta$, and
$\eta^{^{\prime}}$. Thus, we can get the S wave spin single (L = 0 and s =0)
mesons: $\pi^{+}$, $\pi^{0}$, $\pi^{-};$ $\eta,$ $\eta^{\prime},$ $\eta_{c};$
$K^{+},$ $K^{0},$ $K^{-},$ $\overline{K^{0}};$ $\ D^{+},$ $D^{-}, $ $D^{0},$
$\overline{D^{0}};$ $D_{s}^{+},$ $D_{s}^{-};$ B$^{+}$, B$^{-},$ B$^{0},$
$\overline{B^{0}}$, B$_{s}^{0}$, $\overline{B_{s}^{0}},$ B$_{c}^{+}$ =
$($(q$_{c}\overline{q_{b}})_{0}),$ B$_{c}^{-}$ = ((q$_{b}\overline{q_{c}}%
)_{0}$), $\eta_{b}$ = (q$_{b}\overline{q_{b}})_{0}$.

Similarly, we give out the mesons with the angular momentum L = 0 and the spin
s = 1 (the spin three fold state S wave mesons-q$\overline{q},$ $\uparrow
\uparrow$ spin up, up) as follows:%

\begin{equation}%
\begin{tabular}
[c]{llllll}%
1$^{3}S_{1}$ & q$_{u}\uparrow$ & q$_{d}\uparrow$ & q$_{s}\uparrow$ &
q$_{c}\uparrow$ & q$_{b}\uparrow$\\
$\overline{q_{u}}\uparrow$ & (q$_{u}\overline{q_{u}})_{1}$ & $\rho^{-}$ &
$K^{\ast-}$ & D$^{\ast0}$ & B$^{\ast-}$\\
$\overline{q_{d}}\uparrow$ & $\rho^{+}$ & (q$_{d}\overline{q_{d}})_{1}$ &
$\overline{K^{\ast0}}$ & D$^{\ast+}$ & B$^{\ast0}$\\
$\overline{q_{s}}\uparrow$ & $K^{\ast+}$ & K$^{\ast0}$ & (q$_{s}%
\overline{q_{s}})_{1}$ & D$_{s}^{\ast+}$ & B$_{s}^{\ast0}$\\
$\overline{q_{c}}\uparrow$ & $\overline{D^{\ast0}}$ & D$^{\ast-}$ &
D$_{s}^{\ast-}$ & J/$\psi$ & (q$_{b}\overline{q_{c}})_{1}$\\
$\overline{q_{b}}\uparrow$ & B$^{\ast+}$ & $\overline{B^{\ast0}}$ &
$\overline{B_{s}^{\ast0}}$ & (q$_{c}\overline{q_{b}})_{1}$ & $\curlyvee(1S)$%
\end{tabular}
{\LARGE .}\label{,S(J=1) WAVE-MESONS.}%
\end{equation}
(q$_{u}\overline{q_{u}})_{1},$ (q$_{d}\overline{q_{d}})_{1}$, and
(q$_{s}\overline{q_{s}})_{1}$ can be combined into $\rho^{0},$ $\omega$, and
$\phi.$ Thus we can get, $\rho^{+},$ $\rho^{0},$ $\rho^{-};$ $\omega,$ $\phi;$
$K^{\ast+}$, $K^{\ast0},$ $K^{\ast-},$ $\overline{K^{\ast0}}$; $D^{\ast+},$
D$^{\ast-}$, $D^{\ast0}$, $\overline{D^{\ast0}};$ $D_{s}^{\ast+},$
$D_{s}^{\ast-}$, J/$\psi;$ $B^{\ast+},$ B$^{\ast-}$, $B^{\ast0}$,
$\overline{B^{\ast0}}$, $B_{s}^{\ast0}$, $\overline{B_{s}^{\ast0}},$
B$_{c}^{\ast+}$ = ((q$_{c}\overline{q_{b}})_{1}$), B$_{c}^{\ast-}$ =
(q$_{b}\overline{q_{c}})_{1};$ $\curlyvee(1S).$

Similarly, we can get the whole meson spectrum including the angular momentum
L, the spin s, the total angular momentum J, the parity P, the isospin I, and
the electric charge Q. Leaving undiscovered mesons in empty spaces, the most
of the known mesons \cite{Table12. 2.} are listed as following (\textbf{q}%
$_{u}$ = \textbf{u}, \textbf{q}$_{d}$ = \textbf{d}, \textbf{q}$_{s}$ =
\textbf{s}, \textbf{q}$_{c}$ = \textbf{c}, and \textbf{q}$_{b}$ = \textbf{b}.
\ \ \ \ \ \ \ \ \ \ \ \ \ \ \ \ \ \ \ \ \ \ \ \ \ \ \ \ \ \ \ \ \ \ \ \ \ \ \ \ (\ref{5-SAME-QUARKS}%
))%

\begin{equation}%
\begin{tabular}
[c]{lllllll}%
N$^{2s+1}$L$_{J}$ & J$^{PC}$ & u$\overline{d}$,u$\overline{u}$,d$\overline{d}
$ & \ \ \ \ \ \ \ u$\overline{u}$,d$\overline{d},$s$\overline{s}$ &
\ \ c$\overline{c}$ & b$\overline{b}$ & $\overline{s}$u,$\overline{s}$d\\
&  & \ \ I = 1 & \ \ \ \ \ \ \ \ \ I = 0 & I=0 & I=0 & I=1/2\\
1$^{1}$ \ \ S$_{0}$ & 0$^{-+}$ & \ \ \ \ $\pi$ & \ $\ \ \ \ \ \ \ \ \ \eta
$,$\eta^{\prime}$ & \ $\eta_{c}$ &  & \ \ K\\
1$^{3}$ \ \ S$_{1}$ & 1$^{--}$ & \ \ \ \ $\rho$ & $\ \ \ \ \ \ \ \ \ \omega
$,$\phi$ & J/$\psi$(1s) &  & K$^{\ast}$(892)\\
1$^{1}$ \ \ p$_{1}$ & 1$^{+-}$ & b$_{1}$(1235) & h$_{1}$(1170), h$_{1}%
$(1380) & h$_{c}$(1p) &  & K$_{1B}^{\dagger}$\\
1$^{3}$ \ \ p$_{0}$ & 0$^{++}$ & * & * & $\chi_{c0}$(1p) & $\chi_{b0}$(1p) &
K$_{0}^{\ast}$(1430)\\
1$^{3}$ \ \ p$_{1}$ & 1$^{++}$ & a$_{1}$(1260) & f$_{1}$(1285), f$_{1}%
$(1510) & $\chi_{c1}$(1p) & $\chi_{b1}$(1p) & K$_{1A}^{\dagger}$\\
1$^{3}$ \ \ p$_{2}$ & 2$^{++}$ & a$_{2}$(1320) & f$_{2}$(1270), f'$_{2}%
$(1525) & $\chi_{c2}$(1p) & $\chi_{2}$(1p) & K$_{2}^{\ast}$(1430)\\
1$^{1}$ \ \ D$_{2}$ & 2$^{-+}$ & $\pi_{2}$(1670) &  &  &  & K$_{2}$(1770)\\
1$^{3}$ \ \ D$_{1}$ & 1$^{--}$ & $\rho$(1700) & $\omega$(1600) & $\psi
$(3770) &  & K$^{\ast}$(1680)$^{\dagger}$\\
1$^{3}$ \ \ D$_{2}$ & 2$^{--}$ &  &  &  &  & K$_{2}$(1820)\\
1$^{3}$ \ \ D$_{3}$ & 3$^{--}$ & $\rho_{3}$(1690) & $\omega_{3}$%
(1600),$\phi_{3}$(1850) &  &  & K$_{3}^{\ast}$(1780)\\
1$^{3}$ \ \ F$_{4}$ & 4$^{++}$ & $\alpha_{4}$(2040) & f$_{4}$(2050),f$_{4}%
$(2220) &  &  & K$_{4}^{\ast}$(2045)\\
2$^{1}$ \ \ S$_{0}$ & 0$^{-+}$ & $\pi$(1300) & $\eta$(1295) & $\eta_{c}$(2S) &
& K(1460)\\
2$^{3}$ \ \ S$_{1}$ & 1$^{--}$ & $\rho$(1450) & $\omega$(1420),$\phi$(1680) &
$\psi$(2S) & $\curlyvee(2s)$ & K$^{\ast}$(1410)\\
2$^{3}$ \ \ P$_{2}$ & 2$^{++}$ &  & $f_{2}$(1810),f$_{2}$(2010) &  &
$\chi_{b2}$(2p) & K$_{2}^{\ast}$(1980)\\
3$^{1}$ \ \ S$_{0}$ & 0$^{-+}$ & $\pi$(1770) & $\eta$(1760) &  &  & K(1830)\\
&  &  &  &  &  &
\end{tabular}
\label{Meson-A}%
\end{equation}%

\begin{equation}%
\begin{tabular}
[c]{llllll}%
N$^{2s+1}$L$_{J}$ & J$^{PC}$ & c$\overline{u}$,c$\overline{d}$ &
c$\overline{s}$ & $\overline{b}u,\overline{b}d$ & $\overline{b}s$\\
&  & I=1/2 & I = 0 & I = 1/2 & I = 0\\
1$^{1}$ \ \ S$_{0}$ & 0$^{-+}$ & \ \ D & D$_{s}$ & \ \ \ B & B$_{s}$\\
1$^{3}$ \ \ S$_{1}$ & 1$^{--}$ & D$^{\ast}$(2010) & D$_{s}^{\ast}$(2110) &
B$^{\ast}$(5330) & \\
1$^{1}$ \ \ p$_{1}$ & 1$^{-+}$ & D$_{1}$(2420) & D$_{s1}$(2536) &  & \\
1$^{3}$ \ \ p$_{0}$ & 0$^{++}$ &  &  &  & \\
1$^{3}$ \ \ p$_{1}$ & 1$^{++}$ &  &  &  & \\
1$^{3}$ \ \ p$_{2}$ & 2$^{++}$ & D$_{2}^{\ast}$(2460) &  &  &
\end{tabular}
\label{sons-B}%
\end{equation}

Although the SU(N) can give out all quantum numbers of the meson spectrum,
they cannot give out a mass spectrum of the mesons. If SU(4) were an accurate
symmetry group, then the mesons $\pi$, $\eta$, $\eta^{\prime}$, and $\eta_{C}$
would have the same mass, as shown in Fig. 12.1 (a) of \textbf{\cite{Table 12.
1}}. In fact, mass ($\pi)$ = 139, mass ($\eta)$ = 547\ $\approx$
3.95$\times139$\ , mass ($\eta^{\prime})$ = 958\ \ $\approx$ 6.9$\times139$\ ,
mass($\eta_{C})$ = 2980\ \ $\approx$ 21.4$\times139$. If SU(5) were an
accurate symmetry group, the mesons $\rho$, $\omega$, $\phi$, J/$\psi
$(1s),\ and $\curlyvee(1s)$ would have the same mass as shown in Fig. 12.1 (b)
of \textbf{\cite{Table 12. 1}}. In fact, mass ($\rho)$ = 770, mass ($\omega)$
= 782\ \ $\approx$ 770\ , mass ($\phi)$ = 1020\ \ $\approx$ 1.33$\times770$\ ,
mass(J/$\psi$(1s)$)$ = 3097\ \ $\approx$ 4$\times770$\ , mass ($\curlyvee
(1s))$ = 9460 $\approx$ 12.3$\times770$. The above experimental facts clearly
show that the SU(N) (N = 3, 4, and 5) groups are not the best symmetry groups
and the that SU(N) approximations are not the best approximations.

The best approximation is the combined approximation of the cell and the
periodic field. Considering all quarks, it will give the full meson spectrum.
Thus, \textbf{we shall replace the SU(N) group by the body center cubic groups
in the classification of the mesons. }

\subsection{The top quarks \textbf{\ }}

According to the BCC model, there is no excited quark with a mass about 175
Gev (about 185 times the mass of proton) \cite{TOPQUARK}. It may be an energy
band excited state of an electron \cite{Periodic Field}. \ \ \ \ \ \ \ \ \ \ \ \ \ \ \ \ \ \ \ \ \ \ \ \ \ \ \ \ \ \ \ \ 

\subsection{The quarks decay \textbf{\ }}

In the standard model \cite{Higgs}, fermion couplings to the Higgs field not
only determine their masses; they induce a misalignment of quark mass
eigenstates with respect to the eigenstates of the weak charges, thereby
allowing all fermions of heavy families to decay to lighter ones. The BCC
model does not need the Higgs field. We know that the higher mass quarks are
the higher energy excited states of the fundamental quarks (u and d), and that
the lower mass quarks are the lower energy excited states of the same quarks
(u and d). Under the influences of the quark lattice, the higher energy
excited states naturally decay into lower ones, and finally into the ground
states, because they are not truly independent particles.\ In the BCC Model,
the Higgs field is replaced by the periodic field of the quark lattice in the
vacuum. \ \ \ \ \ \ \ \ 

\subsection{Summary \textbf{\ }}

1. The SU(N) (N = 3, 4, and 5) symmetry approximations have the same quarks
(u, d, s, c, and b), the same symmetry group (the SU(3), the SU(4), and the
SU(5)), the same baryon and meson structure (baryons = qqq states and mesons =
q$\overline{q}$ states), and the same sum formulae (\ref{SUM-formula}) (the
quantum numbers and energy of the system are the sum of the quantum numbers
and energies of the constituent quarks \cite{Nonrelat-Model}) , as well as the
same results as the Quark Model (5 quarks). Therefore, the SU(N) approximation
is the Quark Model (with N quarks), and the Quark Model is an approximation of
the BCC model.

2. The BCC model provides the quarks (with the quantum numbers and masses),
the symmetry groups (SU(N)), and the baryon and meson structure molds (baryons
= qqq states and mesons = q$\overline{q}$ states) for the SU(N) symmetry
approximation. These are the physical foundations of the Quark Model. Thus,
the BCC Model provides the physical foundations of the Quark Model.

3. The SU(N) approximations assume that the 5 quarks (u, d, s, c, and b) are
independent particles. This explains why no one can find an united mass
formula for the baryon spectrum or for the meson spectrum. At the same time,
it is difficult to understand why the higher mass quarks decay into lower mass quarks.

4. The SU(N) omits all higher energy band excited quarks, and there are a lot
of such excited quarks, so it cannot explain the whole baryon spectrum and
meson spectrum.

5. The SU(N) omits the accompanying excited cell, it needs the confinement
hypothesis to explain why free quarks cannot be discovered. Confinement is a
very plausible idea but to date its rigorous proof remains outstanding
\cite{CONFINEMENT}.

\section{Comparing The Results}

The combined approximation of the cell and the periodic field is the best
approximation of the BCC model. The results of the combined approximation will
be regarded as the results of the BCC model. The BCC model produces three
spectra: the quark spectrum, the baryon spectrum and the meson spectrum. The
theoretical quark spectrum does not have experimental results with which to
compare, single free quarks have been not found in experiments yet. We will
discuss the meson spectrum in the next paper (Jiao-Lin Xu and Xin Yu, The
Meson Spectrum). Here, we will compare the baryon spectrum only. \ \ \ 

Using Tables 1-6, we compare the theoretical results of the BCC model with the
experimental results \cite{particle}. In the comparison, we do not take into
account the angular momenta of the experimental results. We assume that the
small differences of the masses in the same group of baryons stem from their
different angular momenta. If we ignore this effect, their masses would be
essentially the same. In the comparison, we use the baryon name to represent
the intrinsic quantum numbers as shown in the second column of Table 1. \ 

\ \ \ The ground states of various kinds of baryons are shown in Table 1.
\ These baryons have relatively long lifetimes. They are the most important
experimental results of the baryons. From Table 1, we can see that all
theoretical intrinsic quantum numbers (isospin $I$, strange number $S$,
charmed number $C$, bottom number $b$, and electric charge $Q$) are the same
as those in the experimental results. Also the theoretical mass values are in
very good agreement with the experimental values.

\ \ \ A comparison of the theoretical results with the experimental results of
the unflavored baryons $N$ and $\Delta$ is made in Table 2. From Table 2, we
can see that the intrinsic quantum numbers of the theoretical results are
exactly the same as those of the experimental results.\ Also the theoretical
masses of the baryons $N$ and $\Delta$\ are in very good agreement with the
experimental results. The theoretical results N(1210) and N(1300) are not
found in the experiment. We believe that they are covered up by the
experimental baryon $\Delta(1232)$ because of the following reasons: (1) they
are unflavored baryons with the same S, C, and b; (2) the width (120 Mev) of
$\Delta(1232)$ is very large, and the baryons $N(1210)$ and $N(1300)$ both
fall within the width region of $\Delta(1232)$; (3) the average mass (1255
Mev) of $N(1210)$ and $N(1300)$ is essentially the same as the mass (1232 Mev)
of $\Delta(1232)$ $(\Gamma$ = 120 Mev).

\ \ \ Two kinds of the strange baryons $\Lambda$ and $\Sigma$ are compared in
Table 3. Their theoretical and experimental intrinsic quantum numbers are the
same.\ The theoretical masses of the baryons $\Lambda$ and $\Sigma$ are in
very good agreement with the experimental results.

\ \ \ The theoretical intrinsic quantum numbers of the baryons $\Xi$ and
$\Omega$ are the same as the experimental results (see Table 4). The
theoretical masses of the baryons $\Xi$ and $\Omega$ are compatible with the
experimental results.\ 

\ \ \ The charmed and bottom baryons $\Lambda_{c}^{+}$ and $\Lambda_{b}^{0} $
can be found in Table 5. \ The experimental masses of the charmed baryons
($\Lambda_{c}^{+}$) and bottom baryons ($\Lambda_{b}^{0}$) coincide with the
theoretical results.

\ \ \ Finally, we compare the theoretical results with the experimental
results for the charmed strange baryons $\Omega_{C}$, $\Xi_{c}$ and
$\Sigma_{c}$ in Table 6.\ Their intrinsic quantum numbers are all matched very
well, and their masses are in very good agreement.

\ \ In summary, the BCC model explains all baryon experimental intrinsic
quantum numbers and masses. Virtually no experimentally confirmed baryon is
not included in the model. However, the angular momenta and the parities of
the baryons are not included in the zeroth-order approximation of this paper.
They depend on the wave functions of the energy bands. We will discuss them in
the first order approximation.

\section{Predictions and Discussion}

\subsection{Some New Baryons}

\ \ \ \ \ According to the BCC model, a series of possible baryons may exist.
However, when energy goes higher and higher, the energy bands will become
denser and denser, and the full widths of the baryons will become wider and
wider, making them extremely difficult to be separated. The following new
baryons predicted by the model seem to have a better chance of being
discovered in a not too distant future:
\begin{equation}%
\begin{array}
[c]{cccccc}%
I=\text{ }0\text{ \ } & S=-1 & Q=\text{ \ }0 & \Lambda^{0}(2560) &  &
\text{(\ref{SIGMA_1})}\\
I=\text{ }0\text{\ \ } & C=+1 & Q=+1 & \Lambda_{C}^{+}(6600) & \Lambda_{C}%
^{+}(13800) & \text{(\ref{Baryon-Data-One})}\\
I=\text{ }0\text{ \ } & S=-1 & Q=\text{ \ }0 & \Lambda^{0}(4280) & \Lambda
^{0}(10040) & \text{(\ref{Baryon-Data-One})}\\
I=\text{ }1\text{ \ } & S=-1 & C=+1 & Q=2,1,0 & \Sigma_{C}(2280) &
\text{(\ref{Baryon-Data-One})}\\
I=\text{ }0 & S=-3 & Q=-1 & \Omega^{-}(3720) & \Omega^{-}(7220) &
\text{(\ref{SIGMA_1})}\\
I=\text{ }0 & b=-1 & Q=0 & \Lambda_{b}^{0}(9960) &  & \text{(\ref{SIGMA_1})}%
\end{array}
\end{equation}
in the last column, we give the equation numbers where the baryons are first
deduced in this paper.

\subsection{Discussion}

\begin{enumerate}
\item  From (\ref{Alpha}), we have
\begin{equation}
m_{q}a_{x}^{2}=h^{2}/720\text{ Mev.}%
\end{equation}
Although we do not know the values of $m_{q}$ and $a_{x}$, we find that
$m_{q}a_{x}^{2}$ is a constant. According to the renormalization theory
\cite{renormal}, the bare mass of the quark should be ``infinite'', so that
$a_{x} $ will be ``zero''. Of course, the ``infinite'' and the ``zero'' are
physical concepts in this case. We understand that the ``infinite'' means
$m_{q}$ is huge and the ``zero'' means $a_{x}$ is much smaller than the
nuclear radius. ``$m_{q}$ is huge'' guarantees that we can use the
Schr\"{o}dinger equation instead of the Dirac equation. Since ``$a_{x}$ is
much smaller than the nuclear radius'', the vacuum material looks like a
continuous media, which makes the structure of the vacuum material very
difficult to be discovered.

\item  In a sense, the vacuum material with the body center cubic structure
works like a superconductor.\ There are no electric and mechanical resistances
to any particle and any physical body (with or without electric charge) moving
inside the vacuum material. Since the energy gaps are so large (for an
electron, the energy gap is about 0.5 Mev; for a quark, the energy gap is
about 940 Mev), the vacuum remains unchanged when ordinary physical phenomena
occur. Although when the energy of an individual quark exceeds the gap, the
vacuum state quark will be excited from the vacuum, so the vacuum material
still looks unchanged.

\item  If the vacuum material really exists, it shall not only be super-strong
but also super-dense. As a result, even a hydrogen bomb cannot destroy it. In
fact, when a hydrogen bomb explodes, an individual quark can only receive a
little energy (not to exceed 5 Mev). It is much lower than the vacuum excited
energy (940 Mev) of a quark. Thus, it can not break the vacuum material.

\item  Since the theoretical baryon mass spectrum in the free particle
approximation (V($\vec{r}$) = V$_{o}$ and the wave functions satisfy the body
center cubic periodic symmetries) is very close to the experimental mass
spectrum, the amplitude (A) of the strong interaction periodic field should be
much smaller than the average of the periodic field (V$_{0}$). According to
the BCC model, Dirac's sea concept\cite{diarcsea} is a complete free
approximation for the vacuum periodic field (V($\vec{r}$) = V$_{o}$ and the
wave functions do not have to satisfy the body center cubic periodic
symmetries). Thus, the Dirac's sea concept is a very good approximation of the
BCC Model.
\end{enumerate}

\section{Conclusions\ \ \ }

1. The two quarks (u and d) are the only elemental particles in the quark
family. Other quarks (s, c, b, ...) are all their energy band excited states.
The BCC Model has deduced (not assumed) the intrinsic quantum numbers and
masses of all quarks. Thus, the SU(3) (for the quarks u, d, and s), the SU(4)
(for the quarks u, d, s, and c), and the SU(5) (for the quarks u, d, s, c, and
b) symmetries are the natural extensions of the SU(2) (based on the quarks u
and d).\ The BCC Model has shown that the excited primitive cell (the three
quark system q$^{\ast}u^{\prime}d^{\prime}$) is a baryon. This is the physical
foundation of the assumption that a baryon is made up of three quarks in the
Quark Model. Therefore, the BCC Model \textbf{provides a physical foundation
of the quark model. The Quark Model is a approximation (SU(N)) of the BCC Model.}

2. Although baryons ($\Delta,$ $N,$ $\Lambda,$ $\Sigma,$ $\Xi,$ $\Omega,$
$\Lambda_{C,}$ $\Xi_{C},$ $\Sigma_{C},$ and $\Lambda_{C}$) are quite different
from one another in I, S, C, b, Q, and M, they have the same structure (one
excited quark q$^{\ast}$ and two accompanying excised quarks $u^{\prime}$ and
$d^{\prime}$). The BCC Model has deduced all intrinsic quantum numbers and
masses of all baryons.

3. The vacuum material is a super-superconductor with super-high energy gaps
(proton-939 Mev, electron-0.5 Mev). It has the body center cubic periodic
symmetries. We think that the body center cubic periodic symmetry may be one
of the possible `` have not yet been identified'' symmetries \cite{Wilczek2}.

4. In the BCC Model, the three excited quarks (inside a baryon) are in
different states (one excited quark q$^{\ast}$, one accompanying excited quark
u$^{^{\prime}}$, and one accompanying excited quark d$^{^{\prime}}$). Hence,
the system obeys the \textbf{Pauli exclusion principle \cite{Pauli and Nanbu}. }

5. The confinement concept is not needed. It shall be replaced by the
accompanying excitation concept. According to the accompanying excitation
concept, any excited quark (from the vacuum) is always accompanied by two
accompanying excited quarks $u^{\prime}$ and $d^{\prime}$. They cannot be
separated. Therefore, individual free quarks can never been seen.

6. In the BCC Model, the high mass quarks are the high energy excited states
of the fundamental quark q, the low mass quarks are the low energy excited
states of the same quark q. Thus, the higher mass quarks will decay into the
lower mass quarks.

7. Due to the existence of the vacuum material, all observable particles are
constantly affected by the vacuum material (vacuum state quark lattice). Thus,
some laws of statistics (such as fluctuation) cannot be ignored.

8. The SU(N) symmetries (flavor) shall be replaced by the body center cubic
periodic symmetries. The body center cubic periodic symmetries of the vacuum
material need to be more intensively researched.

9. We need to do first order calculations to find more accurate masses, the
angular momenta, and the parities of the baryons and the mesons, using the
wave functions which satisfy the symmetries of the body center cubic periodic field.

\textbf{Acknowledgment}

I would like to express my heartfelt gratitude to Dr. Xin Yu for checking the
calculations of the energy bands and for helping to writting this paper. I
sincerely thank Professor Robert L. Anderson for his valuable advice. I also
acknowledge\textbf{\ }my indebtedness to Professor D. P. Landau for his help.
I thank Professor W. K. Ge very much for all of his help and for recommending
Wilczek's paper \cite{wilczek}. I thank my friend Z. Y. Wu very much for his
help in preparing this paper. I thank my classmate J. S. Xie very much for
checking the calculations of the energy bands. I thank Professor Y. S. Wu, H.
Y. Guo, and S. Chen \cite{XUarticle} very much for very useful discussions. I
also thank Dr. Fugao Wang very much for helping me post my paper on the web (http://arxiv.org/abs/hep-ph/0010281).

\begin{center}
\bigskip{\LARGE FIGURES}
\end{center}

Fig. 1. \ The first Brillouin zone of the body center cubic lattice. The the
symmetry points and axes are indicated. The axis $\Delta$ is a $4$ fold
rotation axis, the strange number S = 0, the baryon family $\Delta$
($\Delta^{++},$ $\Delta^{+},$ $\Delta^{0},$ $\Delta^{-}$) will appear on the
axis. The axes $\Lambda$ and F are $3$ fold rotation axes, the strange number
S =\ -1, the baryon family $\Sigma$ ($\Sigma^{+},$ $\Sigma^{0},$ $\Sigma^{-})$
will appear on the axes. The axes $\Sigma$ and G are $2$ fold rotation axes,
the strange number S = -2, the baryon family $\Xi$ ($\Xi^{0},$ $\Xi^{-}$) will
appear on the axes. The axis D is parallel to the axis $\Delta$, S = 0. And
the axis is a $2$ fold rotation axis, the baryon family N (N$^{+}$, N$^{0}$)
will be on the axis.

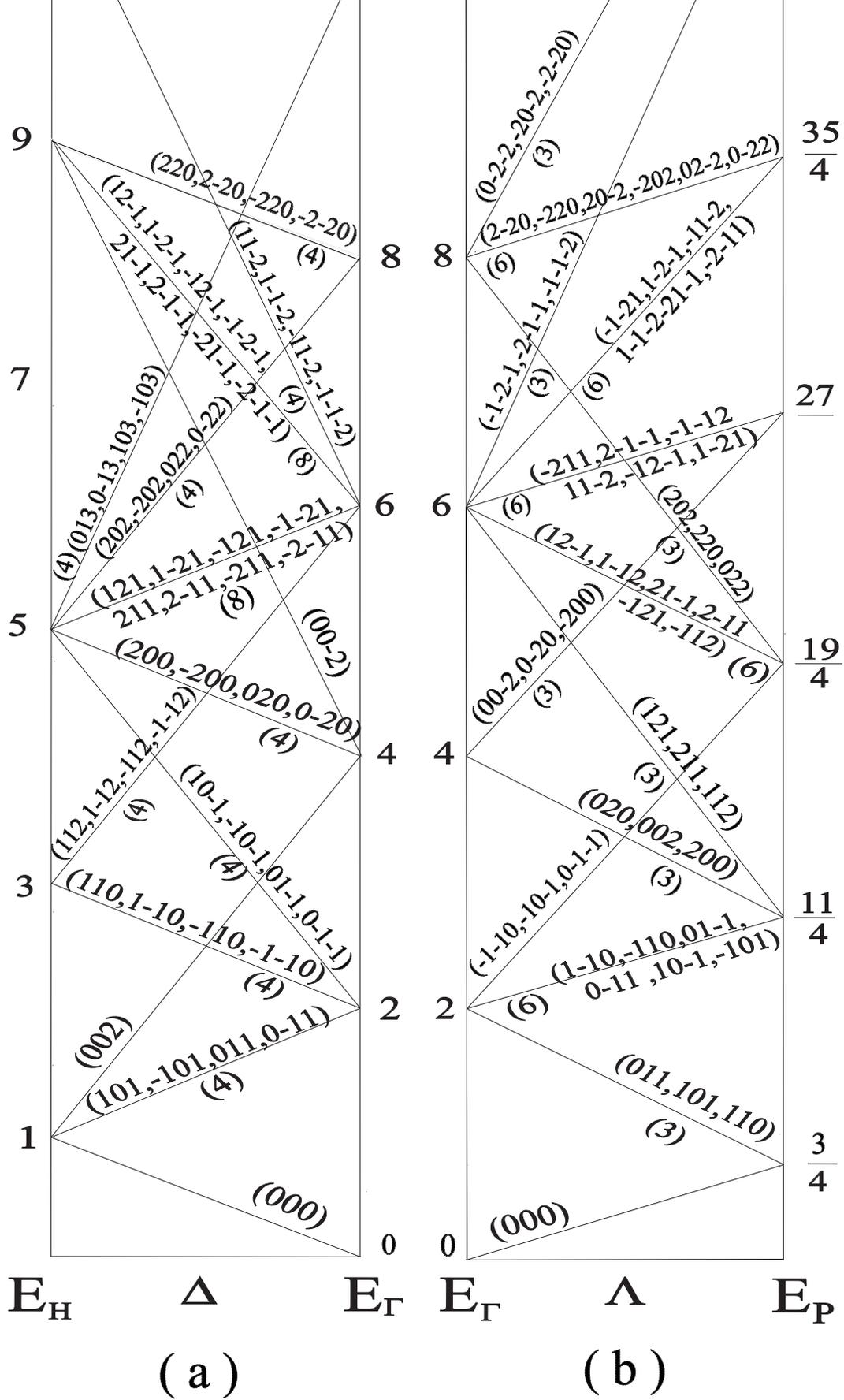
Fig. 2. \ \ (a) The energy bands on the axis $\Delta$. The numbers above the
lines are the values of $\vec{n}$ = ($n_{1}$, $n_{2}$, $n_{3}$). The numbers
under the lines are the fold numbers of the degeneracy. E$_{\Gamma}$ is the
value of E($\vec{k}$, $\vec{n}$) (see Eq. (\ref{energy})) at the end point
$\Gamma,$ while E$_{H}$ is the value of E($\vec{k}$, $\vec{n}$) at other end
point H. \ \ (b) The energy bands on the axis $\Lambda$. E$_{\Gamma}$ is the
value of E($\vec{k}$, $\vec{n}$) (see Eq. (\ref{energy})) at the end point
$\Gamma,$ while E$_{P}$ is the value of E($\vec{k}$, $\vec{n}$) at other end
point P.

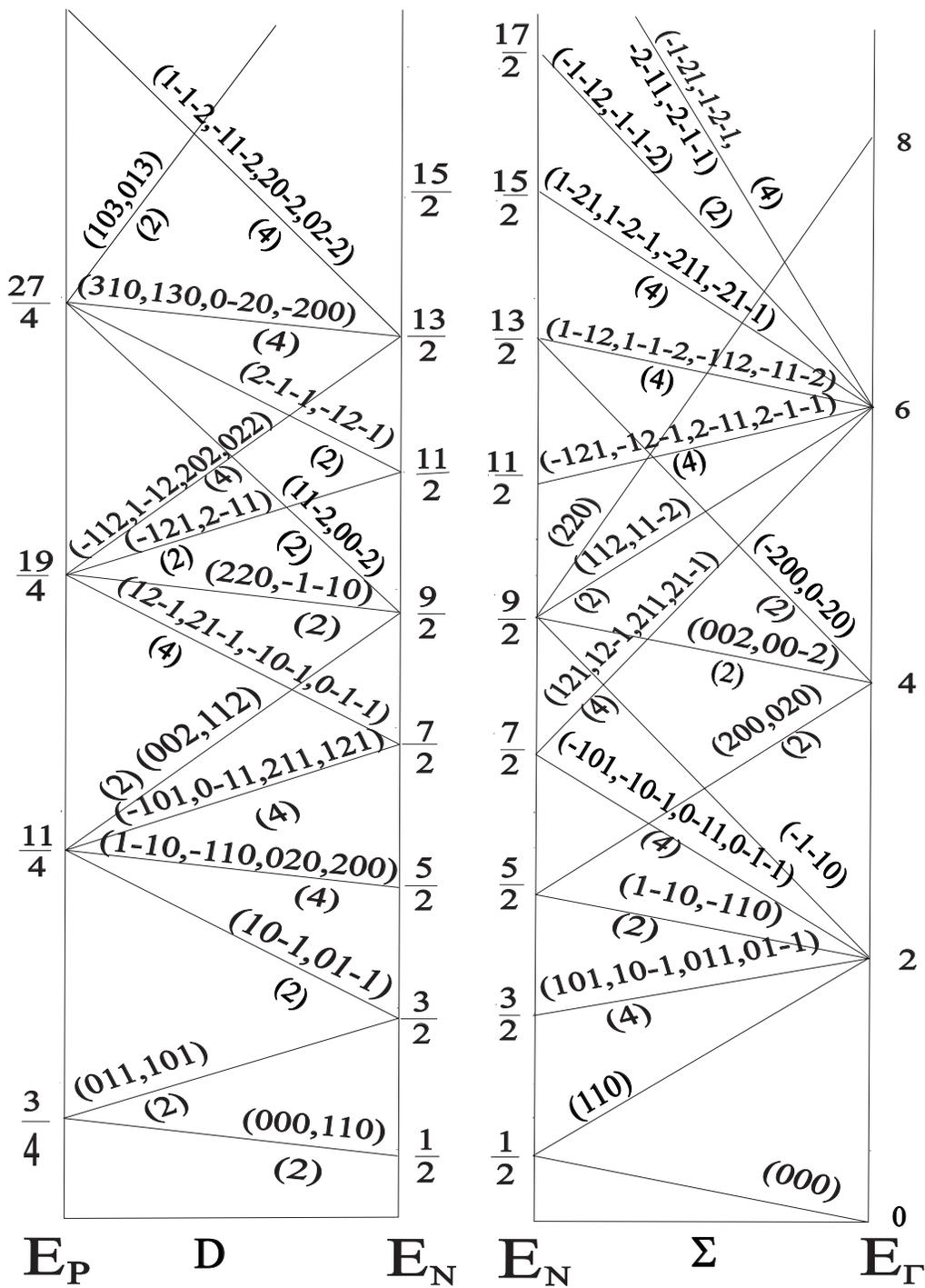
Fig. 3. \ \ (a) The energy bands on the axis $\Sigma$. The numbers above the
lines are the values of $\vec{n}$ = ($n_{1}$, $n_{2}$, $n_{3}$). The numbers
under the lines are the fold numbers of the degeneracy. E$_{\Gamma}$ is the
value of E($\vec{k}$, $\vec{n}$) (see Eq. (\ref{energy})) at the end point
$\Gamma,$ while E$_{N}$ is the value of E($\vec{k}$, $\vec{n}$) at other end
point N. \ (b) The energy bands on the axis $D$. E$_{P}$ is the value of
E($\vec{k}$, $\vec{n}$) (see Eq. (\ref{energy})) at the end point P$,$ while
E$_{N}$ is the value of E($\vec{k}$, $\vec{n}$) at other end point N.

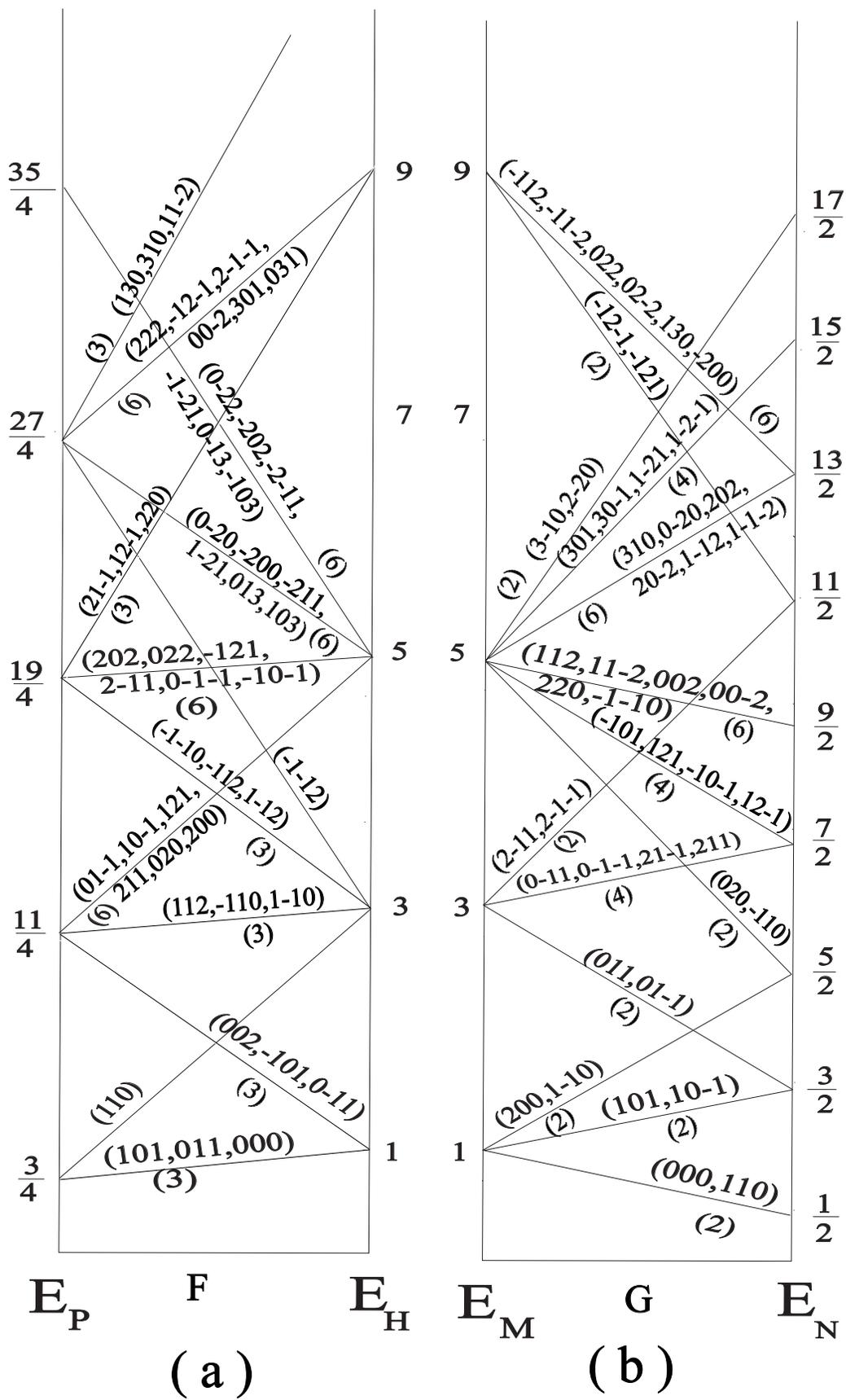
Fig. 4. \ \ (a) The energy bands on the axis $F$. The numbers above the lines
are the values of $\vec{n}$ = ($n_{1}$, $n_{2}$, $n_{3}$). The numbers under
the lines are the fold numbers of the degeneracy. E$_{P}$ is the value of
E($\vec{k}$, $\vec{n}$) (see Eq. (\ref{energy})) at the end point $P,$ while
E$_{H}$ is the value of E($\vec{k}$, $\vec{n}$) at other end point H.\ \ (b)
The energy bands on the axis $G$.\ E$_{M}$ is the value of E($\vec{k}$,
$\vec{n}$) (see Eq. (\ref{energy})) at the end point M$,$ while E$_{N}$ is the
value of E($\vec{k}$, $\vec{n}$) at other end point N.\ \ 

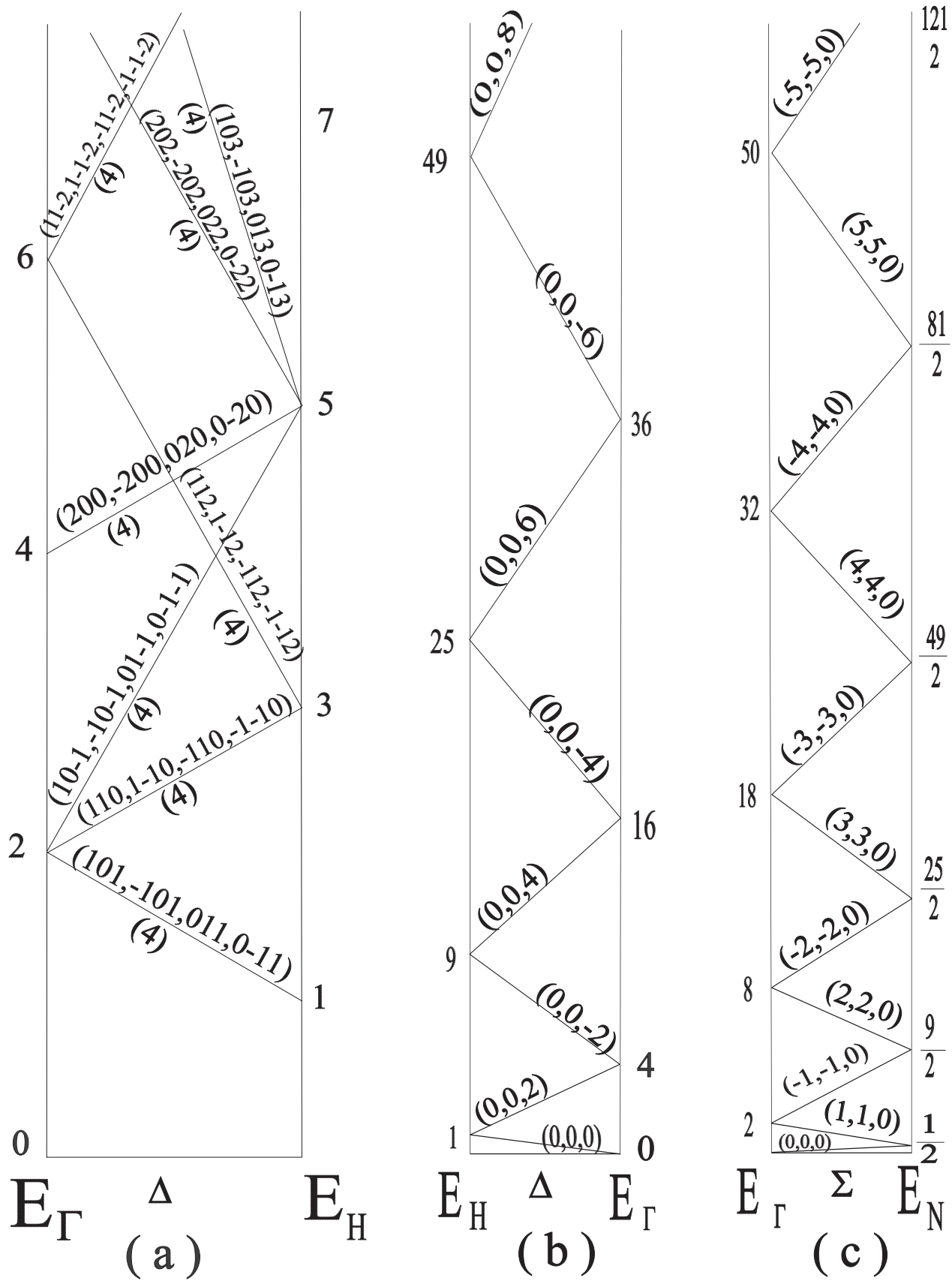
Fig. 5. \ \ (a) The $4$ fold degenerate energy bands (selected from Fig. 2(a))
on the axis $\Delta$. The numbers above the lines are the values of $\vec{n}$
(n$_{1}$, n$_{2}$, n$_{3}$). The numbers under the lines are the numbers of
the degeneracy of the energy bands. \ \ (b) The single energy bands (selected
from Fig. 2(a)) on the axis $\Delta$. The numbers above the lines are the
values of $\vec{n}$ (n$_{1}$, n$_{2}$, n$_{3}$). \ \ (c) The single energy
band (selected from Fig. 3(a)) on the axis $\Sigma$. The numbers above the
lines are the values of $\vec{n}$ (n$_{1}$, n$_{2}$, n$_{3} $)

\newpage

\begin{center}
\bigskip\ \ \ \ \ \ 

{\LARGE TABLE}
\end{center}

\qquad\qquad\qquad\qquad Table 1. \ The Ground States of the Baryons.%

\begin{tabular}
[c]{|l|l|l|l|l|}\hline
{\small Theory} & Quantum. No & {\small Experiment} & R & Life Time\\\hline
Name({\small M}) & \ S, \ C,\ b, \ \ \ I, \ \ Q & Name({\small M}) &  &
\\\hline
N$^{+}$(940) & \ 0,\ \ 0, \ 0, \ {\small 1/2, \ \ }1 & p(938) & 0.2 &
$>$%
10$^{31}years$\\\hline
N$^{0}$(940) & 0, \ 0, \ 0, \ {\small 1/2, \ \ }0 & n(940) & 0.0 &
1.0$\times10^{8}$ s\\\hline
$\Lambda^{0}(1120)$ & -1, \ 0, \ 0, \ \ 0, \ \ 0 & $\Lambda^{0}(1116)$ & 0.4 &
2.6$\times$10$^{-10}s$\\\hline
$\Sigma^{+}(1210)$ & -1, \ 0, \ 0, \ \ 1, \ \ 1 & $\Sigma^{+}(1189)$ & 1.8 &
.80$\times$10$^{-10}s$\\\hline
$\Sigma^{0}(1210)$ & -1, \ 0, \ 0, \ \ 1, \ \ 0 & $\Sigma^{0}(1193)$ & 1.4 &
7.4$\times$10$^{-20}s$\\\hline
$\Sigma^{-}(1210)$ & -1, \ 0, \ 0, \ \ 1, \ -1 & $\Sigma^{-}(1197)$ & 1.1 &
1.5$\times$10$^{-10}s$\\\hline
$\Xi^{0}(1300)$ & -2, \ 0, \ 0, \ {\small 1/2}, \ 0 & $\Xi^{0}(1315)$ & 1.2 &
2.9$\times$10$^{-10}s$\\\hline
$\Xi^{-}(1300)$ & -2, \ 0, \ 0, \ {\small 1/2}, -1 & $\Xi^{-}(1321)$ & 1.6 &
1.6$\times$10$^{-10}s$\\\hline
$\Omega^{-}(1660)$ & -3, \ 0, \ 0. \ \ 0,\ \ -1 & $\Omega^{-}(1672)$ & 0.7 &
.82$\times$10$^{-10}s$\\\hline
$\Lambda_{c}^{+}(2280)$ & 0, \ 1, \ 0, \ \ 0, \ \ \ 1 & $\Lambda_{c}%
^{+}(2285)$ & 0.2 & .21$\times$10$^{-12}s$\\\hline
$\Xi_{c}^{+}(2550)$ & -1, \ 1, \ 0, \ {\small 1/2}, \ 1 & $\Xi_{c}^{+}(2466)$%
& 3.4 & .35$\times$10$^{-12}$\\\hline
$\Xi_{c}^{0}(2550)$ & -1, \ 1, \ 0, \ {\small 1/2}, \ 0 & $\Xi_{c}^{0}(2470)$%
& 3.3 & .10$\times$10$^{-12}s$\\\hline
$\Omega_{c}^{0}(2660)$ & 0, \ 0, -1, \ \ 0, \ \ 0 & $\Omega_{c}(2704)$ & 1.7 &
.64$\times$10$^{-13}s$\\\hline
$\Lambda_{b}^{0}(5540)$ & 0, \ 0, -1, \ \ 0, \ \ 0 & $\Lambda_{b}^{0}(5641) $%
& 1.8 & 1.1$\times$10$^{-12}s$\\\hline
$\Delta^{++}(1210)$ & 0,\ \ 0, \ 0, \ {\small 3/2, \ \ }$2$ & $\Delta
^{++}(1232)$ & 1.8 & $\Gamma$=120 Mev\\\hline
$\Delta^{+}(1210)$ & 0,\ \ 0, \ 0, \ {\small 3/2, \ \ }1 & $\Delta^{+}(1232)$%
& 1.8 & $\Gamma$=120 Mev\\\hline
$\Delta^{0}(1210)$ & 0,\ \ 0, \ 0, \ {\small 3/2, \ \ }0 & $\Delta^{0}(1232)$%
& 1.8 & $\Gamma$=120 Mev\\\hline
$\Delta^{-}(1210)$ & 0,\ \ 0, \ 0, \ {\small 3/2, \ -}1 & $\Delta^{-}(1232) $%
& 1.8 & $\Gamma$=120 Mev\\\hline
\end{tabular}

In the fourth column, R =($\frac{\Delta\text{M}}{\text{M}}$){\small \%.}

\bigskip

\bigskip

\bigskip

\bigskip

\bigskip

\bigskip

\bigskip

\bigskip

\bigskip

\bigskip

\bigskip

\bigskip

\bigskip

\bigskip

\bigskip\newpage\qquad\ \ \ \ \ \ \ \ \ \ Table 2. The Unflavored Baryons $N$
and $\Delta$ ($S$= $C$=$b$ = 0) \ %

\begin{tabular}
[c]{|l|l|l||l|l|l|}\hline
Theory & Experiment & $\frac{\Delta\text{M}}{\text{M}}\%$ & Theory &
Experiment & $\frac{\Delta\text{M}}{\text{M}}\%$\\\hline%
\begin{tabular}
[c]{l}%
$N(1210)$\\
$N(1300)$%
\end{tabular}
&  &  & $%
\begin{array}
[c]{c}%
\Delta(1210)\\
\Delta(1300)
\end{array}
$ & $\Delta(1232)$ & \\\hline
$\mathbf{\bar{N}(1255)}$ &  &  & $\mathbf{\bar{\Delta}(1255)}$ &
$\mathbf{\bar{\Delta}(1232)}$ & \textbf{1.9}\\\hline
$N(1480)$ &
\begin{tabular}
[c]{l}%
$N(1440)$\\
$N(1520)$\\
$N(1535)$%
\end{tabular}
&  &  &  & \\\hline
$\mathbf{\bar{N}(1480)}$ & $\mathbf{\bar{N}(1498)}$ & \textbf{1.2} &  &  &
\\\hline%
\begin{tabular}
[c]{l}%
$N(1660)$\\
$N(1660)$%
\end{tabular}
&
\begin{tabular}
[c]{l}%
$N(1650)$\\
$N(1675)$\\
$N(1680)$\\
$N(1700)$\\
$N(1710)$\\
$N(1720)$%
\end{tabular}
&  &
\begin{tabular}
[c]{l}%
$\Delta(1660)$\\
$\Delta(1660)$%
\end{tabular}
&
\begin{tabular}
[c]{l}%
$\Delta(1600)$\\
$\Delta(1620)$\\
$\Delta(1700)$%
\end{tabular}
& \\\hline
$\mathbf{\bar{N}(1660)}$ & $\mathbf{\bar{N}(1689)\ \ }$ & \textbf{1.7} &
$\mathbf{\bar{\Delta}(1660)}$ & $\mathbf{\bar{\Delta}(1640)}$ & \textbf{1.2}%
\\\hline%
\begin{tabular}
[c]{l}%
$N(1840)$\\
$N(1840)$\\
$N(1930)$\\
$N(1930)$\\
$N(1930)$\\
$N(2020)$%
\end{tabular}
&
\begin{tabular}
[c]{l}%
$N(1900)\ast$\\
$N(1990)\ast$\\
$N(2000)\ast$\\
$N(2080)\ast$%
\end{tabular}
&  &
\begin{tabular}
[c]{l}%
$\Delta(1930)$\\
$\Delta(1930)$\\
$\Delta(1930)$\\
$\Delta(2020)$%
\end{tabular}
&
\begin{tabular}
[c]{l}%
$\Delta(1900)$\\
$\Delta(1905)$\\
$\Delta(1910)$\\
$\Delta(1920)$\\
$\Delta(1930)$\\
$\Delta(1950)$%
\end{tabular}
& \\\hline
$\mathbf{\bar{N}(1915)}$ & $\mathbf{\bar{N}(1923)}$ & \textbf{0.2} &
$\mathbf{\bar{\Delta}(1953)}$ & $\mathbf{\bar{\Delta}(1919)}$ & \textbf{1.8}%
\\\hline%
\begin{tabular}
[c]{l}%
$N(2200)$\\
$N(2200)$%
\end{tabular}
&
\begin{tabular}
[c]{l}%
$N(2190)$\\
$N(2220)$\\
$N(2250)$%
\end{tabular}
&  &  &  & \\\hline
$\mathbf{\bar{N}(2200)}$ & $\mathbf{\bar{N}(2220)}$ & \textbf{0.9} &  &  &
\\\hline
$\mathbf{N(2380)}$ &  &  & $\mathbf{\Delta(2380)}$ & $\mathbf{\Delta(2420)} $%
& \textbf{1.7}\\\hline
$\ N(2560)$ &  &  &  &  & \\\hline
$3N(2650)$ &  &  &  &  & \\\hline
$\overline{\mathbf{N}}\mathbf{(2628)}$ & $\mathbf{N(2600)}$ & \textbf{1.1} &
3$\Delta(2650)$ &  & \\\hline
6$N(2740)$ &  &  & 4$\Delta(2740)$ &  & \\\hline
\end{tabular}

\bigskip*{\small Evidences are fair, they are not listed in the Baryon Summary
Table \cite{particle}}.

\bigskip

\bigskip

\bigskip

\bigskip

\bigskip

\bigskip

\bigskip

\bigskip

\bigskip

\bigskip

\bigskip

\bigskip

\bigskip

\bigskip

\bigskip

\bigskip

\bigskip

\bigskip\newpage\ \ \ \ \ \qquad\ \ \ Table 3. Two Kinds of Strange Baryons
$\Lambda$ and $\Sigma$ ($S=-1$)%

\begin{tabular}
[c]{|l|l|l||l|l|l|}\hline
Theory & Experiment & $\frac{\Delta\text{M}}{\text{M}}\%$ & Theory &
Experiment & $\frac{\Delta\text{M}}{\text{M}}\%$\\\hline
$\mathbf{\Lambda(1120)}$ & $\mathbf{\Lambda(1116)}$ & \textbf{0.36} &
$\mathbf{\Sigma(1210)}$ & $\mathbf{\Sigma(1193)}$ & \textbf{1.4}\\\hline
$\Lambda(1300)$ &  &  & $\Sigma(1300)$ &  & \\\hline
$\Lambda(1400)$ & $\Lambda(1405)$ &  & $\Sigma(1400)$ & $\Sigma(1385)$ &
\\\hline
$\overline{\mathbf{\Lambda}}\mathbf{(1350)}$ & $\overline{\mathbf{\Lambda}%
}\mathbf{(1405)}$ & \textbf{4.6} & $\mathbf{\bar{\Sigma}(1350)}$ &
$\mathbf{\bar{\Sigma}(1385)}$ & \textbf{2.5}\\\hline%
\begin{tabular}
[c]{l}%
$\Lambda(1660)$\\
$\Lambda(1660)$\\
$\Lambda(1660)$%
\end{tabular}
&
\begin{tabular}
[c]{l}%
$\Lambda(1520)$\\
$\Lambda(1600)$\\
$\Lambda(1670)$\\
$\Lambda(1690)$%
\end{tabular}
&  &
\begin{tabular}
[c]{l}%
$\Sigma(1660)$\\
$\Sigma(1660)$\\
$\Sigma(1660)$%
\end{tabular}
&
\begin{tabular}
[c]{l}%
$\Sigma(1660)$\\
$\Sigma(1670)$\\
$\Sigma(1750)$\\
$\Sigma(1775)$%
\end{tabular}
& \\\hline
$\mathbf{\bar{\Lambda}(1660)}$ & $\mathbf{\bar{\Lambda}(1620)}$ & \textbf{2.5}%
& $\mathbf{\bar{\Sigma}(1660)}$ & $\mathbf{\bar{\Sigma}(1714)}$ &
\textbf{3.2}\\\hline%
\begin{tabular}
[c]{l}%
$\Lambda(1930)$\\
$\Lambda(1930)$\\
$\Lambda(1930)$\\
$\Lambda(1930)$\\
$\Lambda(1930)$%
\end{tabular}
&
\begin{tabular}
[c]{l}%
$\Lambda(1800)$\\
$\Lambda(1810)$\\
$\Lambda(1820)$\\
$\Lambda(1830)$\\
$\Lambda(1890)$%
\end{tabular}
&  &
\begin{tabular}
[c]{l}%
$\Sigma(1930)$\\
$\Sigma(1930)$\\
$\Sigma(1930)$\\
$\Sigma(1930)$\\
$\Sigma(1930)$%
\end{tabular}
&
\begin{tabular}
[c]{l}%
$\Sigma(1915)$\\
$\Sigma(1940)$%
\end{tabular}
& \\\hline
$\mathbf{\bar{\Lambda}(1930)}$ & $\mathbf{\bar{\Lambda}(1830)}$ & \textbf{5.5}%
& $\mathbf{\bar{\Sigma}(1930)}$ & $\mathbf{\bar{\Sigma}(1928)}$ &
\textbf{.10}\\\hline%
\begin{tabular}
[c]{l}%
$\Lambda(2020)$\\
$\Lambda(2020)$%
\end{tabular}
&
\begin{tabular}
[c]{l}%
$\Lambda(2100)$\\
$\Lambda(2110)$%
\end{tabular}
&  & $\mathbf{\Sigma(2020)}$ & $\mathbf{\Sigma(2030)}$ & \textbf{.50}\\\hline
$\mathbf{\bar{\Lambda}(2020)}$ & $\mathbf{\Lambda(2105)}$ & \textbf{4.1} &  &
& \\\hline
$%
\begin{tabular}
[c]{l}%
$\Lambda(2380)$\\
$\Lambda(2460)$%
\end{tabular}
$ & $\Lambda(2350)$ &  & $\Sigma(2380)$ &
\begin{tabular}
[c]{l}%
$\Sigma(2250)$\\
$\Sigma(2455)^{\ast}$%
\end{tabular}
& \\\hline
$\mathbf{\bar{\Lambda}(2420)}$ & $\mathbf{\bar{\Lambda}(2350)}$ & \textbf{3.0}%
& $\mathbf{\bar{\Sigma}(2380)}$ & $\mathbf{\bar{\Sigma}(2353)}$ &
\textbf{1.2}\\\hline
$\mathbf{\Lambda(2560)}$ & $\mathbf{\Lambda(2585)}^{\ast}$ & \textbf{1.0} &  &
& \\\hline
\textbf{8}$\mathbf{\Lambda(2650)}$ &  &  & \textbf{7}$\mathbf{\Sigma(2650)} $%
& $\mathbf{\Sigma(2620)}$ & \textbf{1.1}\\\hline
\textbf{6}$\mathbf{\Lambda(2740)}$ &  &  & \textbf{6}$\mathbf{\Sigma(2740)} $%
&  & \\\hline
&  &  &  &  & \\\hline
\end{tabular}
\ \ \ \ \ \ \ \ \ \ \ \ \ \ \ \ \ \ \ \ \ \ \ \ \ \ \ \ \ \ \ 

{\small \ }*Evidences of existence for these baryons are only fair, they are not

listed in the Baryon Summary Table \cite{particle}.

\bigskip

\bigskip\newpage

\qquad\ \ \ \ \ \ \ Table 4. The Baryons $\Xi$ and the Baryons $\Omega$%

\begin{tabular}
[c]{|l|l|l||l|l|l|}\hline
Theory & Experiment & $\frac{\Delta\text{M}}{\text{M}}\%$ & Theory &
Experiment & $\frac{\Delta\text{M}}{\text{M}}\%$\\\hline
\textbf{2}$\mathbf{\Xi(1300)}$ & $\mathbf{\Xi(1318)}$ & \textbf{1.4} &
$\mathbf{\Omega(1660)}$ & $\mathbf{\Omega(1672)}$ & 0\textbf{.7}\\\hline
\textbf{3}$\mathbf{\Xi(1480)}$ & $\mathbf{\Xi(1530)}$ & \textbf{3.3} &
$\Omega(2460)$ &
\begin{tabular}
[c]{l}%
$\Omega(2250)$\\
$\Omega(2380)^{\ast}$\\
$\Omega(2470)^{\ast}$%
\end{tabular}
& \\\hline
\textbf{3}$\mathbf{\Xi(1660)}$ & $\mathbf{\Xi(1690)}$ & \textbf{1.8} &
$\mathbf{\bar{\Omega}(2460)}$ & $\mathbf{\bar{\Omega}(2367)}$ &
\begin{tabular}
[c]{l}%
\textbf{3.9}%
\end{tabular}
\\\hline
\textbf{2}$\mathbf{\Xi(1840)}$ & $\mathbf{\Xi(1820)}$ & \textbf{1.1} &
$\mathbf{\Omega(3080)}$ &  & \\\hline
\textbf{\ \ }$\mathbf{\Xi(1930)}$ & $\mathbf{\Xi(1950)}$ & \textbf{1.1} &
$\mathbf{\Omega(3720)}$ &  & \\\hline
\textbf{3}$\mathbf{\Xi(2020)}$ & $\mathbf{\Xi(2030)}$ & \textbf{1.0} &
$\mathbf{\Omega(7220)}$ &  & \\\hline
\textbf{4}$\mathbf{\Xi(2200)}$ & $\mathbf{\Xi(2250)}^{\ast}$ & \textbf{0.5} &
&  & \\\hline
\textbf{2}$\mathbf{\Xi(2380)}$ & $\mathbf{\Xi(2370)}^{\ast}$ & \textbf{2.2} &
&  & \\\hline
3$\Xi(2560)$ &  &  &  &  & \\\hline
11$\Xi(2740)$ &  &  &  &  & \\\hline
\end{tabular}

\bigskip*Evidences of existence for these baryons are only fair, they are not

listed in the Baryon Summary Table \cite{particle}.

\newpage

{\small \bigskip}

\qquad\qquad Table 5. Charmed \ $\Lambda_{c}^{+}$\ and Bottom $\Lambda
_{b\text{ }}^{0}$ Baryons \ %

\begin{tabular}
[c]{|l|l|l||l|l|l|}\hline
Theory & Experiment & $\frac{\Delta\text{M}}{\text{M}}\%$ & Theory &
Experiment & $\frac{\Delta\text{M}}{\text{M}}\%$\\\hline
$\mathbf{\Lambda}_{c}^{+}\mathbf{(2280)}$ & $\mathbf{\Lambda}_{c}%
^{+}\mathbf{(2285)}$ & \textbf{0.22} & $\mathbf{\Lambda}_{b}^{0}%
\mathbf{(5540)}$ & $\mathbf{\Lambda}_{b}^{0}\mathbf{(5641)}$ & 3.6\\\hline
$%
\begin{tabular}
[c]{l}%
$\Lambda_{c}^{+}(2450)$\\
$\Lambda_{c}^{+}(2540)$\\
$\Lambda_{c}^{+}(2660)$%
\end{tabular}
$ &
\begin{tabular}
[c]{l}%
$\Lambda_{c}^{+}(2593)$\\
$\Lambda_{c}^{+}(2625)$%
\end{tabular}
&  & $\mathbf{\Lambda}_{b}^{0}\mathbf{(9960)}$ &  & \\\hline
$\mathbf{\bar{\Lambda}}_{c}^{+}\mathbf{(2550)}$ & $\mathbf{\bar{\Lambda}}%
_{c}^{+}\mathbf{(2609)}$ & \textbf{2.3} &  &  & \\\hline
$\Lambda_{c}^{+}(2970)$ &  &  &  &  & \\\hline
$\Lambda_{c}^{+}(6600)$ &  &  &  &  & \\\hline
$\Lambda_{c}^{+}(13800)$ &  &  &  &  & \\\hline
\end{tabular}

{\small \bigskip}

\bigskip

\newpage

\qquad\qquad Table 6. Charmed Strange Baryon $\Xi_{c}$, $\Sigma_{c}$ and
$\Omega_{C}$ \ %

\begin{tabular}
[c]{|l|l|l||l|l|l|}\hline
Theory & Experiment & $\frac{\Delta\text{M}}{\text{M}}\%$ & Theory &
Experiment & $\frac{\Delta\text{M}}{\text{M}}\%$\\\hline
$\Xi_{c}(2550)$ &
\begin{tabular}
[c]{l}%
$\Xi_{c}(2468)$\\
$\Xi_{c}(2645)$%
\end{tabular}
&  &
\begin{tabular}
[c]{l}%
$\Sigma_{c}(2280)$\\
$\Sigma_{c}(2450)$\\
$\Sigma_{c}(2540)$%
\end{tabular}
& $%
\begin{tabular}
[c]{l}%
$\Sigma_{c}(2455)$\\
$\Sigma_{c}(2530)^{\ast}$%
\end{tabular}
$ & \\\hline
$\mathbf{\bar{\Xi}}_{c}\mathbf{(2550)}$ & $\mathbf{\bar{\Xi}}_{c}%
\mathbf{(2557)}$ & \textbf{0.3} & $\mathbf{\bar{\Sigma}}_{c}\mathbf{(2423)}$ &
$\mathbf{\bar{\Sigma}}_{c}\mathbf{(2493)}$ & \textbf{2.8}\\\hline
$\Xi_{C}(3170)$ &  &  & $\Sigma_{c}(2970)$ &  & \\\hline
&  &  &  &  & \\\hline
&  &  & $\mathbf{\Omega}_{C}\mathbf{(2660)}$ & $\mathbf{\Omega}_{C}%
\mathbf{(2704)}$ & \textbf{1.6}\\\hline
&  &  & $\Omega_{C}(3480)$ &  & \\\hline
\end{tabular}
\ \ \ 

*Evidences of existence for these baryons are only fair, they are not

listed in the Baryon Summary Table \cite{particle}.

\bigskip

\bigskip\newpage

{\LARGE Appendix A}

For the three symmetry axes which are on the surface of the first Brillouin
zone (the axis $D(P-N)$, the axis $F(P-H)$, the axis $G(M-N)$), the energy
bands in the same degeneracy group may have asymmetric $\vec{n}$ values (see
Fig. 3(b), 4(a) and 4(b)). This may indicate that they belong to different
Brillouin zones. In such cases, even if a sub-degeneracy $d_{sub}\leqslant R$,
it may still be divided further. Finding the criteria for dividing the
degeneracy depends on the structures of the Brillouin zones, the irreducible
representations of the single and double point groups \cite{DoubleGroup}, and
requires a higher order approximation. Hence, it is beyond the scope of this
paper. In order to simplify, we assume: (a) The degeneracy of the energy bands
which are in the first and second Brillouin zones will be divided. (b) If the
energy fluctuation $\Delta\varepsilon\neq0$, an asymmetric sub-degeneracy
should be divided at the point $N$ (low symmetry point, only has $8$ symmetric
operations \cite{SymmetryOp}); at the end point $P$ ($24$ symmetric operations
\cite{SymmetryOp}), may or may not be divided with a possibility of $\ 50\%$;
but should not be divided at the end points $\Gamma,$ $H$ and $M$ (high
symmetry points, $48$ symmetric operations \cite{SymmetryOp}):
\begin{align}
&  \text{if }\Delta\varepsilon\neq0\text{, divided at the end point }%
N\text{;}\nonumber\\
&  \text{not divided at end points }\Gamma,\text{ }H\text{ and }%
M\text{;}\nonumber\\
&  \text{divided at end point }P\text{ with possibility of 50\%.}%
\label{dividing}%
\end{align}
However, if $\Delta\varepsilon=0,$ the asymmetric degeneracy should not be divided.\newpage

{\LARGE Appendix B }\ \ \ \ \ \ \ \ \ \ \ \ \ \ \ \ \ \ \ \ \ \ \ \ \ \ \ \ 

\subsection{ The axis $D(P-N)$\ \ \ \ \ \ \ \ \ \ \ \ \ \ \ \ \ \ \ }

From (\ref{D-S}),the axis D has $S=0$ . For low energy levels, there are $4$
fold degenerate energy bands and $2$ fold degenerate energy bands on the axis
(see Fig. 3(b)).{\LARGE \ \ \ \ \ \ }

\subsubsection{The 4 fold energy bands on the axis $D(P-H)$}

We can see that each $4$ fold degenerate energy band has 4 asymmetric $\vec
{n}$ values. They can be divided into two groups. Each of them has $2$
symmetric $\vec{n}$ values. \ Using (\ref{isomax}), for the two fold
degenerate energy bands, we get $I=1/2,$ S = 0, Q = 2/3, -1/3 from
(\ref{charge}). For the $4$ fold energy bands, we have
\begin{equation}%
\begin{array}
[c]{llll}%
\text{ \ \ \ \ \ }E & \text{ \ \ \ \ \ \ \ \ \ \ \ \ \ \ \ \ \ \ }\vec{n} &  &
q_{N}^{\ast}(m)\\
E_{N}=5/2 & (\text{1-10,-110,020,200}) &  & 1840\\
\text{ \ }\Delta S=0 & (\text{1-10,-110}) &  & q_{N}^{\ast}(1840)\\
\text{ \ }\Delta S=0 & (\text{020,200}) &  & q_{N}^{\ast}(1840)\\
E_{P}=11/4 & (\text{-101,0-11,211,121}) &  & 1930\\
\text{ \ }\Delta S=0 & (\text{-101,0-11}) &  & q_{N}^{\ast}(1930)\\
\text{ \ }\Delta S=0 & (\text{211,121}) &  & q_{N}^{\ast}(1930)\\
E_{N}=7/2 & (\text{12-1,21-1,-10-1,0-1-1}) &  & 2200\\
\text{ \ }\Delta S=0 & (\text{12-1,21-1}) &  & q_{N}^{\ast}(2200)\\
\text{ \ }\Delta S=0 & (\text{-10-1,0-1-1}) &  & q_{N}^{\ast}(2200)\\
E_{P}=19/4 & (\text{-112,1-12,202,022}) &  & 2650\\
\text{ \ }\Delta S=0 & (\text{-112,1-12}) &  & q_{N}^{\ast}(2650)\\
\text{ \ }\Delta S=0 & (\text{202,022}) &  & q_{N}^{\ast}(2650)\\
\ldots &  &  &
\end{array}
\label{Quark-D-4}%
\end{equation}
{\LARGE \ \ \ \ \ }

\subsubsection{The two fold energy bands on the axis $D(P-N)$}

See Fig. 3(b). There are symmetric and asymmetric $\vec{n}$ values in the $2$
fold energy bands. The $2$ fold energy bands with symmetric $\vec{n}$ values
represent the quark family $q_{N}^{\ast}$. However, the case of $2$ fold
energy bands with asymmetric $\vec{n}$ values is not so simple.

At $E_{N}=1/2$, J$_{N}=0$ there are two energy bands with asymmetric $\vec
{n}=(000,110)$. Since the two energy bands are in different Brillouin zones,
they will be divided into $2$ single bands. From (\textbf{\ref{Bands of
First}), we know that the energy band with n = (0, 0, 0)} represents q$_{N}%
$(940). The energy band with $\vec{n}=(110)$ belongs to the second Brillouin
zone, and it represents the excited quark q$_{S}(1120)$ with S = -1, B = 1/3,
I = 0, Q = -1/3, and m = 1120 from (\ref{Strange-Quark}).

At $E_{P}=11/4$, J$_{N}=1,$ $\vec{n}=(002,112),$ which are asymmetric $\vec
{n}$ values. Using (\ref{dividing}), the two energy bands may be divided with
a possibility of $50\%$. However, the fluctuation of energy $\Delta
\varepsilon=0$ from (\ref{fluc}), hence there is not enough energy of
fluctuation for the two bands to be divided. Thus, the two energy bands are
not be divided. They represent a quark family $q_{N}(1930)$.

There are two asymmetric 2 fold energy bands at $E_{N}=9/2$ ($\vec
{n}=(220,-1-10)$ and $\vec{n}=(11-2,00-2)$). Using (\ref{fluc}), the energy
fluctuation for the first 2 energy bands ($\vec{n}=(220,-1-10)$, $J_{N}=1$) is
$\Delta\varepsilon=0$. Hence, they cannot be divided. Similar to $q_{N}%
(1930)$, it represents a quark family $q_{N}(2560)$. However, the energy
fluctuation for the second 2 energy bands ($\vec{n}=(11-2,00-2)$, $J_{N}=2$)
is $\Delta\varepsilon=100(2-1)\Delta S=100\Delta S$. Using (\ref{dividing}),
the 2 fold bands should be divided into 2 single energy bands with $S=0+\Delta
S=\pm1$ from (\ref{strangeflu}). According to (\ref{charmed}), one of the 2
single energy bands has a charmed number $C=\Delta S=+1$ (with $I=0$ and
$Q=+1$), and it represents a Charmed quark $q_{C}^{\ast}$; the other has a
strange number $S=-1$ (with $I=0$ and $Q=0$), and it represents a quark
q$_{S}$. Thus, the second 2 energy bands, after the division, represent
$q_{C}^{\ast}(2660)$ and $q_{S}(2460)$.

Therefore, we have%

\begin{equation}%
\begin{array}
[c]{llll}%
E_{N}=1/2 & \vec{n}=(\text{000,110}) & \varepsilon^{(0)}=1120 & \\
\text{\ \ }J_{N}=0 & \vec{n}=(\text{000}) &  & q_{N}^{\ast}(940)\\
\text{ \ }\Delta\varepsilon=0 & \vec{n}=(\text{110}) & S=-1 & q_{S}^{\ast
}(1120)\\
E_{P}=3/4 & \vec{n}=(\text{101,011}) & \varepsilon^{(0)}=1210 & q_{N}^{\ast
}(1210)\\
E_{N}=3/2 & \vec{n}=(\text{10-1,01-1}) & \varepsilon^{(0)}=1480 & q_{N}^{\ast
}(1480)\\
E_{P}=11/4 & \vec{n}=(\text{002,112}) & \varepsilon^{(0)}=1930 & \\
\text{ \ }J_{P}=1 & \text{ \ }\Delta S=0 & \Delta\varepsilon=0 & q_{N}^{\ast
}(1930)\\
E_{N}=9/2 &  & \varepsilon^{(0)}=2560 & \\
\text{ \ }J_{N}=1 & \vec{n}=(\text{220,-1-10}) & \Delta\varepsilon=0 &
q_{N}^{\ast}(2560)\\
\text{ \ }J_{N}=2 & \vec{n}=(\text{11-2,00-2}) &  & \\
& \text{ \ }\Delta S=+1 & \Delta\varepsilon=+100 & q_{C}^{\ast}(2660)\\
& \text{ \ }\Delta S=-1 & \Delta\varepsilon=-100 & q_{S}^{\ast}(2460)\\
E_{P}=19/4 & \vec{n}=(\text{-121,2-11}) & \varepsilon^{(0)}=2650 & q_{N}%
^{\ast}(2650)\\
E_{N}=11/2 & \vec{n}=(\text{2-1-1,-12-1}) & \varepsilon^{(0)}=2920 &
q_{N}^{\ast}(2920)\\
\ldots &  &  &
\end{array}
\label{Quark-D-2}%
\end{equation}

\subsection{Axis $F(P-H)$ \ \ \ \ \ \ \ \ \ \ \ \ \ \ \ \ \ \ \ }

The axis $F(P-H)$ is a $3$ fold symmetry axis, from (\ref{F-S}), $S=-1$. For
low energy levels, there are $6$ fold energy bands, $3$ fold energy bands, and
single energy bands on the axis (see Fig. 4(a)).

\subsubsection{The single energy bands on the axis $F(P-H)$}

For the single energy band, the strange number $S=-1$, and $I=0$ from
(\ref{isomax}) and Q = - 1/3 from (\ref{charge}). Each single energy band
represents an excited quark q$_{s}^{\ast}$ with S = -1, I = 0, Q = -1/3.
\begin{equation}%
\begin{array}
[c]{lllllll}%
\text{E}_{P}\text{=3/4} & \vec{n}\text{=(110)} & \varepsilon^{(0)}\text{=1210}%
& \text{S=-1} & \text{I=0} & \text{Q=-1/3} & \text{q}_{S}^{\ast}%
\text{(1210)}\\
\text{E}_{H}\text{=3} & \vec{n}\text{=(-1-12)} & \varepsilon^{(0)}\text{=2020}%
& \text{S=-1} & \text{I=0} & \text{Q=-1/3} & \text{q}_{S}^{\ast}%
\text{(2020)}\\
\ldots &  &  &  &  &  &
\end{array}
\end{equation}

\subsubsection{The three fold energy bands on the axis $F(P-H)$}

See Fig. 4(a). Using (\ref{dividing}), we know that the 3 fold degenerate
energy bands (all have asymmetric $\vec{n}$ values) will not be divided at the
point $H$, but may be divided at the point $P$ with $a$ possibility of $50\%$.
The division at point $P$ will result in a single band representing a quark
q$_{S}^{\ast}$, and a 2 fold energy band (with symmetric $\vec{n}$ values)
representing a quark family q$_{N}^{\ast}$ ($\Delta S=+1$) or q$_{\Xi}^{\ast}%
$($\Delta S=-1$) .

At $E_{P}=3/4$, $\vec{n}=(000,101,011)$. They will be divided into a single
band with $\vec{n}=(000)$ (in the first Brillouin zone) and a 2-fold energy
band with $\vec{n}=(101,011)$ (in the second Brillouin zone). From
(\textbf{\ref{Bands of First}), the band with n = (0, 0, 0) } represents
q$_{N}^{\ast}$(940).\textbf{\ } The 2-fold energy band with $\vec
{n}=(101,011)$ represents a quark family q$_{N}^{\ast}(1210)$ from
(\ref{Quark-D-2})

At $E_{P}=11/4$, $\vec{n}=(112,1-10,-110),$ $J_{P}=1\rightarrow\Delta
\varepsilon=0$. Since $\Delta\varepsilon=0,$ the three energy bands will not
be divided (see (\ref{dividing})). Thus, the energy bands represent a quark
family q$_{\Sigma}^{\ast}(1930)$

At $E_{P}=19/4$, $J_{P}=2\rightarrow\Delta\varepsilon=-100\Delta S,$ the three
energy bands with $\vec{n}=($220,21-1,12-1$)$ may be divided with $a$
possibility of $50\%$ (see (\ref{dividing})). Hence the energy bands may
represent q$_{\Sigma}(2650)$ (if not divided), or $q_{s}(2650)$ and
$q_{N}(2550)$ or $q_{\Xi}(2750)$ (if divided).

We have%

\begin{equation}%
\begin{array}
[c]{llll}%
E_{P}=3/4 & \vec{n}=(\text{000,011,101}) & \varepsilon^{(0)}=1210 & \\
\text{ \ }J_{P}=0 & \vec{n}=(\text{000}) & \Delta S=0 & q_{N}^{\ast}(940)\\
\text{ \ }\Delta\varepsilon=0 & \vec{n}=(\text{011,101}) & \Delta S=+1 &
q_{N}^{\ast}(1210)\\
E_{H}=1 & \vec{n}=(\text{002,-101,0-11}) & \varepsilon^{(0)}=1300 & q_{\Sigma
}^{\ast}(1300)\\
E_{P}=11/4 & \vec{n}=(\text{112,1-10,-110}) & \varepsilon^{(0)}=1930 &
q_{\Sigma}^{\ast}(1930)\\
\text{ \ }J_{P}=1 & \Delta\varepsilon=0 &  & \\
E_{H}=3 & \vec{n}=(\text{-1-10,112,1-12}) & \varepsilon^{(0)}=2020 &
q_{\Sigma}^{\ast}(2020)\\
E_{P}=19/4 & \vec{n}=(\text{220,21-1,12-1}) & \varepsilon^{(0)}=2650 &
q_{\Sigma}^{\ast}(2650)\\
\text{\ \ }J_{P}=2 & \vec{n}=(\text{220}) &  & q_{s}^{\ast}(2650)\\
\text{ \ }\Delta\varepsilon=-100 & \vec{n}=(\text{21-1,12-1}) & \Delta S=+1 &
q_{N}^{\ast}(2550)\\
&  & \Delta S=-1 & q_{\Xi}^{\ast}(2750)\\
... &  &  & \\
&  &  & \\
&  &  &
\end{array}
\end{equation}

\subsubsection{The six fold energy bands ($d=6$) on the axis $F(P-H)$}

The $6$ fold energy bands on axis $F$ are a special case. The $6$ asymmetric
$\vec{n}$ values consist of three groups, each of them has $2$ symmetric
$\vec{n}$ values. However, since the axis $F$ has a rotary $R=3$, so there are
two ways to divide the energy bands: A) dividing the energy bands according to
(\ref{degeneracy}) and (\ref{subdegen}); B) dividing the energy bands
according to the symmetry of $\vec{n}$ values.

(A). From (\ref{degeneracy}) and (\ref{subdegen}), each $6$ fold energy band
shall be divided into two $3$ fold energy bands first. They will represent two
quark families q$_{\Sigma}^{\ast}$ . Since the $3$ fold sub-degeneracy band
has asymmetric $\vec{n}$ values, according to (\ref{dividing}), it may be
divided (second division) further at point $P$ with $a$ possibility of $50$\%,
but not be divided at the point $H$. At the point $P$, in order to keep
(\ref{Strangebar}), both of the two $3$ fold sub-degeneracy bands will be
divided, resulting in two $2$ fold bands (one with $\Delta S=+1$, the other
with $\Delta S=-1$) and two single bands. According to (65), the $2$
fold energy band with $\Delta S=+1$ will keep $S$ unchanged while increasing
the Charmed number $C$ by $1$. They are the 2 fold quark excited state
$q_{\Xi_{C}}^{\ast}$ with B = 1/3, S = -1, C = +1, I = 1/2, and Q = 2/3, -1/3.

At $E_{P}=11/4$, $\vec{n}=($01-1,10-1,121,211,020,200$)$, $J_{P}=1$
($E_{P}=3/4,J_{P}=0$), $\Delta\varepsilon=0$ from (\ref{flub}). Since there is
not enough fluctuation energy to divide the two $3$ fold sub-degeneracies
(\ref{dividing}), they are not divided. Thus, the $6$ fold band represents
\begin{equation}
2\times q_{\Sigma}^{\ast}(1930)
\end{equation}

At $E_{P}=19/4$, $\vec{n}=($202,022,-121,2-11,0-1-1,-10-1$)$, $J_{P}=2$.
According to (\ref{flub}), the energy fluctuation $\ \Delta\varepsilon
=-100\times\Delta S$. \ Thus, the two $3$ fold bands represent (with a
possibility of $50$\% to be divided at point $P$)%

\begin{equation}
2\times q_{\Sigma}^{\ast}(2650)+[q_{\Xi_{C}}^{\ast}(2550)+q_{S}^{\ast
}(2650)]+[q_{\Xi}^{\ast}(2750)+q_{S}^{\ast}(2650)]
\end{equation}

It is very important to pay attention to the\textbf{\ baryon }$\Xi_{C}(2550)
$\cite{Kesi-C} \textbf{born here, on the 6 fold energy band, after the second
division from the fluctuation } $\Delta S=+1$ and $\Delta\varepsilon=-100$ Mev.

At $E_{H}=5$, $\vec{n}=($0-20,-200,-211,1-21,013,103$)$. According to
(\ref{dividing}), the two $3$ fold sub-degeneracies are not divided. Thus, the
$6$ fold band represents%

\begin{equation}
2\times q_{\Sigma}^{\ast}(2740)
\end{equation}

At $E_{H}=5$, $\vec{n}=($0-22,-202,-2-11,-1-21,0-13,-103$)$. Similarly, we have%

\begin{equation}
2\times q_{\Sigma}^{\ast}(2740)
\end{equation}

At $E_{P}=27/4,\vec{n}=($-12-1,2-1-1,301,031,222,00-2$),J_{P}=3$. Similar to
the case of $E_{P}=19/4$, $\Delta\varepsilon=-200\times\Delta S$ and the $6$
fold band represents (with $a$ possibility of $50$\% to be divided twice at
the point $P$)
\begin{equation}
2\times q_{\Sigma}^{\ast}(3370)+[q_{\Xi_{C}}^{\ast}(3170)+q_{S}^{\ast
}(3370)]+[q_{\Xi}^{\ast}(3570)+q_{S}^{\ast}(3370)]
\end{equation}
\qquad\qquad\qquad\qquad\newline 

\ \ (B). According to the symmetry values of $\vec{n}$, each $6$ fold
degeneracy can be divided into a $2$ fold sub-degeneracy band and a $4$ fold
sub-degeneracy band. For the division, we get $S=\bar{S}+\Delta S=-1\pm1$ from
(\ref{strangeflu}). To keep (\ref{Strangebar}), the $2$ fold energy\ band may
have $\Delta S=-1$, while the $4$ fold band will have $\Delta S=+1(S=0)$.
(Another possibility is that the $2$ fold energy\ band has $\Delta S=+1(S=0)$,
while the $4$ fold band has $\Delta S=-1$ ($S=-2)$. However, the possibility
of obtaining a $4$ fold band with strange number $S=-2$ is not supported by
experimental results. Hence, it will be ignored here).

Since the symmetric axis is a $3$ rotary one, the $4$ fold energy band will be
divided further. In order to balance the $3$ rotary symmetry of the axis and
the $2$ fold symmetry of the $\vec{n}$ values, the second division should
\textbf{keep the 3 rotary symmetry} (\textbf{may break the }$\mathbf{2}%
$\textbf{\ fold symmetry of }$\vec{n}$ ) because the first division has kept
the $2$ fold symmetry of the $\vec{n}$ values.\ Thus, the $4$ fold energy band
( $S=0$ ) may be divided into ($q_{\Lambda_{C}}^{\ast}+q_{\Sigma}^{\ast}$) or
($q_{S}^{\ast}+q_{\Sigma_{C}}^{\ast}$)\ ( if $\Delta\varepsilon\neq0$ ). But
if $\Delta\varepsilon=0,$ it will be divided into two 2 fold bands (S =
0)$\rightarrow2$ $q_{N}^{\ast}$.

At $E_{P}=11/4$, $\vec{n}=($01-1,10-1,121,211,020,200$)$, $J_{P}=1$, and
$\Delta\varepsilon=0$ from (\ref{flub}).\ Since $\Delta\varepsilon=0$, the $4$
fold energy band will be divided into two 2 fold (S = 0)$\rightarrow2\times
q_{N}^{\ast}(1930)$. We get
\begin{equation}
q_{\Xi}^{\ast}(1930)\text{, 2}\times\text{ }q_{N}^{\ast}(1930)\text{.}%
\end{equation}

At $E_{P}=19/4$, $\vec{n}=($0-1-1,-10-1,-121,2-11,202,022$)$, $J_{P}=2$, the
energy fluctuation $\ \Delta\varepsilon=-100(2-1)\Delta S$ $=-100\times\Delta
S$. After the first division, we have q$_{\Xi}^{\ast}(2750)$, and q$_{\Delta
}^{\ast}(2550)$. For q$_{\Delta}^{\ast}(2550)$, using (\ref{strangeflu}), we
get $S=0+\Delta S==\pm1$. From $\Delta S=\pm1,$ we have q$_{\Delta}^{\ast
}(2550)\rightarrow$\{[q$_{^{C}}^{\ast}(2450)$ ( $\Delta S=+1$ )$+q_{\Sigma
}^{\ast}$ $(2650)$ ($\Delta S=-1$ )] or [$q_{\Sigma_{C}}^{\ast}(2450)$ (
$\Delta C=$ $\ \Delta S=$ $+1$ (see (65)) $+q_{S}^{\ast}(2650)$
($S=-1$)]\}\ to keep the $3$ rotary symmetry. To sum up, the $6$ fold energy
band has the possibility to represent the quark families:
\begin{equation}
\text{ }q_{\Xi}^{\ast}(2750)\text{, }q_{\Sigma}^{\ast}\text{ }(2650)\text{,
}q_{C}^{\ast}(2450)\text{, }q_{\Sigma_{C}}^{\ast}(2450)\text{, }q_{S}^{\ast
}(2650)\text{.}%
\end{equation}

At $E_{H}=5$, $\vec{n}=($0-20, -200, -211, 1-21, 013, 103$)$, $J_{H}=1$ the
energy fluctuation$\ \Delta\varepsilon=0$. We have:
\begin{equation}
q_{\Xi}^{\ast}(2740)\text{, 2}\times\text{ }q_{N}^{\ast}(2740)\text{.}%
\end{equation}

Also at $E_{H}=5$, $\vec{n}$ = $($0-22,-202,-2-11,-1-21,0-13,-103$)$,
$J_{H}=2$, the energy fluctuation$\ \Delta\varepsilon=-100\times\Delta S$.
Similarly to $E_{H}=5$, we have:%

\begin{equation}
\text{ }q_{\Xi}^{\ast}(2840)\text{, }q_{\Sigma}^{\ast}(2740)\text{, }%
q_{C}^{\ast}(2540)\text{, }q_{\Sigma_{C}}^{\ast}(2540)\text{, }q_{S}^{\ast
}(2740)\text{.}%
\end{equation}

At $E_{P}=27/4$, $\vec{n}=($-12-1,2-1-1,301,031,222,00-2$),J_{P}=3$. Similar
to the case of $E_{P}=19/4,$ we have
\begin{equation}
q_{\Xi}^{\ast}(3570),q_{\Sigma}^{\ast}(3700),q_{\Lambda_{C}}^{\ast
}(2970),q_{\Sigma_{C}}^{\ast}(2970),q_{S}^{\ast}(3370)
\end{equation}

\subsection{ The axis $G(M-N)$\ \ \ \ \ \ \ \ \ \ \ \ \ \ \ \ }

The axis $G(M-N)$ is a $2$ fold symmetry axis. From (\ref{G-S}), the strange
number $S=-2$. There are $6$, $4$, and $2$ fold energy bands on the axis (see
Fig. 4(b)).

\subsubsection{The two fold energy bands on the axis $G(M-N)$}

At $E_{N}=1/2$, $J_{N}=0,$ the two energy bands with asymmetric $\vec
{n}=(000,110)$ are in the first and second Brillouin zones, respectively. The
energy band with $\vec{n}=(110)$ represents $q_{S}^{\ast}(1120)$. Another band
with $\vec{n}=(000)$\textbf{\ represents }$q_{N}^{\ast}(940)$.

At $E_{M}=1$, there are two 2 fold energy bands. The 2 fold energy band with
symmetric values $\vec{n}=(101,10-1)$ represents a quark family, $q_{\Xi
}^{\ast}(1300),$ similar to (\ref{2 Fold on SIGMA})$.$ Another $2$ fold energy
band with asymmetric values $\vec{n}=(200,1-10)$ will not be divided (see
(\ref{dividing})), it represents a quark family $q_{\Xi}^{\ast}(1300)$ too.

At $E_{N}=5/2$, $J_{N}=1,$ $\Delta\varepsilon=0.$ Since $\Delta\varepsilon=0$
the 2 fold energy band with $\vec{n}=(020,110)$ should not be divided (see
(\ref{dividing})), it represents a quark family $q_{\Xi}^{\ast}(1840).$

At $E_{M}=5$, the $2$ fold energy band with asymmetric $\vec{n}=($310,2-20$)$
will not be divided (see (\ref{dividing})), it represents quark family
$q_{\Xi}^{\ast}(2740)$.

Thus, we have
\begin{equation}%
\begin{array}
[c]{llll}%
E_{N}=1/2 & \vec{n}=(\text{000,110}) & \varepsilon^{(0)}=1120 & \\
\text{\ \ \ }J_{N}=0 & \vec{n}=(\text{000}) & S=0 & q_{N}^{\ast}(940)\\
& \vec{n}=(\text{110}) & S=-1 & q_{S}^{\ast}(1120)\\
E_{M}=1 & \vec{n}=(\text{101,10-1}) & \varepsilon^{(0)}=1300 & q_{\Xi}^{\ast
}(1300)\\
& \vec{n}=(\text{200,1-10}) & \varepsilon^{(0)}=1300 & q_{\Xi}^{\ast}(1300)\\
E_{N}=3/2 & \vec{n}=(\text{011,01-1}) & \varepsilon^{(0)}=1480 & q_{\Xi}%
^{\ast}(1480)\\
E_{N}=5/2 & \vec{n}=(\text{020,-110}) & \varepsilon^{(0)}=1840 & q_{\Xi}%
^{\ast}(1840)\\
\text{ \ \ }J_{N}=1 & \text{ \ \ \ } & \Delta\varepsilon=0 & \\
E_{M}=3 & \vec{n}=(\text{2-11,2-1-1}) & \varepsilon^{(0)}=2020 & q_{\Xi}%
^{\ast}(2020)\\
E_{M}=5 & \vec{n}=(\text{310,2-20}) & \varepsilon^{(0)}=2740 & q_{\Xi}^{\ast
}(2740)\\
E_{N}=11/2 & \vec{n}=(\text{-121,-12-1}) & \varepsilon^{(0)}=2920 & q_{\Xi
}^{\ast}(2920)\\
\ldots &  &  &
\end{array}
\label{Quark-G-2}%
\end{equation}
{\LARGE \ \ \ }

\subsubsection{The four fold energy bands on the axis $G(M-N)$%
{\protect\LARGE \ \ }}

According to (\ref{degeneracy}) and (\ref{subdegen}), each $4$ fold energy
band will be divided into two $2$ fold energy bands which represent two quark
families $q_{\Xi}^{\ast}$ (S = -2, I = 1/2, Q = -1/3, -4/3) similar to (\ref{4
fold on Segma}):
\begin{equation}%
\begin{array}
[c]{lllll}%
E_{M}\text{=}3 & \vec{n}\text{=(0-11,0-1-1,} & \text{211,21-1)} &
\varepsilon^{(0)}=2020 & \\
\ \ \Delta S\text{=}0 & \vec{n}\text{=(0-11,0-1-1)} & \text{q}_{\Xi}^{\ast
}\text{(2020)} & \vec{n}\text{=(211,21-1)} & \text{q}_{\Xi}^{\ast
}\text{(2020)}\\
E_{N}\text{=}7/2 & \vec{n}\text{=(-101,-10-1,} & \text{121,12-1)} &
\varepsilon^{(0)}=2200 & \\
\ \ \Delta S\text{=}0 & \vec{n}\text{=(-101,-10-1)} & \text{q}_{\Xi}^{\ast
}\text{(2200)} & \vec{n}\text{=(121,12-1)} & \text{q}_{\Xi}^{\ast
}\text{(2200)}\\
E_{M}\text{=}5 & \vec{n}\text{=(301,30-1,} & \text{1-21,1-2-1)} &
\varepsilon^{(0)}=2740 & \\
\ \ \Delta S=0 & \vec{n}\text{=(301,30-1)} & \text{q}_{\Xi}^{\ast
}\text{(2740)} & \vec{n}\text{=(1-21,1-2-1)} & \text{q}_{\Xi}^{\ast
}\text{(2740)}\\
\ldots &  &  &  &
\end{array}
\label{Quark-G-4}%
\end{equation}

\subsubsection{The six fold energy bands on the axis $G(M-N)$}

According to (\ref{degeneracy}) and (\ref{subdegen}), each $6$ fold energy
band will be divided into three $2$ fold energy bands. One of the three $2$
fold sub-degeneracy energy bands has asymmetric $\vec{n}$ values. According to
(\ref{dividing}), the 2 energy bands should be divided further at the point
$N$, but should not be divided at the point $M$. Thus, at point $M$ , each $6$
fold energy band will represent three quark families $q_{\Xi}^{\ast}$ similar
to (\ref{4 fold on Segma}). At the point $N$, each $6$ fold energy band will
represent two $q_{\Xi}^{\ast}$ and two single energy bands .

At $E_{N}=9/2$, $\ \vec{n}=(112,11-2,002,00-2,220,-1-10)$. First, the $6$ fold
energy band will be divided into: $\vec{n}=(112,11-2)$ ($q_{\Xi}^{\ast}%
(2560)$), $\vec{n}=(002,00-2)$ ($q_{\Xi}^{\ast}(2560)$), $\vec{n}%
=(220,-1-10)$. Then the two energy bands with asymmetric $\vec{n}$ values
($\vec{n}=(220,-1-10)$) will be further divided. The fluctuation of energy
associated with this division is $\Delta\varepsilon=100\Delta S$ Mev.
($J_{N}=2$ at $E_{N}=9/2$ for 2-fold asymmetric $\vec{n}$ values, $J_{N}=0$ at
$E_{N}=1/2$ and $J_{N}=1$ at $E_{N}=5/2$ (see (\ref{Quark-G-2})). The
fluctuation of strange number associated with this division is $\Delta
S=\pm1.$ Since this is the second division of the $6$ fold energy band,
according to \textbf{Hypothesis IV. 6, (65), }for $\Delta
S=+1$\textbf{\ }we can get a quark excited states $q_{\Omega_{C}}^{\ast
}(2660)$ with charmed number $C=+1$ while keeping the strange number $S=-2$
unchanged. Hence the $2 $ energy bands will be divided into:
\begin{equation}
\Delta C=\Delta S=+1\text{, \ }q_{\Omega_{C}}^{\ast}(2660)\text{;
\ \ \ \ }\Delta S=-1\text{, \ }q_{\Omega}^{\ast}(2460)\text{.}\label{G_6_C}%
\end{equation}
It is very important to pay attention to \textbf{the quark }$q_{\Omega_{C}%
}^{\ast}$(\textbf{2660}) \cite{OMIGA-C} \textbf{born here}, on the $6$ fold
energy band of the axis G, after the second division with fluctuations $\Delta
S=+1$ and $\Delta\varepsilon=+100$ Mev.

At $E_{M}=5$, $\vec{n}=(202,20-2,1-12,1-1-2,310,0-20)$ will be divided into
three $2$ fold energy bands, representing 3 quark families q$_{\Xi}^{\ast
}(2740)$. The 2 fold band with asymmetric $\vec{n}=(310,0-20)$ will not be
divided further at the point $M$ from (\ref{dividing}). Thus, it represents a
quark family q$_{\Xi}^{\ast}(2740).$

At $E_{N}=13/2$, similar to $E_{N}=9/2$, the $6$ fold degeneracy will be
divided into two 2 fold degeneracies (two quark families $q_{\Xi}^{\ast
}(2920)),$ and 2 single bands. The second division (the 2 single bands) will
result in a quark $q_{\Omega_{C}}^{\ast}(3220)$ and a quark $q_{\Omega}^{\ast
}(2620)$ ($J_{N}=3$ and $\Delta\varepsilon=100(3-1)\Delta S=200\Delta S$ Mev).

We have%

\begin{equation}%
\begin{array}
[c]{llll}%
\begin{array}
[c]{l}%
E_{N}=9/2
\end{array}
&
\begin{array}
[c]{c}%
\vec{n}=\text{(112,11-2,002,}\\
\text{ \ \ \ \ \ \ 00-2,220,-1-10)}%
\end{array}
&
\begin{array}
[c]{l}%
\varepsilon^{(0)}=2560
\end{array}
& \\
\text{ \ \ \ \ }J_{N}=2 & \vec{n}=\text{(112,11-2) \ }q_{\Xi}^{\ast}(2560) &
\vec{n}=\text{(002,00-2)} & \text{\ }q_{\Xi}^{\ast}(2560)\\
\text{ \ (divided)} & \text{single band \ \ \ \ \ }q_{\Omega_{C}}^{\ast
}(2660) & \text{single band} & q_{\Omega}^{\ast}(2460)\\%
\begin{array}
[c]{l}%
E_{M}=5
\end{array}
&
\begin{array}
[c]{c}%
\vec{n}=\text{(202,20-2,1-12,}\\
\text{ \ \ \ \ \ \ 1-1-2,310,0-20)}%
\end{array}
&
\begin{array}
[c]{l}%
\varepsilon^{(0)}=2740
\end{array}
& 3\text{ }q_{\Xi}^{\ast}(2740)\\%
\begin{array}
[c]{l}%
E_{N}=13/2
\end{array}
&
\begin{array}
[c]{c}%
\vec{n}=\text{(-112,-11-2,022,}\\
\text{ \ \ \ \ \ \ 02-2,130,-200)}%
\end{array}
&
\begin{array}
[c]{l}%
\varepsilon^{(0)}=3280
\end{array}
& \\
\text{ \ \ \ \ }J_{N}=3 & \vec{n}=\text{(-112,-11-2) \ \ }q_{\Xi}^{\ast
}(3280) & \vec{n}=\text{(022,02-2)} & q_{\Xi}^{\ast}(3280)\\
\text{ \ (divided)} & \text{single band \ \ \ \ \ \ \ \ }q_{\Omega_{C}}^{\ast
}(3480) & \text{single band} & q_{\Omega}^{\ast}(3080)\\
\ldots & \text{ \ \ \ \ \ \ \ \ \ \ \ \ \ \ \ \ \ \ \ \ } & \text{
\ \ \ \ \ \ \ \ \ \ \ \ \ \ \ \ \ } & \text{
\ \ \ \ \ \ \ \ \ \ \ \ \ \ \ \ \ }%
\end{array}
\end{equation}

\ \ \ \ \ \ \ \ \ \ \ \ \ \ \ \ \ 
\end{document}